\documentclass[preprint]{aastex62}
\usepackage{graphicx}
\usepackage{natbib}
\usepackage{amsmath}
\usepackage{bm}
\usepackage{gensymb}
\usepackage{float}
\usepackage{lineno}
\usepackage{xcolor}
\begin{document}

\title{\bf Photometric analyses of Saturn's small moons: \\  Aegaeon, Methone and Pallene are dark; Helene and Calypso are bright.}

\author{M. M. Hedman}
\affiliation{Department of Physics, University of Idaho, Moscow, ID 83844}
\author{P. Helfenstein}
\affiliation{Cornell Center for Astrophysics and Planetary Science, Cornell University, Ithaca NY 14853}
\author{R. O. Chancia}
\affiliation{Department of Physics, University of Idaho, Moscow, ID 83844}
\affiliation{Center for Imaging Science, Rochester Institute of Technology, Rochester NY 14623}
\author{P. Thomas}
\affiliation{Cornell Center for Astrophysics and Planetary Science, Cornell University, Ithaca NY 14853}
\author{E. Roussos}
\affiliation{Max Planck Institute for Solar System Research, G\"ottingen, Germany 37077}
\author{C. Paranicas}
\affiliation{APL, John Hopkins University, Laurel MD 20723}
\author{A. J. Verbiscer}
\affiliation{Department of Astronomy, University of Virginia, Charlottesville, VA 22904}

\begin{abstract}
We examine the surface brightnesses of Saturn's smaller satellites using a photometric model that explicitly accounts for their elongated shapes and thus facilitates comparisons among different moons. Analyses of Cassini imaging  data with this model reveals that the moons Aegaeon, Methone and Pallene are darker than one would expect given trends previously observed among the nearby mid-sized satellites. On the other hand, the trojan moons Calypso and Helene have substantially brighter surfaces than their co-orbital companions Tethys and Dione. These observations are inconsistent with the moons' surface brightnesses being entirely controlled by the local flux of E-ring particles, and therefore strongly imply that other phenomena are affecting their surface properties. The darkness of Aegaeon, Methone and Pallene is correlated with the fluxes of high-energy protons, implying that high-energy radiation is responsible for darkening these small moons. Meanwhile, Prometheus and Pandora appear to be brightened by their interactions with nearby dusty F ring, implying that enhanced dust fluxes are most likely responsible for Calypso's and Helene's excess brightness. However, there are no obvious structures in the E ring  that would preferentially brighten these two moons,  so there must either be something subtle in the E-ring particles' orbital properties that leads to asymmetries in the relevant fluxes, or something happened recently to  temporarily increase these moons' brightnesses. 
\end{abstract}

\section{Introduction}

{Saturn's regular moons exhibit a very diverse range of spectral and photometric properties, with Enceladus having an extremely bright and ice-rich surface, while the leading side of Iapetus is exceptionally dark due to being covered with a layer of debris derived from Saturn's irregular satellites  \citep[for recent reviews, see][]{Hendrix18, Thomas18, Verbiscer18}. For the mid-sized moons orbiting between Titan and the rings (i.e. Mimas, Enceladus, Tethys, Dione and Rhea), many of the observed spectral and photometric  trends have been attributed to interactions with the E ring. This broad and diffuse ring consists of dust-sized, ice-rich particles generated by Enceladus' geological activity that impact the surfaces of different moons at different rates. The brightness and density of the E ring is strongly correlated with the geometric albedos of these satellites at visible \citep{Verbiscer07} and radio wavelengths \citep{Ostro10, LeGall19}, as well as  spectral features like the depth of water-ice absorption bands \citep{Filacchione12, Filacchione13}. Furthermore, there are brightness asymmetries between the leading and trailing sides of these moons that can be attributed to differences in the fluxes of E-ring particles \citep{Buratti98, Schenk11, Scipioni13, Scipioni14, Hendrix18, LeGall19}.}

However, there are several smaller moons orbiting within the E ring whose {surface scattering properties} deviate from the trends observed among the larger moons. On the one hand, four extremely small moons (Aegaeon, Anthe, Methone and Pallene) are found within the E-ring's inner flank. Aegaeon is situated between Janus and Mimas in the G ring, while Methone, Anthe and Pallene orbit between Mimas and Enceladus. One might therefore expect these moons to follow the same trend as Janus, Mimas and Enceladus, but in fact they appear to be somewhat darker than this trend would predict \citep{Thomas18, Verbiscer18}.  On the other hand, the small moons Telesto and Calypso share Tethys' orbit, while Helene and Polydeuces share Dione's orbit, but they do not all appear to have the same {spectrophotometric properties} as their larger companions \citep{Verbiscer07, Verbiscer18, Filacchione12, Filacchione13}.

A major challenge in interpreting existing brightness estimates of these objects is that many of them are significantly elongated, and so their brightness varies substantially depending on their orientation relative to the observer and the Sun, leading to large scatter in the disk-integrated brightness estimates at any given phase angle. Fortunately, all these objects are spin-locked and so their orientation relative to the Sun and  the spacecraft can be securely predicted. Furthermore, {almost all of these moons were observed at high enough resolution to determine their shapes, and  Aegaeon, Methone and Pallene in particular are very close to perfect ellipsoids.} It should therefore be possible to model their shape-related brightness variations to a fair degree of accuracy, and thereby derive estimates of their surface brightnesses that can be compared to those of the larger moons.

This paper describes a new investigation of Saturn's moons that uses a photometric model to obtain precise and comparable measurements of the surface brightnesses for both the small and mid-sized moons orbiting interior to Titan. The brightness estimates for the mid-sized moons, along with the smaller moons Janus and Epimetheus, are consistent with previous studies in that they are clearly correlated with the local flux of E-ring particles. However, this work also confirms that many small moons deviate from this basic trend, implying that other processes influence the moons' surface brightnesses. Specifically, we find that (1)  Aegaeon is exceptionally dark,  (2) Methone and Pallene are darker than one would expect given their locations between Mimas and Enceladus, (3) Calypso and Helene are brighter than their larger companions, and (4) Prometheus and Pandora are brighter than moons orbiting nearby like Atlas, Janus and Epimetheus. Aegaeon, Methone, and Pallene are most likely dark because they occupy belts of high energy proton fluxes. While the mechanism by which this radiation darkens their surfaces is still obscure, we find that the overall brightnesses of these moons are probably determined by the ratio of the energetic proton flux to the E-ring particle flux. On the other hand, the excess brightness of Helene, Calypso, Prometheus and Pandora is most likely due to some localized increase in the particle flux. For Prometheus and Pandora, the additional flux of particles from the F ring  is probably responsible for increasing their brightness. However, there is no obvious particle source that would preferentially brighten Calypso and Helene more than their co-orbital companions, so the brightness of these moons either involves a previously-unknown asymmetry in the E-ring particle flux or is a transient phenomenon due to a recent event like an impact.

Section~\ref{data} below describes the Cassini imaging data used in this study, and how it is transformed into estimates of the disk-integrated brightness of Saturn's various moons. Section~\ref{model} then describes the photometric model we use to convert these disk-integrated brightness estimates into estimates of the moons' surface brightness while accounting for the moons' variably elongated shapes. Finally, Section~\ref{trends} describes the trends among these shape-corrected brightness estimates and their implications, while Section~\ref{summary} summarizes our findings. 

\section{Observations and Preliminary Data Reduction}
\label{data}

The disk-integrated brightness estimates considered in this study are derived from images obtained by the Narrow Angle Camera (NAC) of the Imaging Science Subsystem (ISS) onboard the Cassini Spacecraft \citep{Porco04}. These images were all calibrated using version 3.9 of the Cisscal package  to remove dark current and instrumental electronic noise, apply flat-field corrections, and convert the raw data numbers to values of radiance factors $I/F$, a standard dimensionless measure of reflectance that is unity for an illuminated Lambertian surface viewed at normal incidence and emission angles \citep{Porco04, West10}. Here $I$ is the scattered intensity of light and ${\pi F}$ is the specific solar flux over the camera filter bandpass. 

While this analysis focuses on the photometry of the small moons, in order to facilitate comparisons with the mid-sized satellites we will consider data for all the satellites interior to Titan except for Daphnis (whose location within a narrow gap in the main rings would have required specialized algorithms). We searched for images containing all moons obtained by the NAC through its clear filters using the OPUS search tool on the PDS ring-moon system node ({\tt  https://pds-rings.seti.org/search}). For the small moons Aegaeon, Anthe, Methone and Pallene, as well as the trojan moons Telesto, Calypso, Helene and Polydueces we considered all images where the moon was observed at phase angles below 80$^\circ$ (at higher phase angles the signal-to-noise ratio for these moons was often too poor to be useful), while for Pan, Atlas, Prometheus, Pandora, Janus, Epimetheus, Mimas, Enceladus, Tethys, Dione and Rhea we only considered images where the moon was at phase angles between 20$^\circ$ and 40$^\circ$. This more restricted phase range corresponds to conditions with the best signal-to-noise ratio data for the smaller moons, and reduces the number of images that needed to be analyzed to a manageable level. 

Since nearly all of the images of the small moons were unresolved, this analysis will only consider disk-integrated brightness estimates for the moons, which were computed following approaches similar to those described in \citet{Hedman10}. This process begins by selecting a region within each image that contains the entire signal from the moon based on visual inspection. Then, instrumental and ring backgrounds are removed from this region using one of three different procedures depending on the moon:
\begin{itemize}
\item {\bf Aegaeon and Pan } For these moons the dominant backgrounds come from the nearby rings (the A ring for Pan and the G ring for Aegaeon). Each image was therefore geometrically navigated based on the positions of stars in the field of view and the appropriate SPICE kernels \citep{Acton96} {using a variant of the CAVIAR software package {\tt (https://www.imcce.fr/recherche/equipes/pegase/caviar)}.} We then used regions extending 10 pixels on either side of the selected region along the moon's orbit to determine the mean background image brightness as a function of ringplane radius. This profile was then interpolated onto the pixels in the selected region containing the moon and removed from that region.
\item{\bf  Atlas, Anthe, Methone, Pallene, Telesto, Calypso, Helene and Polydeuces.} For these moons,  instrumental backgrounds dominate,  and these are typically a stronger function of row than sample number \citep{West10}. Hence we used regions 10 columns wide on either side of the region containing the moon to define a background brightness level as a function of row number which was then removed from all the pixels in the selected regions.
\item{\bf Prometheus, Pandora, Janus, Epimetheus, Mimas, Enceladus, Tethys, Dione and Rhea.} These larger moons were often resolved and so the signal-to-noise ratio was much higher than for the smaller moons. Hence we used regions 10 columns wide on either side of the region containing the moon to define a mean background brightness level that was removed from all the pixels in the selected regions.
\end{itemize}

After removing the backgrounds, the total brightness of the object in each image was  computed and expressed in terms of an effective area, $ A_{\rm eff}$, which is the equivalent area of material with $ I/F = 1$ that would be required to account for the observed brightness:
\begin{equation}
A_{\rm eff} = \sum_{x} \sum_{y} I/F_{x,y}\times \Omega_{pixel}\times D^2
\label{aeff}
\end{equation}
where $x$ and $y$ are the row and column numbers of the pixels in the
selected region, $I/F_{x,y}$ is the (background-subtracted) brightness in the $x,y$ pixel, $\Omega_{pixel} =$ (6  $\mu$rad$)^2$ is the assumed solid angle subtended by a NAC pixel, and $D$ is the distance between the spacecraft and the object during the observation {(designated as ``Range'' in Tables~\ref{aegaeontab}-~\ref{methonecoltab} of Appendix C).} The assumed values for $D$ are derived from the appropriate SPICE kernels \citep{Acton96}.  This approach deviates from traditional integrated disk measurement convention in that an effective whole-disk area is measured at each phase angle rather than the magnitude equivalent of an average whole-disk reflectance. Two advantages of this approach for our study are that (1)  it requires no \textit{a priori} knowledge of the object's average size and (2)  it easily accommodates target blur in which subpixel-sized objects are smeared across several pixels due to spacecraft motion and/or long camera exposures.  We also estimate the statistical uncertainly on $A_{\rm eff}$ based on the standard deviation of the brightness levels in the second region after any radial trends have been removed. Note that this procedure underestimates the true uncertainty in the measurements, but is still a useful way to identify images with low signal-to-noise ratios. For the smaller objects, we also computed their mean position in the field of view by computing the coordinates (in pixels) of the streak’s center of light $x_c$ and $y_c$:
\begin{equation} x_c = \frac{\sum_x \sum_y x\times I/F_{x,y}}{\sum_x \sum_y I/F_{x,y}}
\end{equation}
\begin{equation} y_c = \frac{\sum_x \sum_y y\times I/F_{x,y}}{\sum_x \sum_y I/F_{x,y}} \end{equation}
These numbers are not used directly for any part of this study. However, they are useful for identifying images where the background levels were not removed properly, since in those images the computed center of light would fall outside the image of the moon.

For the larger moons, the signal-to-noise ratio for every image was sufficiently high that all unsaturated images yielded useful estimates of $A_{\rm eff}$. After excluding any moons with saturated pixels, we had the following numbers of data points for these moons: 40 for Prometheus, 40 for Pandora, 53 for Janus, 47 for Epimetheus, 76 for Mimas, 82 for Enceladus, 67 for Tethys, 67 for Dione and 74 for Rhea {(see Tables~\ref{prometheustab}-~\ref{rheatab} in Appendix C)}.  For the smaller moons, we visually inspected the regions containing the moons and excluded any images where these regions were obviously corrupted by bad pixels or cosmic rays, or where the computed center of light was noticeably displaced from the bright pixels containing the signal from the moon. For Aegaeon, Anthe, Methone and Pallene, we also excluded any images where $A_{\rm eff}$ was negative or more than 10 times the median value among all the images, as well as any  images where the brightest pixel in the relevant region was more than 10 times the median pixel brightness times the number of pixels in the region (this removed images contaminated by a bad pixel or cosmic ray that was not obvious upon visual inspection). Finally, we excluded any images of Aegaeon, Anthe, Methone and Pallene with exposures less than 0.5 seconds because these usually had poor signal-to-noise ratios, and any images of Anthe, Methone and Pallene with exposures longer than  2 seconds, where unresolved images could be saturated. Similarly, we excluded images of Polydeuces with exposures longer than 0.68 seconds and any images of Telesto, Calypso and Helene with exposures longer than 0.15 seconds. After these selections, the final number of measurements for the small moons were: 168 for Aegaeon, 167  for Anthe, 187 for Methone, 159 for Pallene, 41 for Pan, 71 for Atlas, 142 for Telesto, 157 for Calypso, 122 for Helene and 136 for Polydueces {(see Tables~\ref{aegaeontab}-~\ref{atlastab} in Appendix C)}. Estimates of $A_{\rm eff}$ for these images, along with relevant geometric parameters like the sub-solar and sub-observer latitudes and longitudes, are provided in {Tables~\ref{aegaeontab}-~\ref{rheatab} in Appendix C.}

\begin{figure*}
\resizebox{6.5in}{!}{\includegraphics{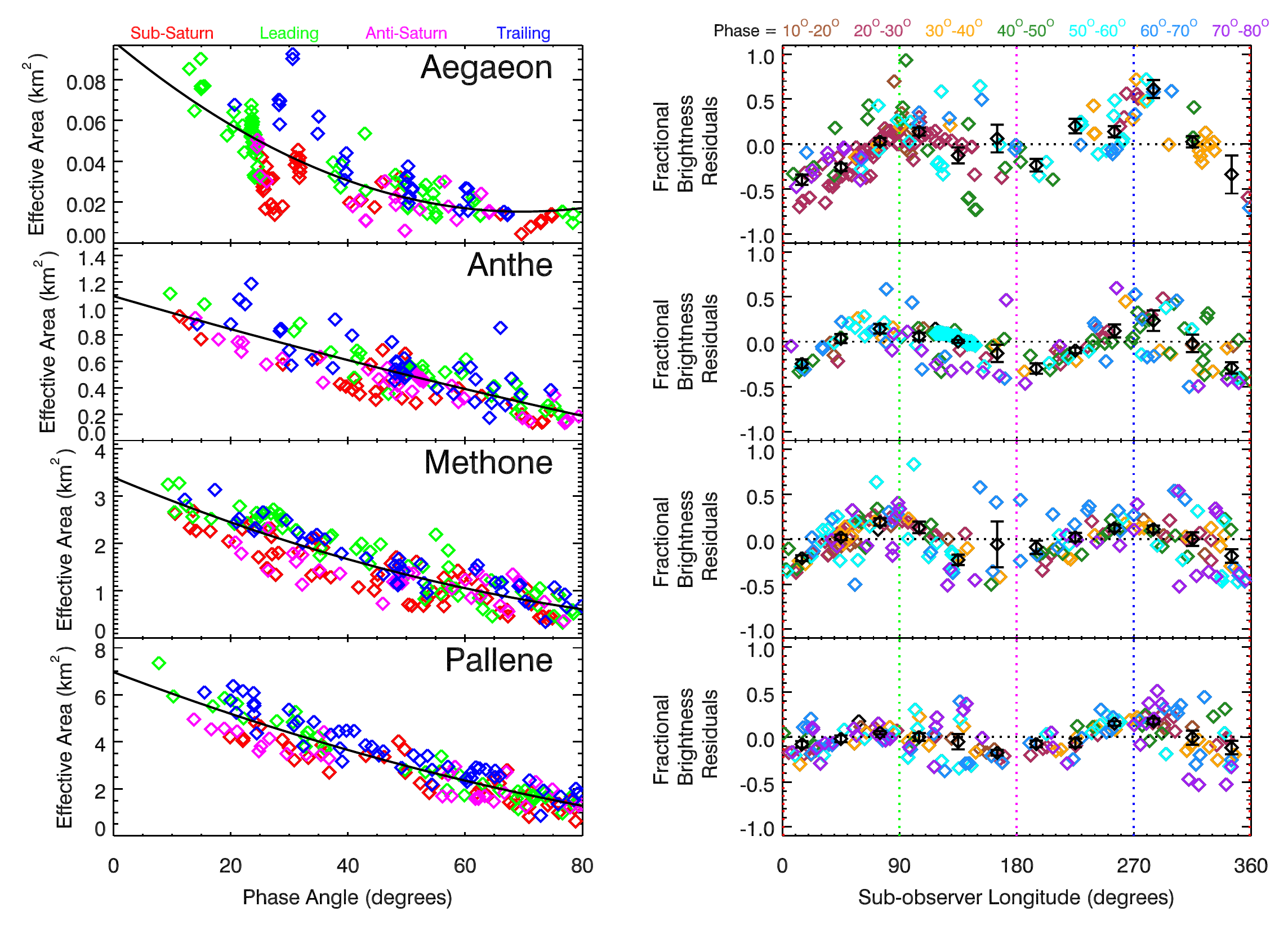}}
\caption{Summary of the brightness measurements for Aegaeon, Anthe, Methone and Pallene. The left-hand  plots show the effective areas $A_{\rm eff}$ of the moons as a function of phase angle, with the data points color coded by the quadrant viewed by the spacecraft. The right-hand plots show the fractional residuals of the brightness estimates relative to the quadratic trend shown as a solid line in the left-hand plots. Note that all these objects are consistently brighter when their leading or trailing sides are viewed than when their sub-Saturn or anti-Saturn sides are viewed.}
\label{phaserot}
\end{figure*}

\begin{figure*}
\resizebox{6.5in}{!}{\includegraphics{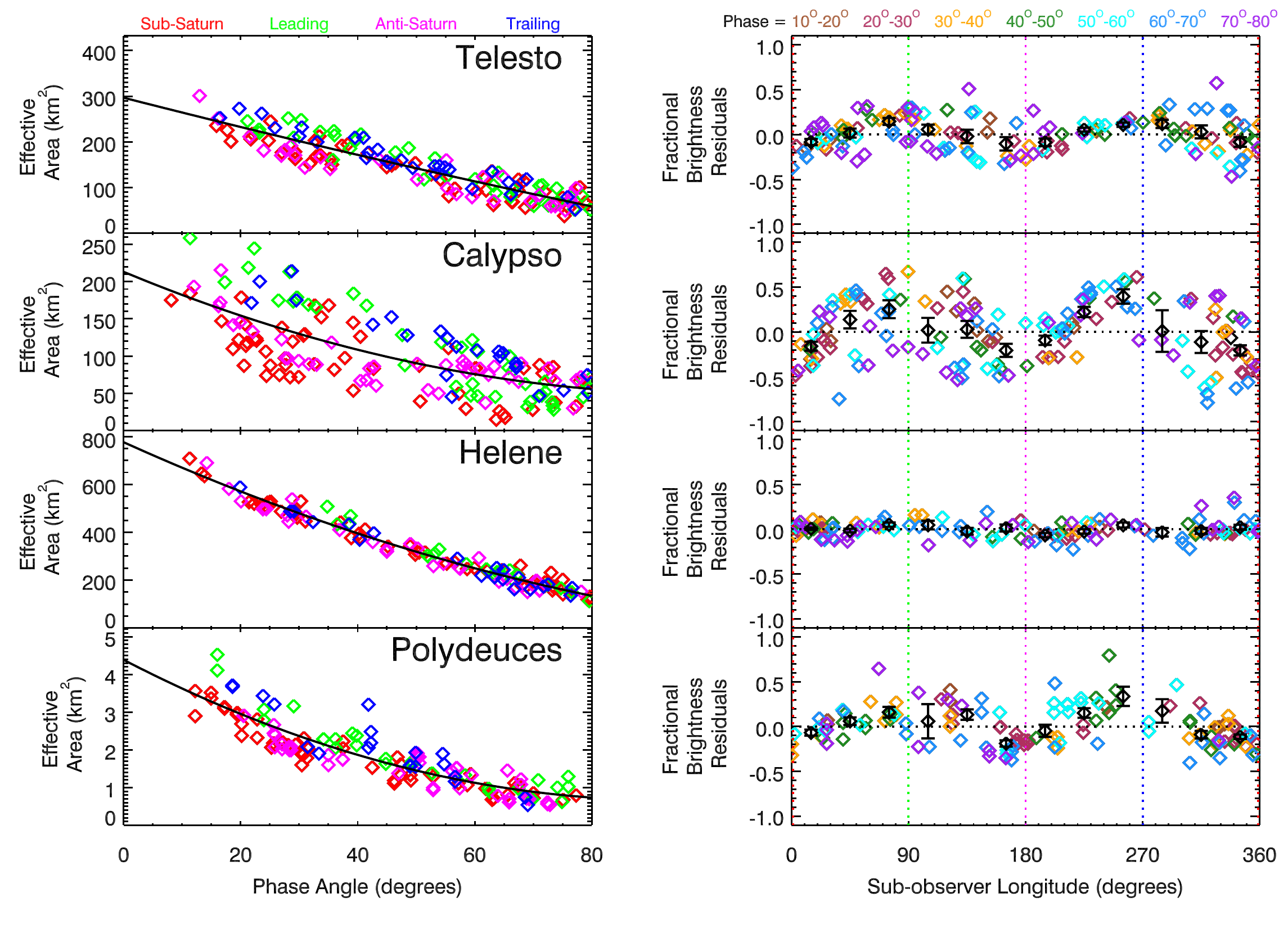}}
\caption{Summary of the brightness measurements for the trojan moons Telesto, Calypso, Helene and Polydeuces. The left-hand plots show the effective areas $A_{\rm eff}$ of the moons as a function of phase angle, with the data points color coded by the quadrant viewed by the spacecraft. The right-hand plots show the fractional residuals of the brightness estimates relative to the quadratic trend shown as a solid line in the left-hand plots. Note that these objects are consistently brighter when their leading or trailing sides are viewed than they are when their sub-Saturn or anti-Saturn sides are viewed.}
\label{tphaserot}
\end{figure*}

\begin{figure*}
\resizebox{6.5in}{!}{\includegraphics{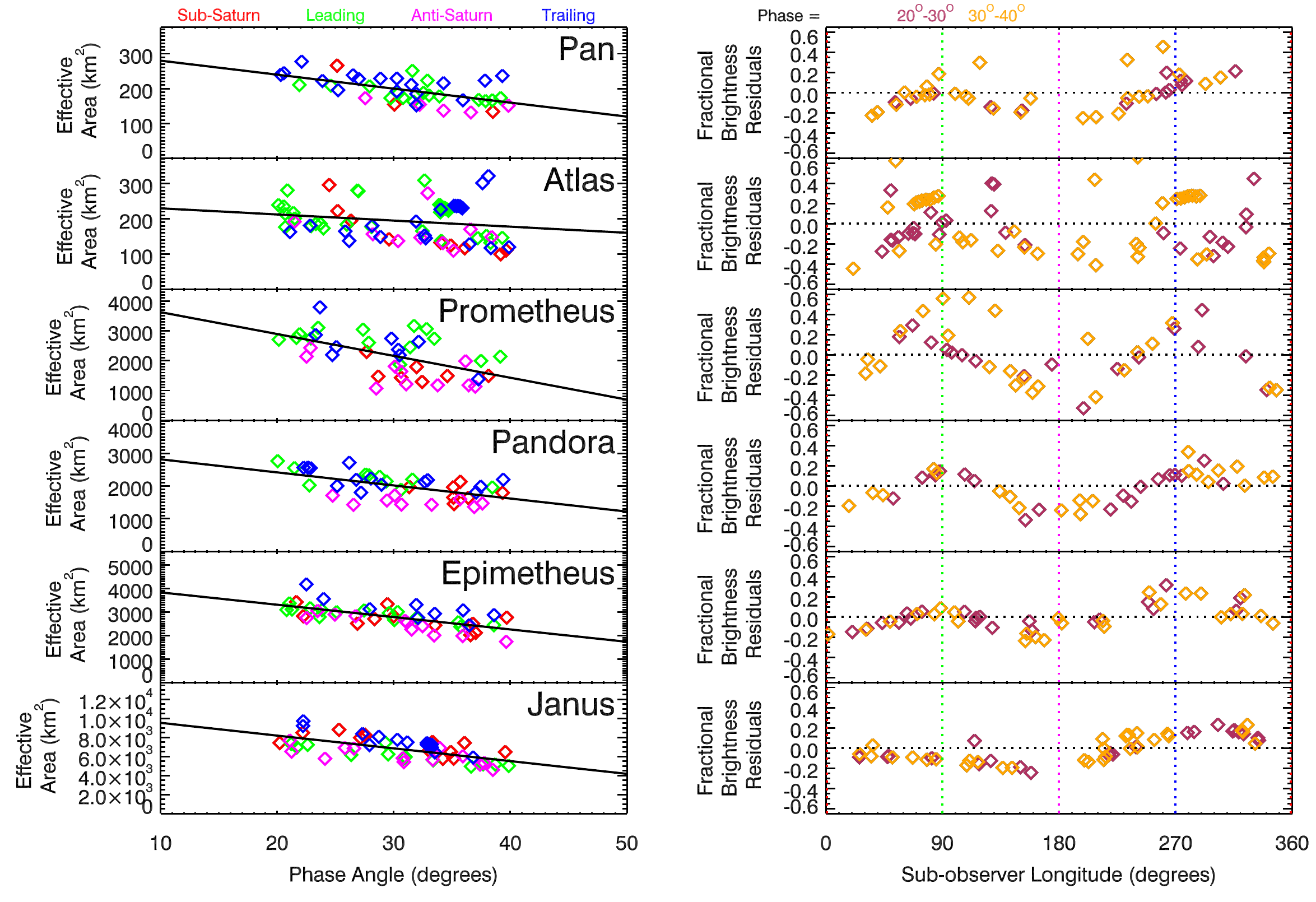}}
\caption{Summary of the brightness measurements for the  ring moons. The left-hand plots show the effective areas $A_{\rm eff}$ of the moons as a function of phase angle, with the data points color coded by the quadrant viewed by the spacecraft. The right-hand plots show the fractional residuals of the brightness estimates relative to the lienar trend shown as a solid line in the left-hand plots. Note that these objects are often brighter when their leading or trailing sides are viewed than they are when their sub-Saturn or anti-Saturn sides are viewed.}
\label{rphaserot}
\end{figure*}

\begin{deluxetable*}{lccccccc}
\tablecaption{Satellite shapes from resolved images \citep[and Appendix B]{Thomas10, Thomas13, Thomas19}
\label{shapes}}
\tablehead{
\colhead{Object} & \colhead{$a$ (km)} & \colhead{$b$ (km)} & \colhead{$c$ (km)} & \colhead{$b/a$} & \colhead{$c/b$} & \colhead{$R_m$ (km)}  \\}
\startdata
Aegaeon & $0.70\pm0.05$ & $0.25\pm0.06$ & $0.20\pm 0.08$ & $0.36\pm0.09$ & $0.80\pm 0.37$ & $0.33\pm0.06$  \\
Methone & $1.94\pm0.02$ & $1.29\pm0.04$ & $1.21\pm0.02$ & $0.66\pm0.02$ & $0.94\pm0.03$ & $1.45\pm0.03$  \\
Pallene & $2.88\pm0.07$ & $2.08\pm0.07$ & $1.84\pm0.07$ & $0.72\pm0.03$ & $0.89\pm0.04$ & $2.23\pm0.07$ \\ \hline
Mimas & 207.8$\pm$0.5 & 196.7$\pm$0.5 & 190.6$\pm$0.3 & 0.947$\pm$0.003 & 0.969$\pm$0.003  & 198.2$\pm$0.4 \\
Enceladus & 256.6$\pm$0.3 & 251.4$\pm$0.2 & 248.3$\pm$0.2 & 0.980$\pm$0.002 & 0.988$\pm$0.001 & 252.1$\pm$0.2 \\
Tethys & 538.4$\pm$0.3 & 528.3$\pm$1.1 & 526.3$\pm$0.6 & 0.981$\pm$0.002 & 0.996$\pm$0.002 & 531.0$\pm$0.6  \\ 
Dione & 563.4$\pm$0.6 & 561.3$\pm$0.5 & 559.6$\pm$0.4 & 0.996$\pm$0.001 & 0.997$\pm$0.001 & 561.4$\pm$0.4 \\
Rhea  &765.0$\pm$0.7 & 763.1$\pm$0.6 & 762.4$\pm$0.6 & 0.998$\pm$0.001 & 0.999$\pm$0.001 & 763.5$\pm$0.6 \\ \hline
Pan & 17.3$\pm$0.2 & 14.1$\pm$0.2 & 10.5$\pm$0.5 & 0.815$\pm$0.015 & 0.745$\pm$0.037 & 13.7$\pm$0.3 \\
Atlas & 20.4$\pm$0.1 & 17.7$\pm$0.2 & 9.3$\pm$0.3 & 0.868$\pm$0.011 & 0.525$\pm$0.018 & 14.9$\pm$0.2 \\
Prometheus & 68.5$\pm$0.5 & 40.5$\pm$1.4 & 28.1$\pm$0.4 & 0.591$\pm$0.021 & 0.694$\pm$0.026 & 42.8$\pm$0.7 \\
Pandora & 51.5$\pm$0.3 & 39.5$\pm$0.3 & 31.5$\pm$0.2 & 0.767$\pm$0.007 & 0.797$\pm$0.008 & 40.0$\pm$0.3  \\ 
Epimetheus & 64.8$\pm$0.3 & 58.1$\pm$0.2 & 53.5$\pm$0.2 & 0.897$\pm$0.005 & 0.921$\pm$0.005 & 58.6$\pm$0.3  \\
Janus  & 101.7$\pm$0.9 & 92.9$\pm$0.3 & 74.5$\pm$0.3 & 0.913$\pm$0.009 & 0.802$\pm$0.004 & 89.0$\pm$0.5 \\ \hline
Telesto & 16.6$\pm$0.3 & 11.7$\pm$0.3 & 9.6$\pm$0.2 & 0.705$\pm$0.022 & 0.821$\pm$0.027 & 12.3$\pm$0.3\\ 
Calypso & 14.7$\pm$0.3 & 9.3$\pm$0.9 & 6.4$\pm$0.3 & 0.632$\pm$0.062 & 0.688$\pm$0.073 & 9.5$\pm$0.4\\
Helene & 22.6$\pm$0.2 & 19.6$\pm$0.3 & 13.3$\pm$0.2& 0.867$\pm$0.015 & 0.679$\pm$0.015 & 18.1$\pm$0.2 \\
Polydeuces & 1.75$\pm$0.2 & 1.55$\pm$0.2 & 1.31$\pm$0.2 & 0.89$\pm$0.15 & 0.85$\pm$0.17 & 1.53$\pm$0.2 \\ \hline
%Pan & 17.2$\pm$1.7 & 15.4$\pm$1.2 & 10.4$\pm$0.9 & 0.89$\pm$0.11 & 0.91$\pm$0.09 & 14.0$\pm$1.2 \\
%Atlas & 20.5$\pm$0.9 & 17.8$\pm$0.7 & 9.4$\pm$0.8 & 0.86$\pm$0.05 & 0.53$\pm$0.05 & 15.1$\pm$0.8 \\
%Prometheus & 68.2$\pm$0.8 & 41.6$\pm$1,8 & 28. 2$\pm$0.8 & 0.61$\pm$0.03 & 0.68$\pm$0.04 & 43.1$\pm$1.2 \\
%Pandora & 52.2$\pm$1.8 & 40.8$\pm$2.0 & 31.5$\pm$0.9 & 0.78$\pm$0.05 & 0.77$\pm$0.04 & 40.6$\pm$1.5  \\ 
%Epimetheus & 64.9$\pm$1.3 & 57.3$\pm$2.5 & 53.0$\pm$0.5 & 0.88$\pm$0.04 & 0.92$\pm$0.04 & 58.2$\pm$1.2  \\
%Janus  & 101.7$\pm$1.6 & 93.0$\pm$0.7 & 76.3$\pm$0.4 & 0.91$\pm$0.02 & 0.82$\pm$0.01 & 76.3$\pm$0.4 \\ \hline
%Telesto & 16.3$\pm$0.5 & 11.8$\pm$0.3 & 9.8$\pm$0.3 & 0.72$\pm$0.03 & 0.83$\pm$0.03 & 12.4$\pm$0.4\\ 
%Calypso & 15.3$\pm$0.6 & 9.3$\pm$2.2 & 6.3$\pm$0.6 & 0.61$\pm$0.14 & 0.68$\pm$0.17 & 9.6$\pm$0.6\\
%Helene & 22.5$\pm$0.5 & 19.6$\pm$19.6 & 13.3$\pm$0.2& 0.87$\pm$0.02 & 0.68$\pm$0.01 & 18.0$\pm$0.4 \\
%Polydeuces & $1.5\pm0.6$ & $1.2\pm0.4$ & $1.0\pm0.2$ & $0.36\pm_{0.07}^{0.05}$ & $0.80\pm _{1.3}^{0.4}$ & $0.33\pm 0.06$ & PT13 
\enddata
\end{deluxetable*}

\begin{figure*}
\resizebox{6.5in}{!}{\includegraphics{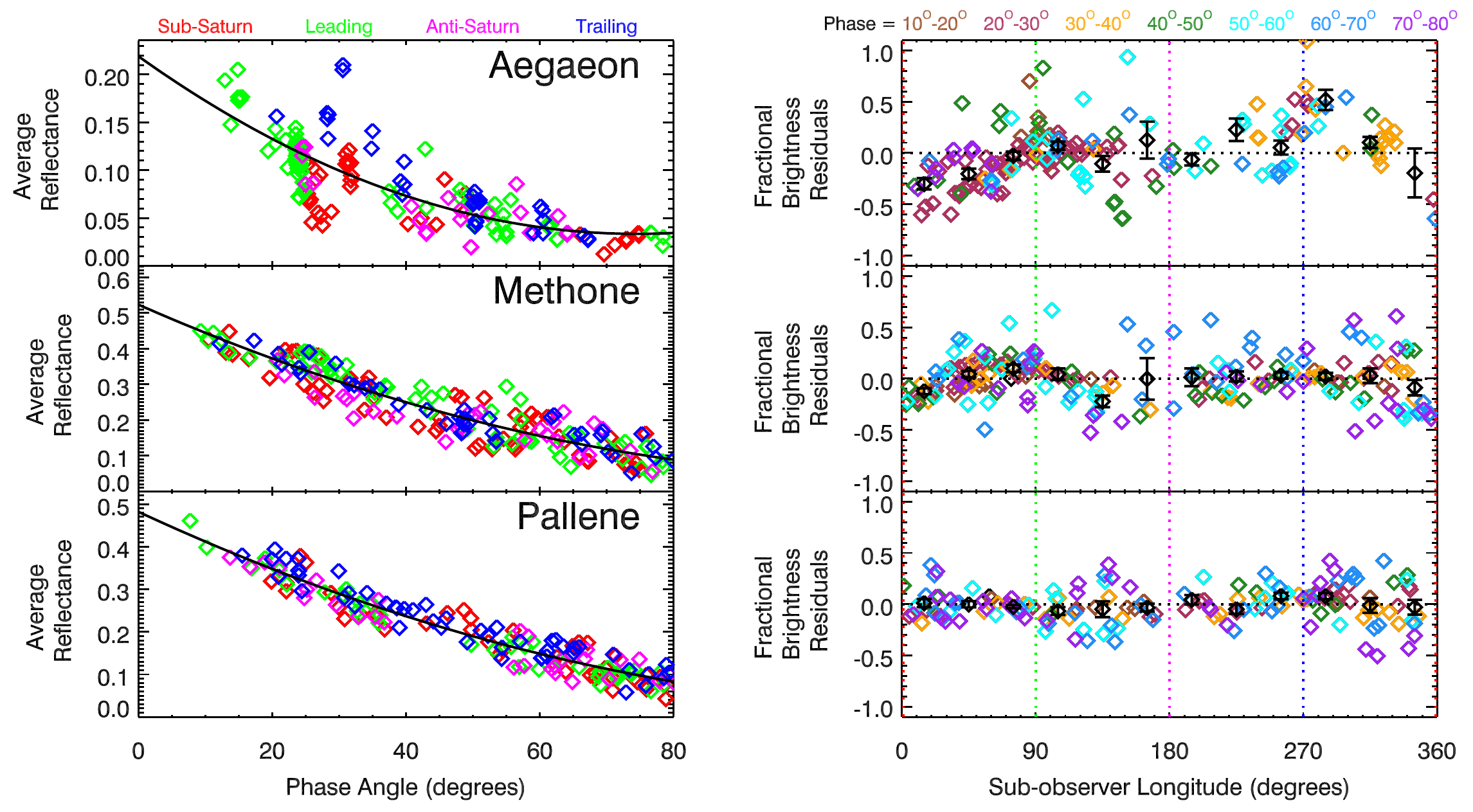}}
\caption{Average reflectances of Aegaeon, Methone and Pallene. The left-hand plots show the average reflectances of the moons as a function of phase angle, with the data points color coded by the quadrant viewed by the spacecraft. The right-hand plots show the fractional residuals of the brightness estimates relative to the quadratic trend shown as a solid line in the left-hand plots. While the longitudinal brightness variations are reduced compared to the effective areas shown in Figure~\ref{phaserot}, they are not completely removed, especially at higher phase angles.}
\label{phaserotarea}
\end{figure*}

\begin{figure*}
\resizebox{6.5in}{!}{\includegraphics{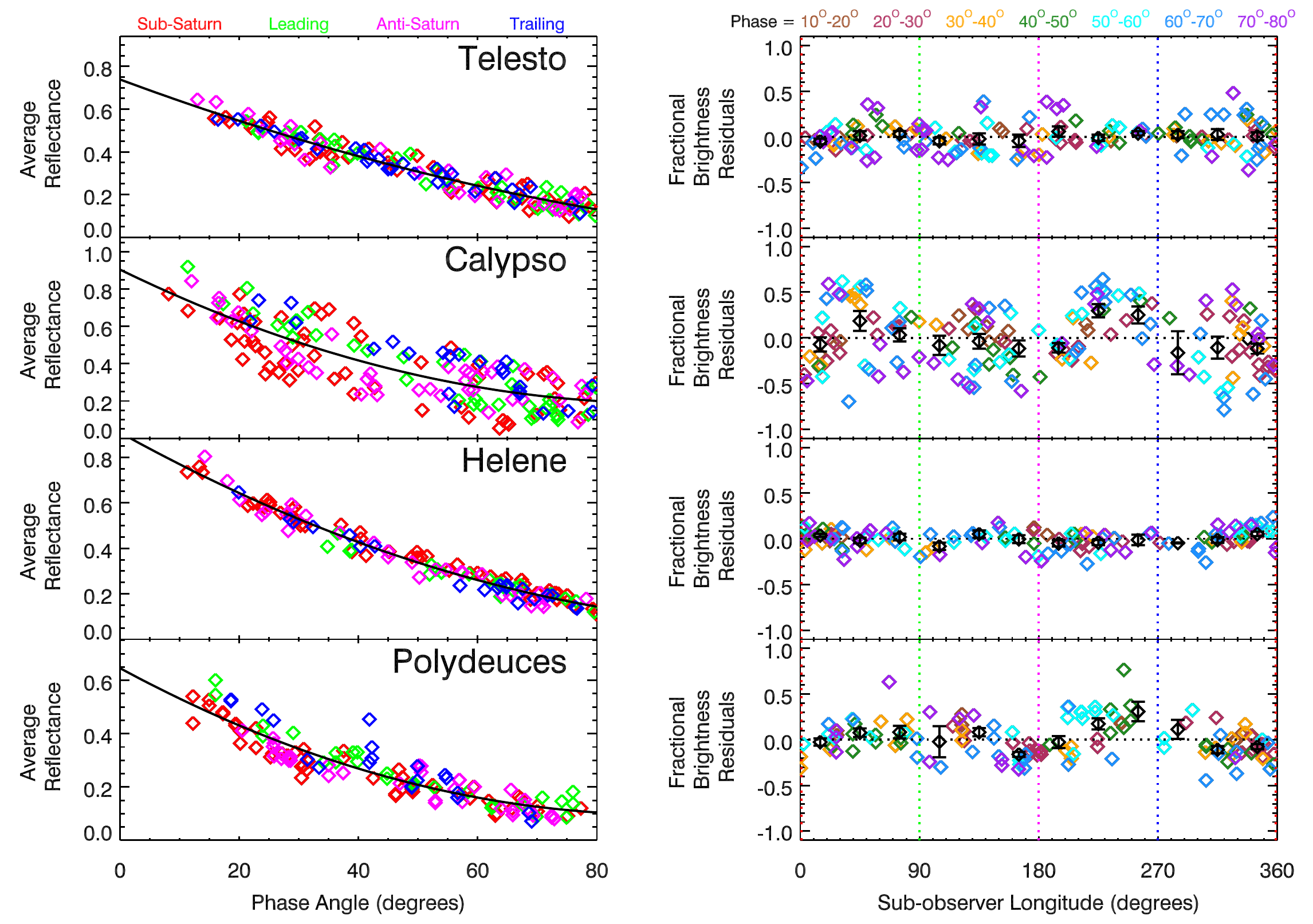}}
\caption{Average reflectances of of Telesto, Calypso, Helene and Polydeuces with their nominal shapes. The left-hand plots show the average reflectances of the moons as a function of phase angle, with the data points color coded by the quadrant viewed by the spacecraft. The right-hand plots show the fractional residuals of the average reflectances relative to the quadratic trend shown as a solid line in the left-hand plots.  While the longitudinal brightness variations are reduced compared to the effective areas shown in Figure~\ref{tphaserot}, they are not completely removed, especially at higher phase angles.}
\label{tphaserotarea}
\end{figure*}

\begin{figure*}
\resizebox{6.5in}{!}{\includegraphics{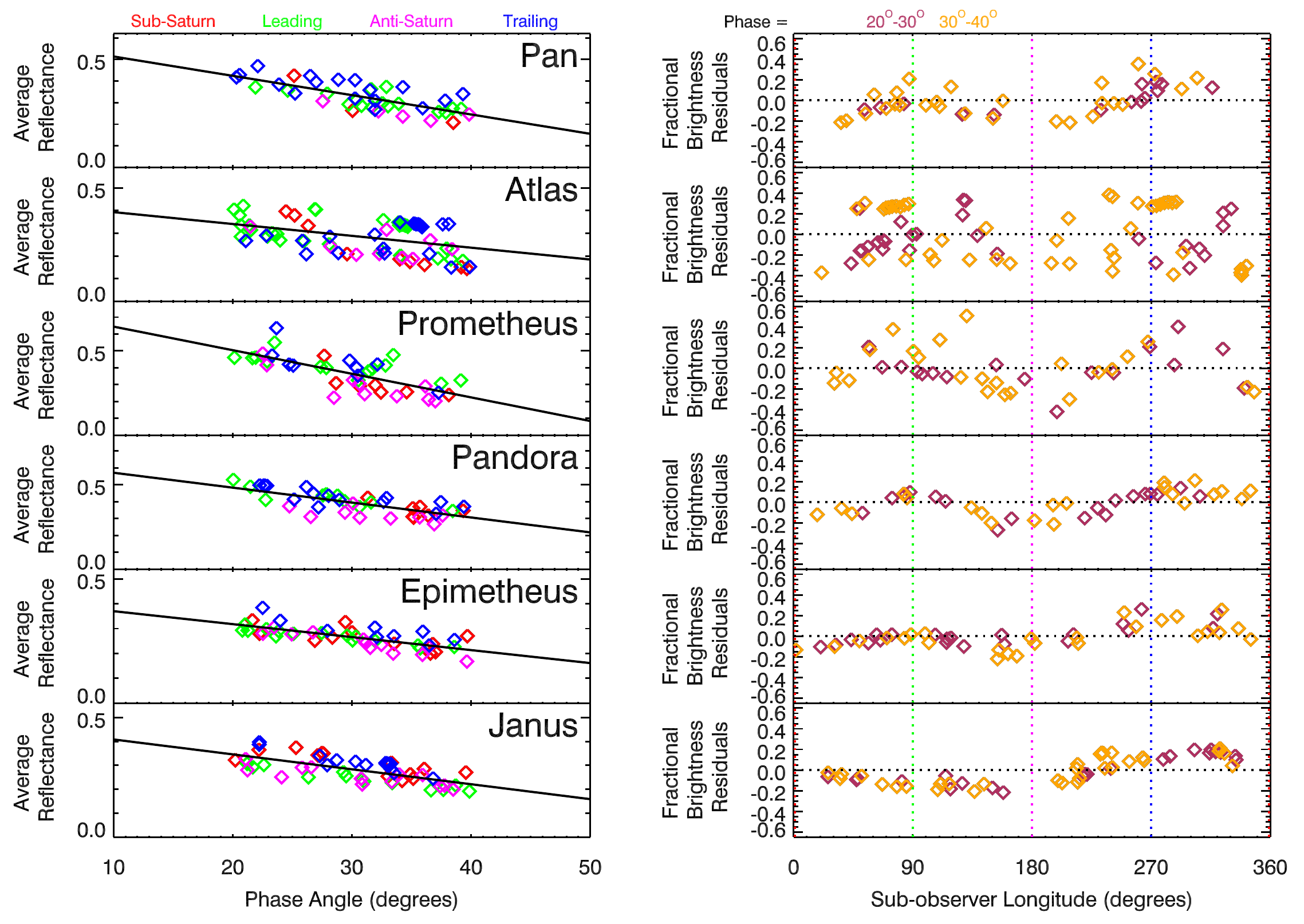}}
\caption{Average reflectances of of the ring moons with their nominal shapes. The left-hand plots show the average reflectances of the moons as a function of phase angle, with the data points color coded by the quadrant viewed by the spacecraft. The right-hand plots show the fractional residuals of the brightness estimates relative to the linear trend shown as a solid line in the left-hand plots. Again, while the longitudinal brightness variations are slightly reduced compared to the effective areas shown in Figure~\ref{rphaserot}, they are not completely removed.}
\label{rphaserotarea}
\end{figure*}

\clearpage

\section{ A photometric model for elongated bodies}
\label{model}

The challenges associated with photometric analyses of small moons are best seen in the left-hand panels of Figures~\ref{phaserot}-\ref{rphaserot}, which show the effective area estimates for these moons as functions of the observed phase angle. While there is a clear trend of decreasing brightness with increasing phase angle, the data points also show a relatively large amount of scatter around this basic trend. This scatter cannot be entirely attributed to measurement errors because it is strongly correlated with the viewing geometry. In particular,  the moons' sub-Saturn and anti-Saturn quadrants are systematically lower in $A_{\rm eff}$ than their leading and trailing  quadrants. These variations can be more clearly documented by plotting the fractional residuals from the mean trend (where $A_{\rm eff}$ is assumed to be a linear or quadratic function of phase angle) as functions of the sub-observer longitude. With the exception of Helene, Pan and Atlas, these plots all show clear sinusoidal patterns with maxima around $90^\circ$ and $270^\circ$ and minima around $0^\circ$ and $180^\circ$. 

These trends with sub-observer longitude arise because these small moons have ellipsoidal shapes and are tidally locked so that their long axes point towards Saturn.  Resolved images of  these moons show that they are all significantly elongated (see Table~\ref{shapes}, note that Anthe was never observed with sufficient resolution to determine its shape and size accurately). Furthermore, the relative magnitudes of the longitudinal brightness variations shown in Figures~\ref{phaserot}-\ref{rphaserot} are generally consistent with the relative $a/b$ ratios of these different moons, with Aegaeon being the most elongated object and the one with the largest longitudinal brightness variations, followed by Calypso, Promethus and Methone.

%Since the dimensions of these objects are reasonably well measured, and the viewing and illumination geometries for each image is known, it should be possible to account for these longitudinal brightness variations. However, it turns out that accomplishing this is not as easy as it appears.

\subsection{Variations in projected area do not adequately account for the moons' photometric properties}

The simplest way to account for these shape-related brightness variations is to divide the observed $A_{\rm eff}$ by the plane-projected area of the object $A_{\rm phys}$, which for an ellipsoidal object is given by the following formula \citep{Vickers96}:
\begin{equation}
\begin{aligned}
A_{\rm phys}= \pi  \left[b^2c^2\cos^2\lambda_O\cos^2\phi_O+ a^2c^2\cos^2\lambda_O\sin^2\phi_O+ a^2b^2\sin^2\lambda_O\right]^{1/2} \\
\end{aligned}
\end{equation}
where $a$, $b$, and $c$ are the dimensions of the ellipsoid, while $\lambda_O$ and $\phi_O$ are the sub-observer latitude and longitude for the object. The resulting ratio then yields the disk-averaged reflectance of the object:
\begin{equation} <I/F> = \frac{A_{\rm eff}}{A_{\rm phys}}. \end{equation}
Tables~\ref{aegaeontab}-~\ref{methonecoltab} include estimates of $A_{\rm phys}$ computed assuming that these bodies have the shape parameters given in Table~\ref{shapes}, and Figures~\ref{phaserotarea}-\ref{rphaserotarea} show the resulting estimates of $<I/F>$ as functions of phase and sub-observer longitude. While the fractional brightness residuals around the mean phase curve are somewhat smaller for $<I/F>$ than they are for $A_{\rm eff}$,  the dispersion is still rather large. For Aegaeon, Prometheus and Pandora there are still clear maxima at $90^\circ$ and $270^\circ$, while for Methone, Pallene, Telesto and Calypso the dispersion at larger phase angles has a similar magnitude for the two parameters. 

\subsection{A Generic Ellipsoid Photometric Model (GEPM) }

The major problem with using $<I/F>$ is that this parameter only accounts for the viewing geometry, but not how the moon is illuminated by the Sun, which for very elongated bodies can strongly affect the observed brightness \citep{HV89}.  \citet{ML15} provide analytical formula for elongated bodies assuming the surface follows a Lommel-Seeliger scattering law, which is appropriate for dark surfaces. However, the applicability of such a model for bright objects like many of Saturn's moons is less clear. We therefore use a more generic numerical method to translate whole-disk brightness data into information about the object's global average reflectance behavior, which we call the  \textit{ Generic Ellipsoid Photometric Model (GEPM)}.  

The GEPM predicts an ellipsoidal object's $A_{\rm eff}$ in any given image assuming its surface reflectance $R$ follows a generic scattering law that is a function of the cosines of the incidence and emission angles $\cos i$ and $\cos e$. The predicted effective area of an object whose surface obeys these scattering laws is given by the following integral:
\begin{equation}
A_{\rm pred}=\int_{++} (\cos e) R  r_{\rm eff}^2  \cos\lambda d\lambda d\phi
\label{apred1}
\end{equation}
where $\lambda$ and $\phi$  are the latitude and longitude on the object, the $++$ sign indicates that the integral is performed only over the part of the object that is both illuminated and visible (i.e. where $\cos i$ and $\cos e$ are both positive), the factor of $\cos e$ in this integral accounts for variations in the projected size of the area element, and $r_{\rm eff}$ is the effective radius of the object at a given latitude and longitude, which is given by the following expression:
\begin{equation}
r_{\rm eff}^2 =\left[b^2c^2\cos^2\lambda\cos^2\phi+ a^2c^2\cos^2\lambda\sin^2\phi+ a^2b^2\sin^2\lambda\right]^{1/2} .
\end{equation}

The factors of $\cos i$ and $\cos e$ in the above integral are also functions of $\lambda$ and $\phi$ that depend upon the viewing and illumination geometry, as well as the object's shape. To obtain these functions, we first use the sub-observer latitude $\lambda_O$ and longitude $\phi_O$ to define a unit vector pointing from the center of the body towards the spacecraft:
\begin{equation}
\hat{\bf o}=\cos\lambda_O\cos\phi_O\hat{\bf x}+\cos\lambda_O\sin\phi_O\hat{\bf y}+\sin\lambda_O\hat{\bf z}
\end{equation} 
where $\hat{\bf x}$ points from the moon towards Saturn, $\hat{\bf y}$ points in the direction of orbital motion and $\hat{\bf z}$ points along the object's rotation axis. Similarly, we use the sub-solar latitude $\lambda_S$ and longitude $\phi_S$ to define a unit vector pointing from the center of the body towards the Sun:
\begin{equation}
\hat{\bf s}=\cos\lambda_S\cos\phi_S\hat{\bf x}+\cos\lambda_S\sin\phi_S\hat{\bf y}+\sin\lambda_S\hat{\bf z}
\end{equation} 
Finally, for each latitude $\lambda$ and longitude $\phi$ on the surface, we compute the surface normal for the body, which depends on the shape parameters $a$, $b$ and $c$.
\begin{equation}
\hat{\bf n}=\frac{\frac{\cos\lambda\cos\phi}{a}\hat{\bf x}+\frac{\cos\lambda\sin\phi}{b}\hat{\bf y}+\frac{\sin\lambda}{c}\hat{\bf z}}{\sqrt{\frac{\cos^2\lambda\cos^2\phi}{a^2}+\frac{\cos^2\lambda\sin^2\phi}{b^2}+\frac{\sin^2\lambda}{c^2}}}
\end{equation} 
The cosines of the incidence and emission angles are then given by the standard expressions $\cos i =\hat{\bf n}\cdot\hat{\bf s}$ and  $\cos e =\hat{\bf n}\cdot\hat{\bf o}$.

The above expressions allow us to evaluate $A_{\rm pred}$ for any given scattering law $R$, {including Lunar-Lambert functions or Akimov functions \citep{McEwen91, Shkuratov11, Schroeder14}. However, for the sake of concreteness, we will here assume that $R$ is given by the Minnaert function} \citep{Minnaert41}:
\begin{equation} R_M  = B (\cos i)^{k}(\cos e)^{1-k},
\end{equation}
where the quantities  $k$ and $B$ will be assumed to be constants for each image. Note that if we assume that $k=1$ the above expression reduces to the Lambert photometric function:
\begin{equation} R_L  = B \cos i.
\end{equation}
For a generic Minnaert function, Equation~\ref{apred1}  becomes:
\begin{equation}
A_{\rm pred}=B \int_{++} (\cos i)^k (\cos e)^{2-k}  r^2_{\rm eff} \cos\lambda d\lambda d\phi =B a_{\rm pred}
\end{equation}
The factor of $B$ moves outside the integral, and the remaining factor $a_{\rm pred}$ can be numerically evaluated for any specified viewing and illumination geometry, so long as we also assume values for the object's shape parameters $a$, $b$ and $c$, as well as the photometric parameter $k$. A Python function which evaluates $a_{\rm pred}$ given these parameters is provided in Appendix A. From this, we can estimate the ``brightness coefficient" parameter $B$ to be simply the ratio $A_{\rm eff}/a_{\rm pred}$. Note that $B$ is not equal to the mean reflectance $<I/F>$, even for  spherical bodies, because it includes corrections for the varying incidence and emission angles across the disk.

\begin{figure}
\centerline{\resizebox{3in}{!}{\includegraphics{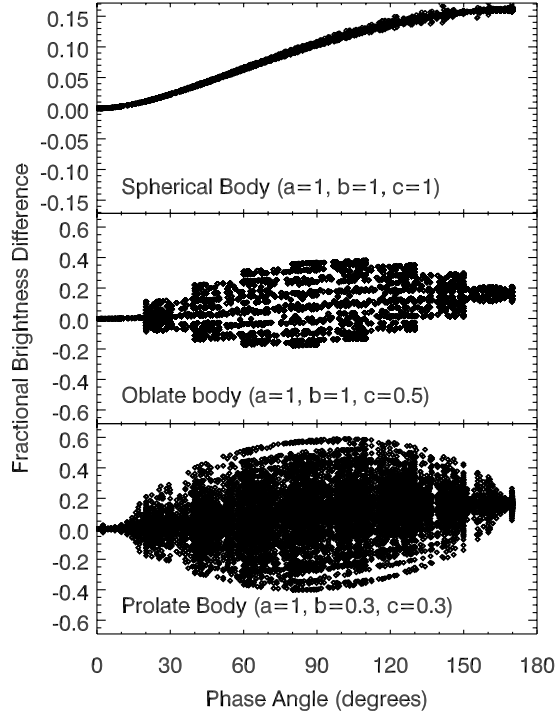}}}
\caption{The fractional difference between the predicted values of $a_{\rm pred}$ between a Minnaert model with {$k=0.5$} and a Lambertian model for three different shapes. Each point corresponds to a different combination of viewing and illuminate geometries scattered uniformly over the objects' surface.}
\label{minlam}
\end{figure}

Of course, the estimated value of $B$ depends on the assumed values of $a$, $b$, $c$ and $k$, which are typically estimated from resolved images.  While only a few images are needed to obtain useful estimates of $a$, $b$ and $c$, the parameter $k$ can depend on the observed phase angle and location on the object, and so $k$ is far harder to estimate reliably for the full range of observation geometries. {Previous studies of resolved Voyager images of the mid-sized moons found that Mimas, Enceladus, Tethys, Dione and Rhea had $k=0.65, 0.78, 0.63, 0.53-0.56$ and 0.52-0.54 respectively at phase angles between 5$^\circ$ and $17^\circ$ \citep{Buratti84}, while a study of resolved Cassini images of Methone found that $k=0.887-0.003\alpha$  ($\alpha$ being the solar phase angle) based on images obtained at phase angles between 45$^\circ$ and $65^\circ$ \citep{Thomas13}. Thus this parameter likely varies among these moons and with observation geometry and terrain for each moon.} 

{Figure~\ref{minlam} shows the fractional differences in the predicted values of $a_{\rm pred}$ between a model where $k=0.5$ and a model where $k=1$ (that is, a Lambertian surface). Each data point corresponds to a different combination of sub-observer and sub-solar locations scattered over the entire object. For the spherical object there is a trend where the $k=0.5$ model becomes 15\% brighter than the Lambertian model at high phase angles. Very prolate and oblate objects show a similar overall trend, but with considerably more scatter depending on the exact viewing and illumination geometry. This suggests that one could constrain $k$ by minimizing the $rms$ scatter in the estimated brightness coefficients from unresolved images of elongated bodies at each phase angle. In practice, the $rms$ scatter in the brightness coefficients for the various moons are very weak functions of $k$, most likely because real surface albedo variations and/or instrumental noise dominate the dispersion in the brightness coefficients. We therefore expect that a thorough analysis of resolved images would be needed to constrain $k$ reliably for each of the moons, and such an analysis is well beyond the scope of this paper. }

{Fortunately, it turns out that the results presented below are  insensitive to the exact value of $k$. We  bracket the range of possible $k$-values by considering cases where $k=0.5, 0.75$ and 1.0 when estimating the brightnesses of the various moons. Estimates of $a_{\rm pred}$ derived from all three cases are provided in Tables~\ref{aegaeontab}-~\ref{methonecoltab}. However, since the brightness differences among these different scenarios are subtle, we will normally only plot brightness coefficients computed assuming the surface follows a Lambertian scattering law (i.e. $k=1$).}

\subsection{Validation of General Ellipsoid Photometric Model with Saturn's small moons}

\begin{figure*}
\centerline{\resizebox{5.5in}{!}{\includegraphics{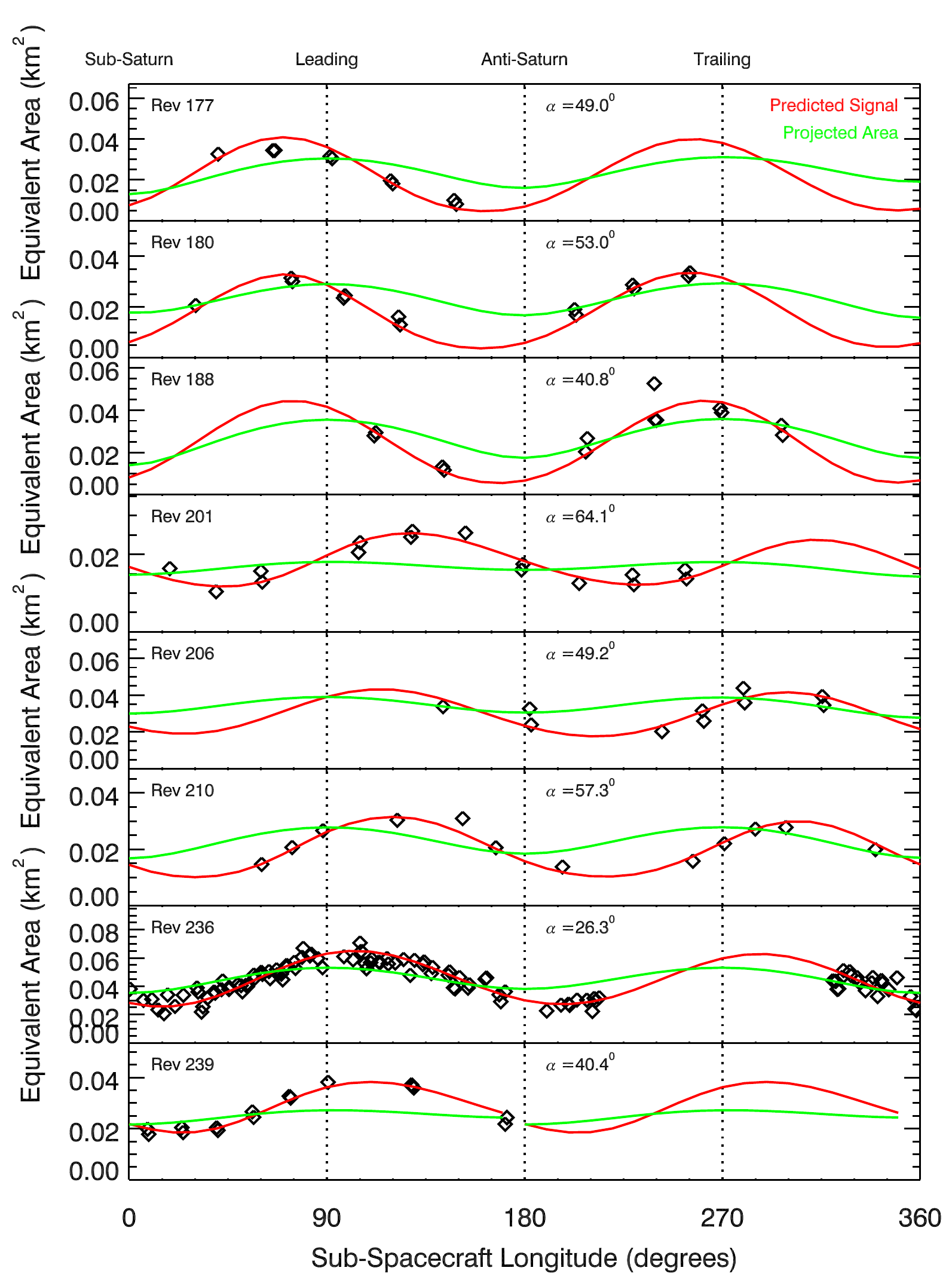}}}
\caption{Observed and predicted rotation light curves for Aegaeon. Each plot shows data from one particular ARCORBIT observation of Aegaeon, which measured the brightness of Aegaeon as a function of sub-observer longitude. The data points are the measured data, while the green curves show the variations in the projected area of the moon, and the red curves show the predicted variations in $a_{\rm pred}$ for a Lambertain model with Aegaeon's nominal shape {(Minnaert models with $k=0.5$ and $k=0.75$ show nearly the same trends)}. The predicted curves are both scaled to best match the median observed signal. The GEPM  model predictions clearly match the data much better than the effective areas.  }
\label{lightcurve}
\end{figure*}

\begin{table}
\caption{{Cassini ISS NAC images belonging to the various ARCORBIT sequences}}
\label{arcorbittab}
\resizebox{6.5in}{!}{\begin{tabular}{l l}\hline
Sequence & Images \\ \hline \hline
Rev 177 & 
N1735282418
N1735287671
N1735287895
N1735292924
N1735293148
N1735298177
N1735298401
N1735303430
N1735303654 \\  \hline
Rev 180 &
N1738781138
N1738781312
N1738785881
N1738786055
N1738790624
N1738790798
N1738804853
N1738805027
N1738809596
N1738809770 \\ &
N1738814339
N1738814513
N1738842971 \\ \hline
Rev 188 &
N1746467285
N1746467459
N1746472918
N1746473092
N1746484184
N1746484358
N1746489747
N1746489817
N1746489991
N1746495450 \\ &
N1746495624
N1746501083
N1746501257 \\ \hline
Rev 201 &
N1769540993
N1769545670
N1769549999
N1769550173
N1769559005
N1769559179
N1769563508
N1769563682
N1769568011
N1769572514 \\ &
N1769572688
N1769577191
N1769581520
N1769581694
N1769586023
N1769586197 \\ \hline
Rev 206 & 
N1783325708
N1783332836
N1783333010
N1783343963
N1783347440
N1783347614
N1783351091
N1783351265
N1783358393
N1783358567 \\ \hline
Rev 210 &
N1796027549
N1796033113
N1796035895
N1796041459
N1796052587
N1796055369
N1796058151
N1796060933
N1796069279
N1796085971 \\ &
N1796088753
N1796091535 \\ \hline
Rev 236 &
N1843274411
N1843274619
N1843274723
N1843274810
N1843274984
N1843275331
N1843275435
N1843276039
N1843276143
N1843276747 \\ &
N1843276851
N1843277559
N1843278163
N1843278267
N1843278739
N1843279092
N1843279196
N1843279800
N1843280612
N1843281924 \\ &
N1843282326
N1843282500
N1843282957
N1843284269
N1843285081
N1843285685
N1843286261
N1843286614
N1843287322
N1843288134 \\ &
N1843289446
N1843289550
N1843289848
N1843290022
N1843290479
N1843291083
N1843291187
N1843291791
N1843291895
N1843292499 \\ &
N1843292603
N1843293311
N1843293609
N1843293783
N1843294136
N1843294240
N1843294844
N1843294948
N1843295552
N1843295656 \\ &
N1843296260
N1843296364
N1843296968
N1843297072
N1843297370
N1843297544
N1843297897
N1843298001
N1843298605
N1843298709 \\ &
N1843299313
N1843299417
N1843300021
N1843300125
N1843300729
N1843301131
N1843303074
N1843303886
N1843304490
N1843304594 \\ &
N1843304892
N1843305066
N1843305419
N1843305523
N1843306127
N1843306231
N1843306835
N1843306939
N1843307543
N1843308251 \\ &
N1843308827
N1843309180
N1843309888
N1843309992
N1843310596
N1843310700
N1843312012
N1843312116
N1843312414
N1843312588 \\ &
N1843312941
N1843313649
N1843313753
N1843315065
N1843315169
N1843316175
N1843316349
N1843316702
N1843320110
N1843321275 \\ &
N1843321879
N1843321983
N1843322587
N1843323399
N1843323864
N1843324006
N1843324110
N1843324422 \\ \hline
Rev 239 &
N1848936800
N1848936974
N1848940466
N1848940640
N1848944132
N1848944306
N1848947798
N1848947972
N1848951464
N1848951638 \\ &
N1848955130
N1848962392
N1848962462
N1848962636
N1848969794
N1848969968 
N1848969968 \\
\hline
\end{tabular}}
\end{table}

The utility of this model can be most clearly demonstrated by taking a closer look at the data for Aegaeon. {Aegaeon is not only the most elongated moon, it was also observed in multiple ARCORBIT sequences where the spacecraft repeatedly imaged the moon as it moved around a significant fraction of its orbit, yielding true rotational light curves}. {The images from eight of  these sequences are listed in Table~\ref{arcorbittab}, while the values of $A_{\rm eff}$ are shown in Figure~\ref{lightcurve}, plotted as a function of the sub-observer longitude}. All show clear brightness variations that can be attributed to the changing viewing geometry of the moon, with the lowest signals seen when the sub-observer longitude is near 0$^\circ$ or 180$^\circ$, and the highest signals seen when the sub-observer longitude is close to $90^\circ$ or 270$^\circ$. However, it is also clear that the maxima and minima are not precisely aligned with these four longitudes (this is most clearly seen in the Rev 201, 210 and 236 data). This result clearly demonstrates that the projected area is not the only thing affecting the moon's projected brightness. Indeed, the green curves show the variations in the projected area, scaled to match the mean signal in the data. These variations are both too subtle to match the observations, and the observed and predicted peaks and troughs do not line up. By contrast, the GEPM predictions are a much better fit to the data in all cases. This demonstrates that accounting for the lighting geometry not only causes the predicted locations of peaks and troughs to shift into alignment with the data, but also increases the predicted fractional brightness variations.

\begin{figure*}
\resizebox{6.5in}{!}{\includegraphics{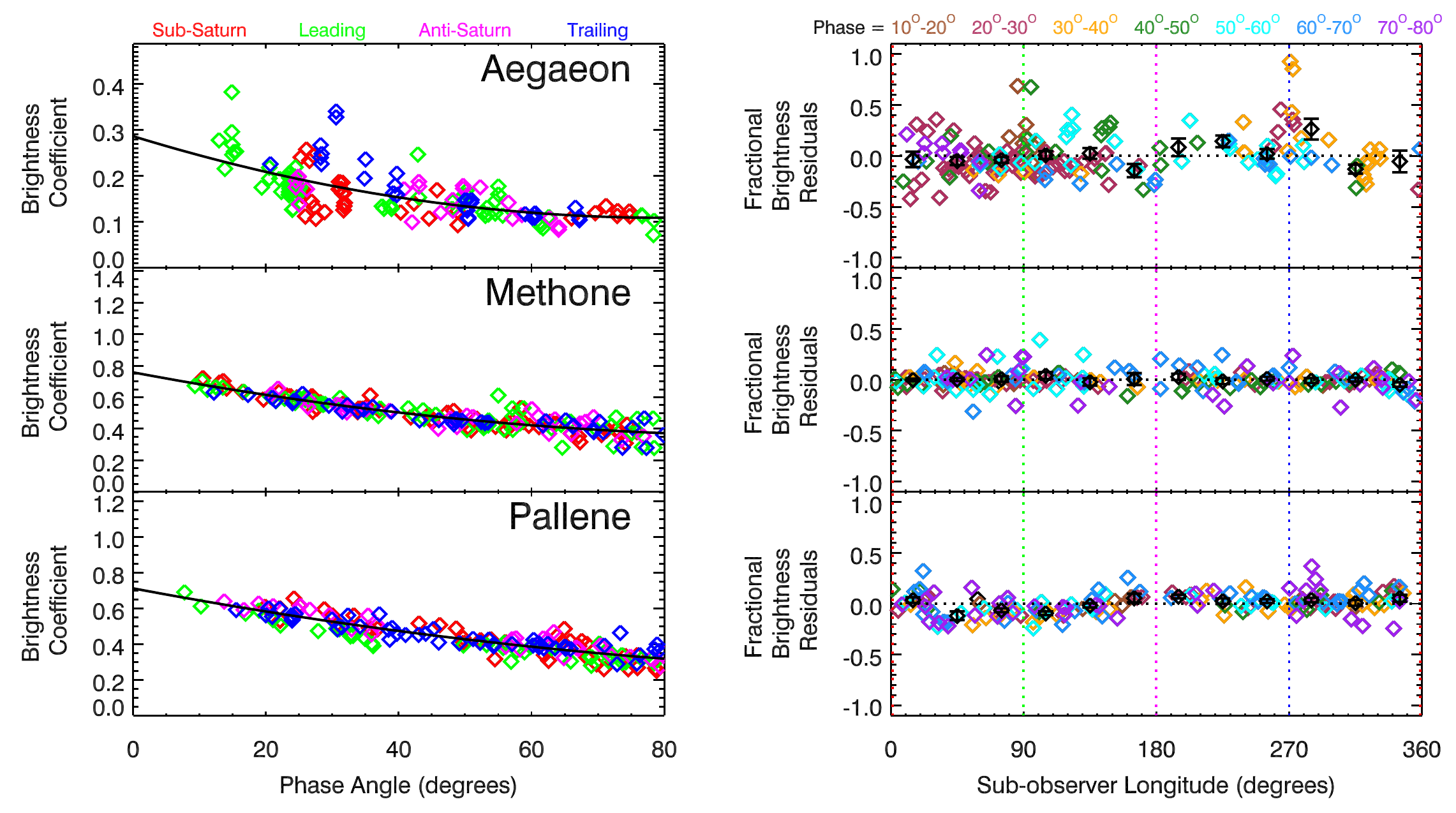}}
\caption{Estimated brightness coefficients for Lambertian models of Aegaeon, Methone and Pallene with their nominal shapes. The left plots show the brightness coefficients of the moons as a function of phase angle, with the data points color coded by the quadrant viewed by the spacecraft. The right-hand plots show the fractional residuals of the brightness estimates relative to the quadratic trend shown as a solid line in the left-hand plots. The longitude dependent brightness variations are almost completely removed, and those that remain, like Pallene's leading-trailing asymmetry, are probably real surface features.}
\label{phaserotpred}
\end{figure*}

\begin{figure*}
\resizebox{6.5in}{!}{\includegraphics{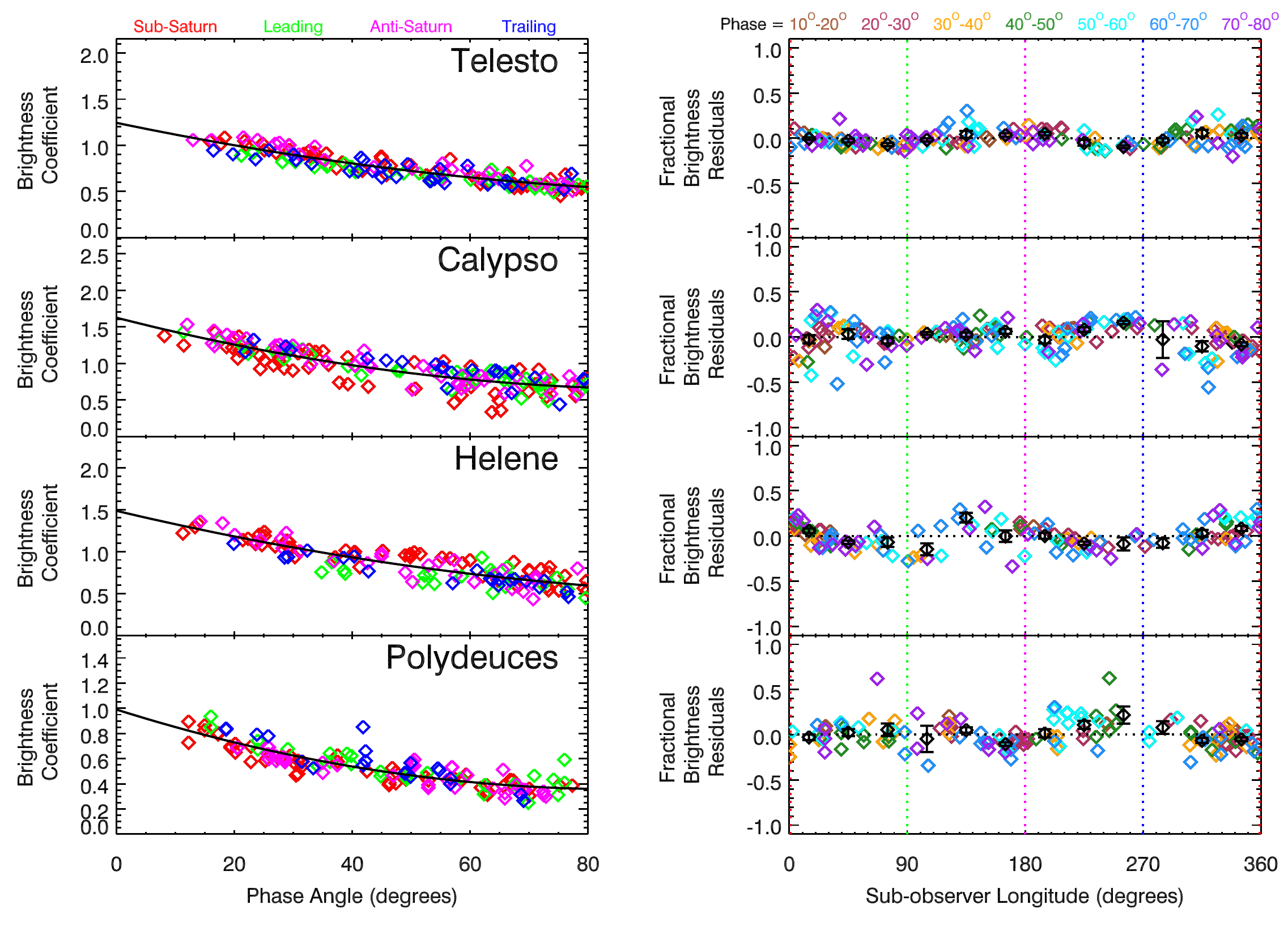}}
\caption{Estimated brightness coefficients for Lambertian models of Telesto, Calypso, Helene and Polydeuces with their nominal shapes. The left plots show the brightness coefficients of the moons as a function of phase angle, with the data points color coded by the quadrant viewed by the spacecraft. The right-hand plots show the fractional residuals of the brightness estimates relative to the quadratic trend shown as a solid line in the left-hand plots.  For Telesto, Calypso and Polydeuces, this model yields reduces the dispersion associated with the viewing geometry. However, for Helene the dispersion of the measurements around the mean trend  has actually increased.}
\label{tphaserotmod}
\end{figure*}

\begin{figure*}
\resizebox{6.5in}{!}{\includegraphics{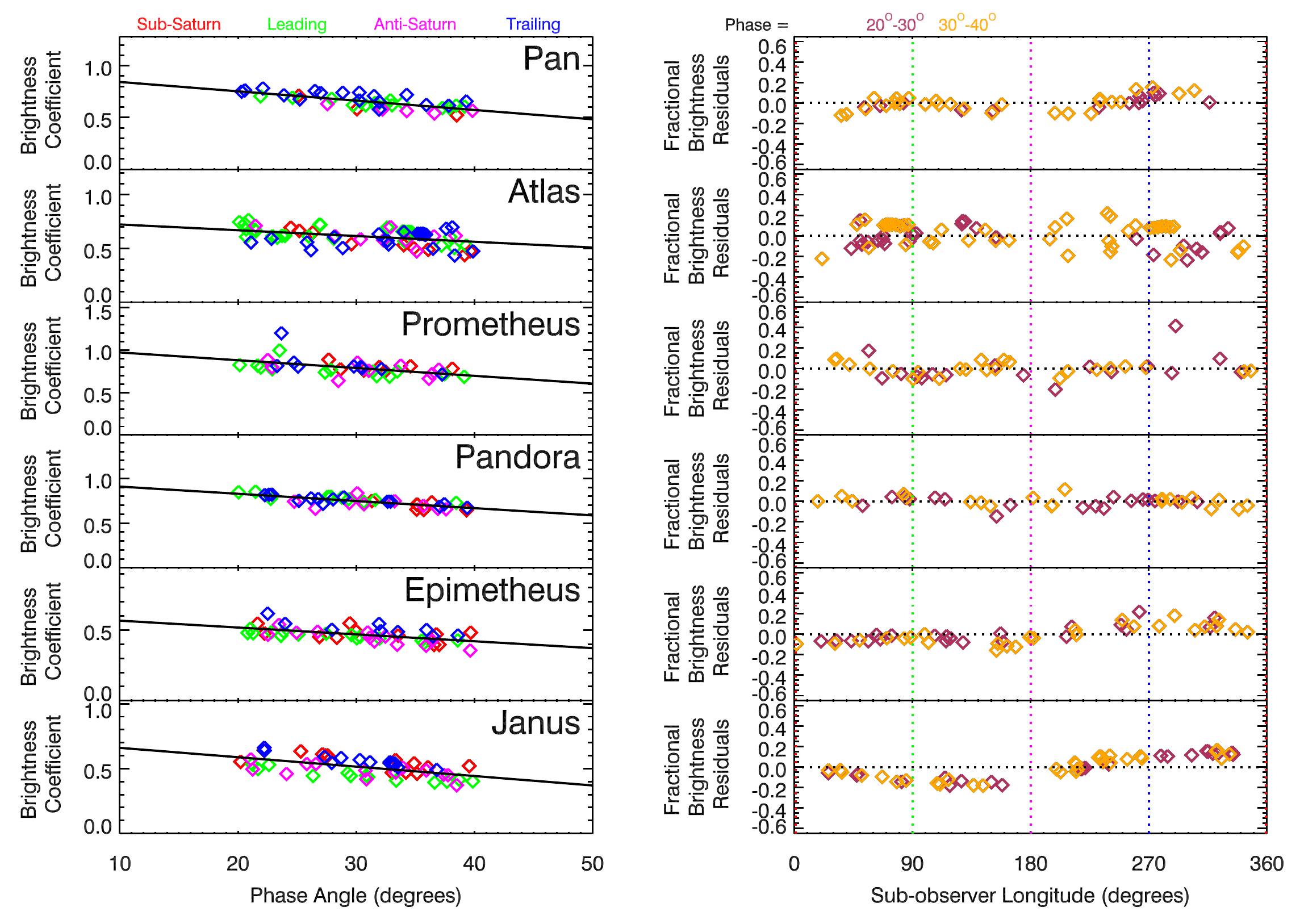}}
\caption{Estimated brightness coefficients for Lambertian models of the ring moons with their nominal shapes. The left plots show the brightness coefficients of the moons as a function of phase angle, with the data points color coded by the quadrant viewed by the spacecraft. The right-hand plots show the fractional residuals of the brightness estimates relative to the linear trend shown as a solid line in the left-hand plots. This model again reduces the dispersion around the phase trend, although for Atlas the dispersion is rather high compared to the other moons. Also note that for Janus and Epimethus the leading hemisphere is clearly darker than the trailing hemisphere.}
\label{rphaserotmod}
\end{figure*}

\begin{figure*}
\resizebox{6.5in}{!}{\includegraphics{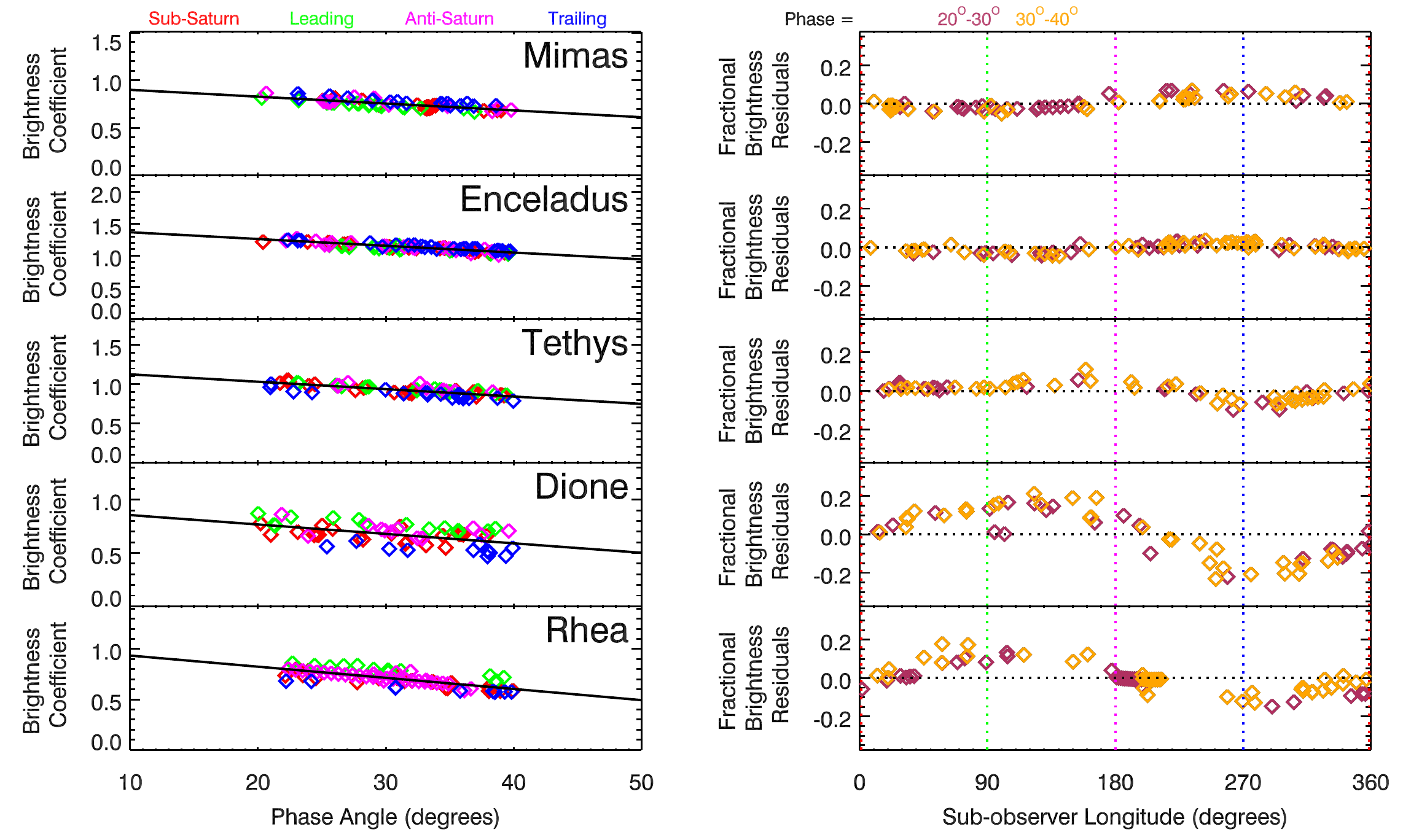}}
\caption{Estimated brightness coefficients for Lambertian models of the mid-sized moons with their nominal shapes. The left plot show the brightness coefficients of the moons as a function of phase angle, with the data points color coded by the quadrant viewed by the spacecraft. The right-hand plots show the fractional residuals of the brightness estimates relative to the linear trend shown as a solid line in the left-hand plots. Since all of these moons are close to spherical, the model does not affect the trends with longitude very much. Instead, this model simply enables the data for these moons to be compared with that from smaller moons. Note the longitudinal brightness asymmetries seen in these data are consistent with previous measurements and are mostly due to asymmetries in the E-ring flux.}
\label{mphaserotmod}
\end{figure*}

More generally, Figures~\ref{phaserotpred}-\ref{rphaserotmod} show the estimated brightness coefficients for the small moons derived assuming  that they have the nominal shapes given in Table~\ref{shapes} and surfaces that follow a Lambertian scattering law {(assuming the surfaces follow Minnaert scattering laws with k=0.5 and 0.75 yield nearly identical results)}. Compared with Figures~\ref{phaserotarea}-\ref{rphaserotarea}, the scatter around the mean trends with phase angle are much tighter for almost all of the moons. The exceptions to this trend are Helene, Polydeuces, Janus and Epimetheus, for which the dispersions in $B$ and $<I/F>$ are comparable. These exceptions are likely because these four moons are the most spherical in shape, so the corrections included in the GEPM are less important. There are some outlying data points for Aegaeon, Atlas and Prometheus, but these can attributed to the low signals from Aegaeon and misestimated background levels for Atlas and Prometheus. 

Besides the reduction in dispersion, the fractional brightness residuals no longer show double-peaked trends with co-rotating longitude. Instead, the most obvious remaining longitudinal trends are that the leading sides of Pallene, Epimetheus and Janus are 10-20\% darker than their trailing sides. Such leading-trailing asymmetries are reminiscent of the longitudinal brightness variations seen in ground-based, voyager and Cassini images of the mid-sized icy moons \citep{Noland74, Franz75, BV84, Verbiscer89, Buratti90, Verbiscer92, Verbiscer94, Buratti98, Pitman10, Schenk11}. Figure~\ref{mphaserotmod} shows the brightness coefficients for these larger moons derived assuming the same Lambertian model as was used for the smaller moons.\footnote{Due to the nearly spherical shape of these moons, the dispersions around the mean trends in these cases do not change much between $A_{\rm eff}$, $<I/F>$ and $B$.}  Consistent with prior work  \citep{Buratti98, Schenk11, Hendrix18, Verbiscer18}, we find that Tethys, Dione and Rhea have brighter leading sides, while Mimas has a brighter trailing side. This overall trend is thought to arise because the particles in the E ring preferentially strike the leading sides of moons orbiting exterior to Enceladus and the trailing sides of moons orbiting interior to Enceladus \citep{HB94}. The longitudinal brightness asymmetries observed on Pallene, Janus and Epimetheus are consistent with this pattern, but it is worth noting  that Aegaeon, Methone and the trojan moons do not appear to have such asymmetries. The implications of these findings will be discussed further in Section~\ref{trends} below.

\begin{figure*}
\centerline{\resizebox{5.5in}{!}{\includegraphics{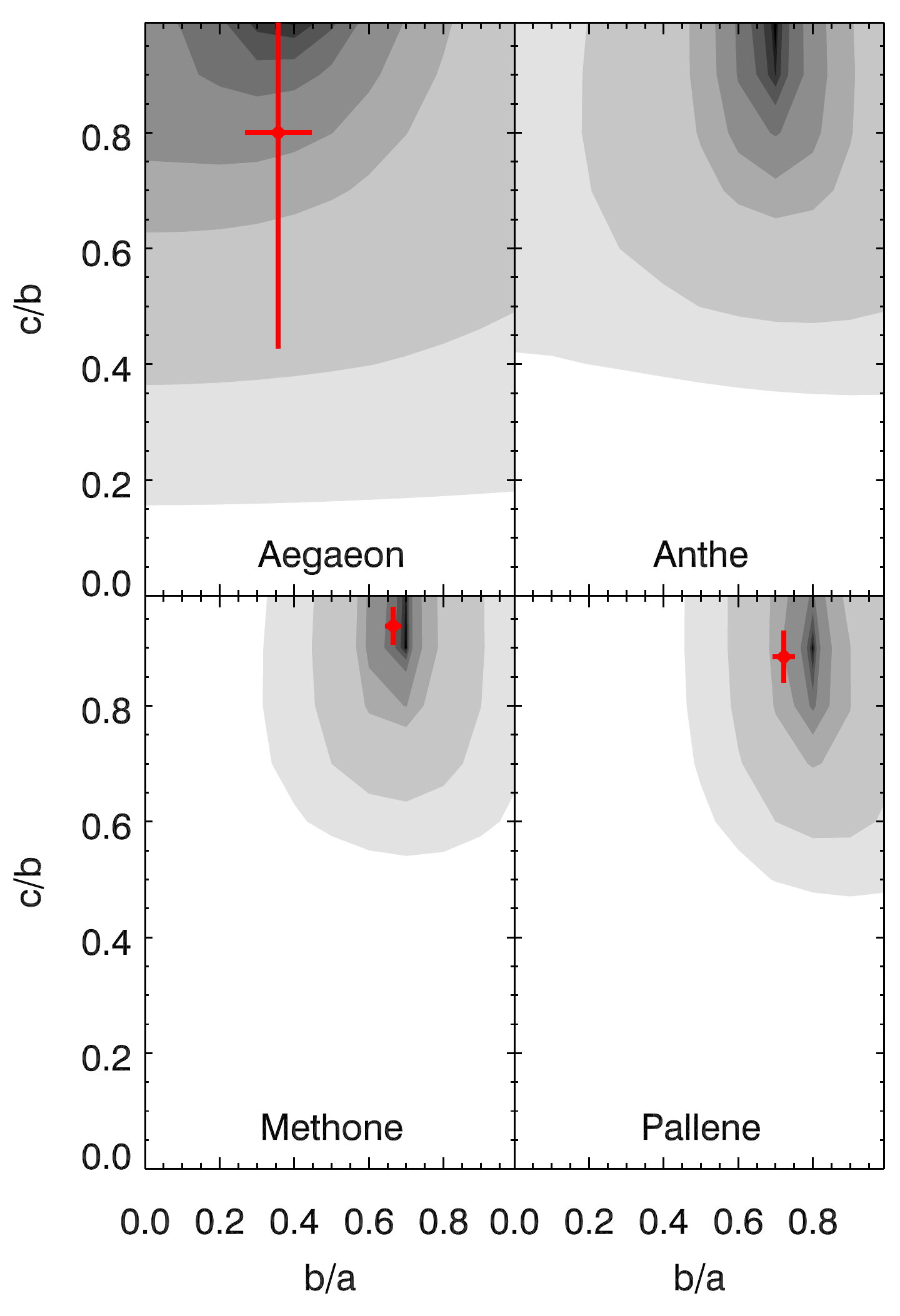}}}
\caption{Photometric analysis of the shapes of Aegaeon, Anthe, Methone and Pallene. Each panel shows the $rms$ dispersion of the residuals between the photometric data and a Lambertian GEPM model with different shape parameters $a/b$ and $b/c$ and a quadratic $B(\alpha)$ function.  The red data points are the estimates of these parameters derived from resolved images. For Aegaeon, Methone and Pallene the best-fit photometric models are reasonably consistent with the observed shapes. For Pallene, the best-fit model has a slightly higher $b/a$ than was observed, but this can probably be attributed to the longitudinal brightness variations in this moon. These data also indicate that Methone and Anthe have very similar aspect ratios.}
\label{shapefig}
\end{figure*}

\begin{figure*}
\centerline{\resizebox{5.5in}{!}{\includegraphics{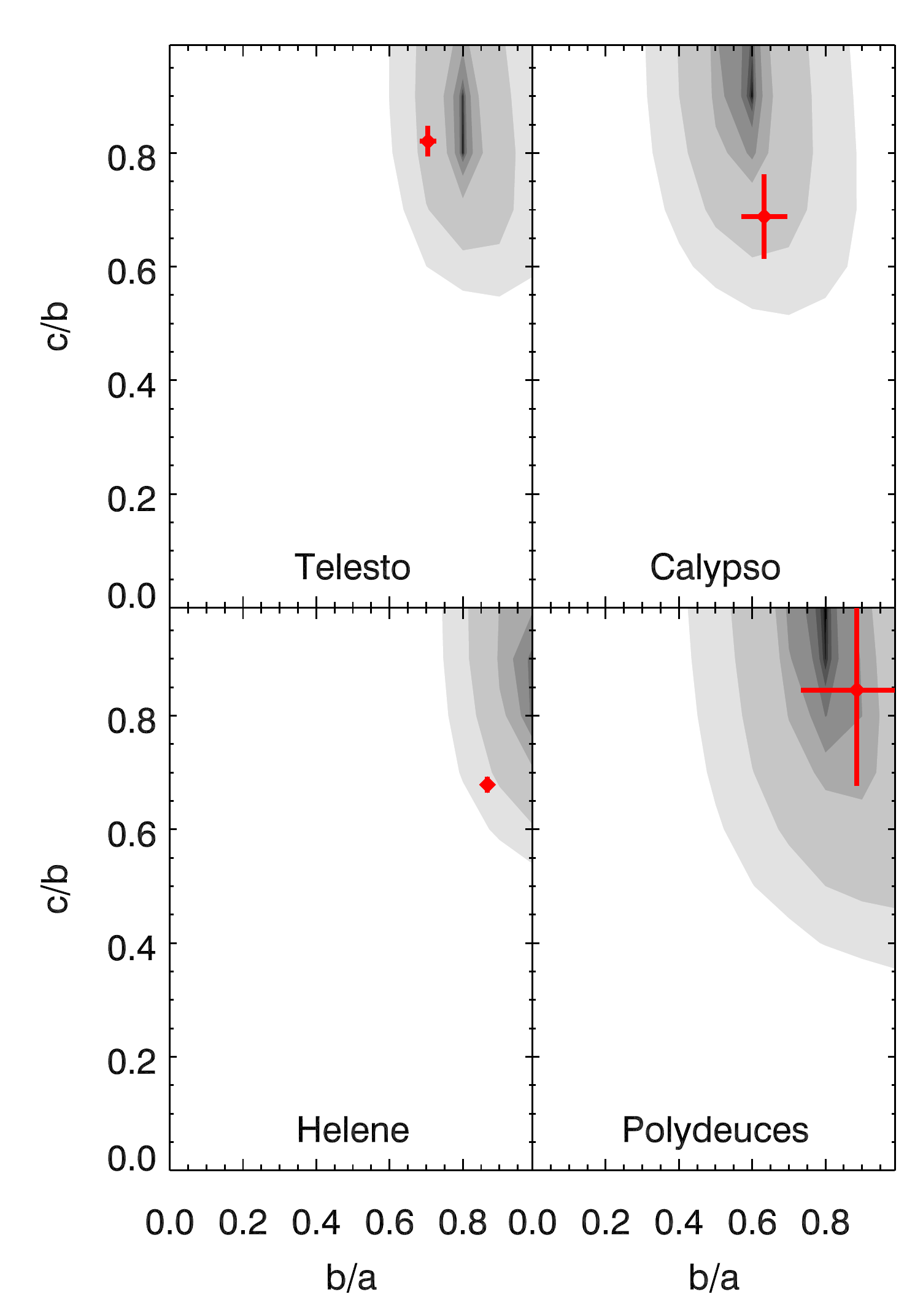}}}
\caption{Photometric analysis of the shapes of Telesto, Calypso, Helene and Polydeuces. Each panel shows the $rms$ dispersion of the residuals between the photometric data and a Lambertian GEPM model with different shape parameters $a/b$ and $b/c$ and a quadratic $B(\alpha)$ function.  The red data points are the estimates of these parameters derived from resolved images. For Calypso and Polydeuces the best-fit photometric models are reasonably consistent with the observed shapes. For Telesto, the best-fit model has a slightly higher $b/a$ than was observed, and for Helene the best-fit model has a $b/a=1$. }
\label{shapefigt}
\end{figure*}

Finally, we may consider what happens if we relax the assumption that the shapes of these moons equal  the best-fit values from resolved images. Figures~\ref{shapefig} and~\ref{shapefigt} show the $rms$ residuals as functions of the aspect ratios $b/a$ and $c/b$ for Aegaeon, Anthe, Methone, Pallene and the trojan moons, along with the aspect ratios derived from the resolved images. For the smaller moons shown in Figure~\ref{shapefig}, the two methods for determining the shape agree fairly well. The best-fit solution for Aegeaon falls where $b \simeq c$ and $b \simeq 0.3 a$, consistent with the constraints from the resolved images.  For Methone, the two methods yield best-fit solutions with the same value of $c/b \simeq 0.95$ and $b/a \simeq 0.65$, although the photometry favors a slightly higher value for $b/a$. Similarly, for Pallene we find both methods give $c/b \simeq 0.9$, but the resolved images favor $b/a \simeq 0.7$, while the photometry prefers 0.8. This discrepancy is likely related to the variations in the surface brightness with longitude mentioned above. More sophisticated photometric models could potentially resolve this difference, but even this level of consistency gives us some confidence in this method. Also, the photometric data for Anthe clearly favor a shape with $b/a \simeq 0.7$ and $c/b \simeq 0.95$, which implies that Anthe has a shape similar to Methone. 

The shapes of the trojan moons derived from the photometry and resolved images are somewhat more discrepant. For Polydeuces, both methods yield $b/a \simeq 0.8$ and $c/b \simeq 1$ with rather large uncertainties. For Calypso, both methods agree that $b/a\simeq0.6$, but the photometry favors $c/b \simeq 0.9$ while the resolved images prefer $c/b \simeq 0.7$. For Telesto, both methods agree that $c/b\simeq0.8$, but the photometry favors $b/a \simeq 0.8$ while the resolved images prefer $b/a \simeq 0.7$. 
The biggest discrepancies are found with Helene, where the photometry prefers a much more spherical shape than found with the resolved images. These differences most likely arise because the shapes of these objects are not simple ellipsoids like the smaller moons appear to be. Again, a more sophisticated photometric model that could accommodate more complex shapes would likely reduce these discrepancies. Even so,  the relatively simple GEPM clearly reduces the dispersion in the brightness estimates for Telesto, Polydeuces, and especially Calypso, and so it provides a useful basis for comparing the brightnesses of all these satellites.

\clearpage

\begin{table}
\caption{Brightness Coefficients for Saturn's Moons based on data between 20$^\circ$ and 40$^\circ$ phase assuming a Lambertian surface scattering law. The first error bars are statistical, second are systematic errors due to uncertainties in the object's average size.}
\label{brighttab}
\centerline{\resizebox{6.5in}{!}{\begin{tabular}{lccccc}\hline
$B$ at $\alpha=30^\circ$ & All Data & Sub-Saturn Quadrant & Leading Quadrant & Anti-Saturn Quadrant & Trailing Quadrant \\
$C_B=dB/d\alpha$ (per degree) & & & & & \\ \hline
Pan & 0.6492$\pm$0.0058$\pm$0.0142 & 0.6398$\pm$0.0233$\pm$0.0140 & 0.6314$\pm$0.0056$\pm$0.0138 & 0.6293$\pm$0.0126$\pm$0.0138 & 0.6744$\pm$0.0089$\pm$0.0148 \\ 
 & -0.0071$\pm$0.0011$\pm$0.0002 & -0.0135$\pm$0.0041$\pm$0.0003 & -0.0066$\pm$0.0011$\pm$0.0001 & -0.0056$\pm$0.0022$\pm$0.0001 & -0.0042$\pm$0.0016$\pm$0.0001 \\ \hline
Atlas & 0.6254$\pm$0.0084$\pm$0.0084 & 0.5960$\pm$0.0092$\pm$0.0080 & 0.6475$\pm$0.0100$\pm$0.0087 & 0.6334$\pm$0.0242$\pm$0.0085 & 0.5898$\pm$0.0180$\pm$0.0079 \\ 
 & -0.0055$\pm$0.0014$\pm$0.0001 & -0.0151$\pm$0.0015$\pm$0.0002 & -0.0051$\pm$0.0016$\pm$0.0001 & -0.0070$\pm$0.0046$\pm$0.0001 &  0.0026$\pm$0.0031$\pm$0.0000 \\ \hline
Prometheus & 0.7928$\pm$0.0127$\pm$0.0130 & 0.8153$\pm$0.0184$\pm$0.0133 & 0.7550$\pm$0.0164$\pm$0.0123 & 0.7924$\pm$0.0236$\pm$0.0130 & 0.8129$\pm$0.0420$\pm$0.0133 \\ 
 & -0.0083$\pm$0.0025$\pm$0.0001 & -0.0038$\pm$0.0048$\pm$0.0001 & -0.0084$\pm$0.0027$\pm$0.0001 & -0.0062$\pm$0.0047$\pm$0.0001 & -0.0168$\pm$0.0090$\pm$0.0003 \\ \hline
Pandora & 0.7600$\pm$0.0053$\pm$0.0057 & 0.7638$\pm$0.0320$\pm$0.0057 & 0.7704$\pm$0.0052$\pm$0.0058 & 0.7570$\pm$0.0225$\pm$0.0057 & 0.7550$\pm$0.0052$\pm$0.0057 \\ 
 & -0.0072$\pm$0.0010$\pm$0.0001 & -0.0089$\pm$0.0054$\pm$0.0001 & -0.0069$\pm$0.0009$\pm$0.0001 & -0.0049$\pm$0.0049$\pm$0.0000 & -0.0073$\pm$0.0009$\pm$0.0001 \\ \hline
Janus & 0.5305$\pm$0.0082$\pm$0.0027 & 0.5726$\pm$0.0138$\pm$0.0029 & 0.4654$\pm$0.0055$\pm$0.0024 & 0.4989$\pm$0.0120$\pm$0.0026 & 0.5814$\pm$0.0043$\pm$0.0030 \\ 
 & -0.0075$\pm$0.0015$\pm$0.0000 & -0.0081$\pm$0.0024$\pm$0.0000 & -0.0069$\pm$0.0009$\pm$0.0000 & -0.0057$\pm$0.0020$\pm$0.0000 & -0.0104$\pm$0.0010$\pm$0.0001 \\ \hline
Epimetheus & 0.4628$\pm$0.0060$\pm$0.0026 & 0.4665$\pm$0.0141$\pm$0.0026 & 0.4480$\pm$0.0038$\pm$0.0025 & 0.4422$\pm$0.0080$\pm$0.0025 & 0.5188$\pm$0.0103$\pm$0.0029 \\ 
 & -0.0049$\pm$0.0010$\pm$0.0000 & -0.0041$\pm$0.0024$\pm$0.0000 & -0.0042$\pm$0.0006$\pm$0.0000 & -0.0079$\pm$0.0015$\pm$0.0000 & -0.0076$\pm$0.0019$\pm$0.0000 \\ \hline
Aegaeon & 0.1793$\pm$0.0050$\pm$0.0326 & 0.1648$\pm$0.0084$\pm$0.0300 & 0.1582$\pm$0.0042$\pm$0.0288 & 0.1352$\pm$0.0921$\pm$0.0246 & 0.2437$\pm$0.0145$\pm$0.0443 \\ 
 & -0.0009$\pm$0.0009$\pm$0.0002 & -0.0019$\pm$0.0008$\pm$0.0003 & -0.0034$\pm$0.0006$\pm$0.0006 & -0.0086$\pm$0.0183$\pm$0.0016 & -0.0056$\pm$0.0022$\pm$0.0010 \\ \hline
Mimas & 0.7546$\pm$0.0028$\pm$0.0015 & 0.7533$\pm$0.0040$\pm$0.0015 & 0.7340$\pm$0.0016$\pm$0.0015 & 0.7622$\pm$0.0073$\pm$0.0015 & 0.7862$\pm$0.0034$\pm$0.0016 \\ 
 & -0.0072$\pm$0.0006$\pm$0.0000 & -0.0093$\pm$0.0009$\pm$0.0000 & -0.0081$\pm$0.0004$\pm$0.0000 & -0.0079$\pm$0.0012$\pm$0.0000 & -0.0075$\pm$0.0007$\pm$0.0000 \\ \hline
Methone & 0.5558$\pm$0.0033$\pm$0.0115 & 0.5631$\pm$0.0081$\pm$0.0117 & 0.5529$\pm$0.0042$\pm$0.0114 & 0.5542$\pm$0.0083$\pm$0.0115 & 0.5498$\pm$0.0066$\pm$0.0114 \\ 
 & -0.0058$\pm$0.0006$\pm$0.0001 & -0.0048$\pm$0.0003$\pm$0.0001 & -0.0060$\pm$0.0009$\pm$0.0001 & -0.0073$\pm$0.0015$\pm$0.0002 & -0.0049$\pm$0.0011$\pm$0.0001 \\ \hline
Pallene & 0.5228$\pm$0.0066$\pm$0.0164 & 0.5556$\pm$0.0126$\pm$0.0174 & 0.4660$\pm$0.0057$\pm$0.0146 & 0.5519$\pm$0.0114$\pm$0.0173 & 0.5267$\pm$0.0078$\pm$0.0165 \\ 
 & -0.0071$\pm$0.0011$\pm$0.0002 & -0.0051$\pm$0.0005$\pm$0.0002 & -0.0076$\pm$0.0012$\pm$0.0002 & -0.0085$\pm$0.0021$\pm$0.0003 & -0.0052$\pm$0.0012$\pm$0.0002 \\ \hline
Enceladus & 1.1530$\pm$0.0029$\pm$0.0009 & 1.1451$\pm$0.0034$\pm$0.0009 & 1.1215$\pm$0.0043$\pm$0.0009 & 1.1580$\pm$0.0049$\pm$0.0009 & 1.1711$\pm$0.0034$\pm$0.0009 \\ 
 & -0.0107$\pm$0.0005$\pm$0.0000 & -0.0104$\pm$0.0006$\pm$0.0000 & -0.0095$\pm$0.0010$\pm$0.0000 & -0.0126$\pm$0.0009$\pm$0.0000 & -0.0103$\pm$0.0006$\pm$0.0000 \\ \hline
Tethys & 0.9363$\pm$0.0046$\pm$0.0011 & 0.9408$\pm$0.0050$\pm$0.0011 & 0.9550$\pm$0.0035$\pm$0.0011 & 0.9704$\pm$0.0106$\pm$0.0011 & 0.8903$\pm$0.0067$\pm$0.0010 \\ 
 & -0.0094$\pm$0.0008$\pm$0.0000 & -0.0103$\pm$0.0008$\pm$0.0000 & -0.0087$\pm$0.0007$\pm$0.0000 & -0.0086$\pm$0.0022$\pm$0.0000 & -0.0081$\pm$0.0011$\pm$0.0000 \\ \hline
Telesto & 0.9028$\pm$0.0100$\pm$0.0220 & 0.9365$\pm$0.0148$\pm$0.0228 & 0.8383$\pm$0.0087$\pm$0.0204 & 0.9671$\pm$0.0147$\pm$0.0236 & 0.8494$\pm$0.0216$\pm$0.0207 \\ 
 & -0.0125$\pm$0.0019$\pm$0.0003 & -0.0088$\pm$0.0006$\pm$0.0002 & -0.0084$\pm$0.0014$\pm$0.0002 & -0.0151$\pm$0.0034$\pm$0.0004 & -0.0112$\pm$0.0042$\pm$0.0003 \\ \hline
Calypso & 1.0999$\pm$0.0176$\pm$0.0463 & 1.0628$\pm$0.0224$\pm$0.0448 & 1.1331$\pm$0.0239$\pm$0.0477 & 1.1645$\pm$0.0240$\pm$0.0490 & 1.2138$\pm$0.0712$\pm$0.0511 \\ 
 & -0.0144$\pm$0.0032$\pm$0.0006 & -0.0123$\pm$0.0011$\pm$0.0005 & -0.0107$\pm$0.0045$\pm$0.0004 & -0.0251$\pm$0.0061$\pm$0.0011 & -0.0028$\pm$0.0134$\pm$0.0001 \\ \hline
Dione & 0.6785$\pm$0.0099$\pm$0.0005 & 0.6669$\pm$0.0105$\pm$0.0005 & 0.7634$\pm$0.0094$\pm$0.0005 & 0.7027$\pm$0.0151$\pm$0.0005 & 0.5561$\pm$0.0144$\pm$0.0004 \\ 
 & -0.0088$\pm$0.0017$\pm$0.0000 & -0.0039$\pm$0.0020$\pm$0.0000 & -0.0064$\pm$0.0015$\pm$0.0000 & -0.0062$\pm$0.0033$\pm$0.0000 & -0.0049$\pm$0.0021$\pm$0.0000 \\ \hline
Helene & 1.0251$\pm$0.0148$\pm$0.0113 & 1.0807$\pm$0.0187$\pm$0.0119 & 0.8373$\pm$0.1644$\pm$0.0093 & 1.0544$\pm$0.0308$\pm$0.0117 & 0.9522$\pm$0.0289$\pm$0.0105 \\ 
 & -0.0196$\pm$0.0027$\pm$0.0002 & -0.0086$\pm$0.0007$\pm$0.0001 & -0.0074$\pm$0.0224$\pm$0.0001 & -0.0169$\pm$0.0058$\pm$0.0002 & -0.0001$\pm$0.0064$\pm$0.0000 \\ \hline
Polydeuces & 0.6075$\pm$0.0089$\pm$0.0794 & 0.6034$\pm$0.0120$\pm$0.0789 & 0.6476$\pm$0.0205$\pm$0.0846 & 0.5945$\pm$0.0090$\pm$0.0777 & 0.6266$\pm$0.0121$\pm$0.0819 \\ 
 & -0.0071$\pm$0.0016$\pm$0.0009 & -0.0073$\pm$0.0006$\pm$0.0010 & -0.0072$\pm$0.0030$\pm$0.0009 & -0.0094$\pm$0.0017$\pm$0.0012 & -0.0300$\pm$0.0029$\pm$0.0039 \\ \hline
Rhea & 0.7144$\pm$0.0057$\pm$0.0006 & 0.6967$\pm$0.0066$\pm$0.0005 & 0.7937$\pm$0.0046$\pm$0.0006 & 0.7146$\pm$0.0042$\pm$0.0006 & 0.6321$\pm$0.0031$\pm$0.0005 \\ 
 & -0.0110$\pm$0.0011$\pm$0.0000 & -0.0109$\pm$0.0011$\pm$0.0000 & -0.0093$\pm$0.0009$\pm$0.0000 & -0.0120$\pm$0.0011$\pm$0.0000 & -0.0065$\pm$0.0004$\pm$0.0000 \\ \hline
%Iapetus & 0.1806$\pm$0.0205$\pm$0.0007 & 0.1448$\pm$0.0185$\pm$0.0006 & 0.0985$\pm$0.0018$\pm$0.0004 & 0.0000$\pm$0.0000$\pm$0.0000 & 0.4593$\pm$0.0035$\pm$0.0018 \\ 
% &  0.0030$\pm$0.0033$\pm$0.0000 &  0.0052$\pm$0.0034$\pm$0.0000 &  0.0044$\pm$0.0002$\pm$0.0000 &  0.0000$\pm$0.0000$\pm$0.0000 & -0.0018$\pm$0.0013$\pm$0.0000 \\ \hline
\end{tabular}}}
\end{table}

\begin{table}
{\caption{Brightness Coefficients for Saturn's Moons based on data between 20$^\circ$ and 40$^\circ$ phase assuming a Minnaert surface scattering law with $k=0.75$. The first error bars are statistical, second are systematic errors due to uncertainties in the object's average size.}}
\label{brighttab75}
\centerline{\resizebox{6.5in}{!}{\begin{tabular}{lccccc}\hline
$B$ at $\alpha=30^\circ$ & All Data & Sub-Saturn Quadrant & Leading Quadrant & Anti-Saturn Quadrant & Trailing Quadrant \\
$C_B=dB/d\alpha$ (per degree) & & & & & \\ \hline
Pan & 0.6379$\pm$0.0061$\pm$0.0140 & 0.6289$\pm$0.0284$\pm$0.0138 & 0.6190$\pm$0.0060$\pm$0.0136 & 0.6166$\pm$0.0144$\pm$0.0135 & 0.6633$\pm$0.0091$\pm$0.0145 \\ 
 & -0.0080$\pm$0.0011$\pm$0.0002 & -0.0152$\pm$0.0050$\pm$0.0003 & -0.0072$\pm$0.0012$\pm$0.0002 & -0.0063$\pm$0.0025$\pm$0.0001 & -0.0053$\pm$0.0016$\pm$0.0001 \\ \hline
Atlas & 0.6122$\pm$0.0088$\pm$0.0082 & 0.5863$\pm$0.0108$\pm$0.0079 & 0.6347$\pm$0.0110$\pm$0.0085 & 0.6126$\pm$0.0215$\pm$0.0082 & 0.5787$\pm$0.0187$\pm$0.0078 \\ 
 & -0.0064$\pm$0.0015$\pm$0.0001 & -0.0174$\pm$0.0018$\pm$0.0002 & -0.0058$\pm$0.0017$\pm$0.0001 & -0.0074$\pm$0.0041$\pm$0.0001 &  0.0016$\pm$0.0032$\pm$0.0000 \\ \hline
Prometheus & 0.7702$\pm$0.0127$\pm$0.0126 & 0.8051$\pm$0.0208$\pm$0.0132 & 0.7404$\pm$0.0184$\pm$0.0121 & 0.7668$\pm$0.0236$\pm$0.0125 & 0.7848$\pm$0.0398$\pm$0.0128 \\ 
 & -0.0108$\pm$0.0025$\pm$0.0002 & -0.0114$\pm$0.0054$\pm$0.0002 & -0.0089$\pm$0.0031$\pm$0.0001 & -0.0108$\pm$0.0047$\pm$0.0002 & -0.0198$\pm$0.0086$\pm$0.0003 \\ \hline
Pandora & 0.7498$\pm$0.0052$\pm$0.0056 & 0.7639$\pm$0.0299$\pm$0.0057 & 0.7577$\pm$0.0061$\pm$0.0057 & 0.7397$\pm$0.0220$\pm$0.0055 & 0.7481$\pm$0.0059$\pm$0.0056 \\ 
 & -0.0075$\pm$0.0010$\pm$0.0001 & -0.0102$\pm$0.0051$\pm$0.0001 & -0.0073$\pm$0.0011$\pm$0.0001 & -0.0048$\pm$0.0048$\pm$0.0000 & -0.0078$\pm$0.0011$\pm$0.0001 \\ \hline
Janus & 0.5257$\pm$0.0084$\pm$0.0027 & 0.5692$\pm$0.0142$\pm$0.0029 & 0.4595$\pm$0.0054$\pm$0.0024 & 0.4924$\pm$0.0115$\pm$0.0025 & 0.5777$\pm$0.0051$\pm$0.0030 \\ 
 & -0.0077$\pm$0.0015$\pm$0.0000 & -0.0085$\pm$0.0025$\pm$0.0000 & -0.0070$\pm$0.0009$\pm$0.0000 & -0.0058$\pm$0.0019$\pm$0.0000 & -0.0107$\pm$0.0012$\pm$0.0001 \\ \hline
Epimetheus & 0.4588$\pm$0.0060$\pm$0.0026 & 0.4651$\pm$0.0145$\pm$0.0026 & 0.4433$\pm$0.0039$\pm$0.0025 & 0.4375$\pm$0.0078$\pm$0.0025 & 0.5136$\pm$0.0105$\pm$0.0029 \\ 
 & -0.0051$\pm$0.0011$\pm$0.0000 & -0.0043$\pm$0.0024$\pm$0.0000 & -0.0045$\pm$0.0006$\pm$0.0000 & -0.0083$\pm$0.0015$\pm$0.0000 & -0.0078$\pm$0.0019$\pm$0.0000 \\ \hline
Aegaeon & 0.1758$\pm$0.0049$\pm$0.0320 & 0.1620$\pm$0.0079$\pm$0.0295 & 0.1512$\pm$0.0042$\pm$0.0275 & 0.1323$\pm$0.0939$\pm$0.0241 & 0.2386$\pm$0.0138$\pm$0.0434 \\ 
 & -0.0008$\pm$0.0008$\pm$0.0001 & -0.0016$\pm$0.0008$\pm$0.0003 & -0.0037$\pm$0.0006$\pm$0.0007 & -0.0099$\pm$0.0186$\pm$0.0018 & -0.0054$\pm$0.0021$\pm$0.0010 \\ \hline
Mimas & 0.7498$\pm$0.0028$\pm$0.0015 & 0.7506$\pm$0.0038$\pm$0.0015 & 0.7285$\pm$0.0015$\pm$0.0015 & 0.7576$\pm$0.0076$\pm$0.0015 & 0.7796$\pm$0.0037$\pm$0.0016 \\ 
 & -0.0073$\pm$0.0006$\pm$0.0000 & -0.0094$\pm$0.0008$\pm$0.0000 & -0.0082$\pm$0.0004$\pm$0.0000 & -0.0082$\pm$0.0013$\pm$0.0000 & -0.0076$\pm$0.0008$\pm$0.0000 \\ \hline
Methone & 0.5508$\pm$0.0035$\pm$0.0114 & 0.5663$\pm$0.0087$\pm$0.0117 & 0.5488$\pm$0.0043$\pm$0.0114 & 0.5423$\pm$0.0073$\pm$0.0112 & 0.5415$\pm$0.0074$\pm$0.0112 \\ 
 & -0.0070$\pm$0.0007$\pm$0.0001 & -0.0050$\pm$0.0003$\pm$0.0001 & -0.0064$\pm$0.0009$\pm$0.0001 & -0.0090$\pm$0.0014$\pm$0.0002 & -0.0064$\pm$0.0013$\pm$0.0001 \\ \hline
Pallene & 0.5181$\pm$0.0064$\pm$0.0163 & 0.5537$\pm$0.0127$\pm$0.0174 & 0.4622$\pm$0.0050$\pm$0.0145 & 0.5468$\pm$0.0097$\pm$0.0172 & 0.5210$\pm$0.0065$\pm$0.0164 \\ 
 & -0.0077$\pm$0.0011$\pm$0.0002 & -0.0052$\pm$0.0005$\pm$0.0002 & -0.0078$\pm$0.0010$\pm$0.0002 & -0.0095$\pm$0.0018$\pm$0.0003 & -0.0058$\pm$0.0010$\pm$0.0002 \\ \hline
Enceladus & 1.1462$\pm$0.0029$\pm$0.0009 & 1.1391$\pm$0.0035$\pm$0.0009 & 1.1146$\pm$0.0043$\pm$0.0009 & 1.1511$\pm$0.0048$\pm$0.0009 & 1.1639$\pm$0.0035$\pm$0.0009 \\ 
 & -0.0110$\pm$0.0005$\pm$0.0000 & -0.0108$\pm$0.0006$\pm$0.0000 & -0.0099$\pm$0.0010$\pm$0.0000 & -0.0130$\pm$0.0009$\pm$0.0000 & -0.0106$\pm$0.0006$\pm$0.0000 \\ \hline
Tethys & 0.9311$\pm$0.0046$\pm$0.0011 & 0.9360$\pm$0.0049$\pm$0.0011 & 0.9494$\pm$0.0035$\pm$0.0011 & 0.9645$\pm$0.0108$\pm$0.0011 & 0.8849$\pm$0.0065$\pm$0.0010 \\ 
 & -0.0096$\pm$0.0008$\pm$0.0000 & -0.0105$\pm$0.0008$\pm$0.0000 & -0.0090$\pm$0.0007$\pm$0.0000 & -0.0087$\pm$0.0023$\pm$0.0000 & -0.0083$\pm$0.0010$\pm$0.0000 \\ \hline
Telesto & 0.8915$\pm$0.0103$\pm$0.0217 & 0.9298$\pm$0.0150$\pm$0.0227 & 0.8245$\pm$0.0067$\pm$0.0201 & 0.9545$\pm$0.0162$\pm$0.0233 & 0.8341$\pm$0.0185$\pm$0.0203 \\ 
 & -0.0134$\pm$0.0019$\pm$0.0003 & -0.0085$\pm$0.0006$\pm$0.0002 & -0.0094$\pm$0.0011$\pm$0.0002 & -0.0169$\pm$0.0037$\pm$0.0004 & -0.0107$\pm$0.0036$\pm$0.0003 \\ \hline
Calypso & 1.0928$\pm$0.0196$\pm$0.0460 & 1.0672$\pm$0.0257$\pm$0.0449 & 1.1091$\pm$0.0336$\pm$0.0467 & 1.1360$\pm$0.0286$\pm$0.0478 & 1.2102$\pm$0.0790$\pm$0.0510 \\ 
 & -0.0143$\pm$0.0035$\pm$0.0006 & -0.0125$\pm$0.0012$\pm$0.0005 & -0.0126$\pm$0.0064$\pm$0.0005 & -0.0226$\pm$0.0073$\pm$0.0010 & -0.0026$\pm$0.0148$\pm$0.0001 \\ \hline
Dione & 0.6747$\pm$0.0098$\pm$0.0005 & 0.6631$\pm$0.0104$\pm$0.0005 & 0.7591$\pm$0.0093$\pm$0.0005 & 0.6987$\pm$0.0151$\pm$0.0005 & 0.5530$\pm$0.0142$\pm$0.0004 \\ 
 & -0.0090$\pm$0.0017$\pm$0.0000 & -0.0041$\pm$0.0020$\pm$0.0000 & -0.0066$\pm$0.0015$\pm$0.0000 & -0.0063$\pm$0.0032$\pm$0.0000 & -0.0051$\pm$0.0020$\pm$0.0000 \\ \hline
Helene & 1.0163$\pm$0.0150$\pm$0.0112 & 1.0758$\pm$0.0184$\pm$0.0119 & 0.8208$\pm$0.2026$\pm$0.0091 & 1.0479$\pm$0.0289$\pm$0.0116 & 0.9473$\pm$0.0256$\pm$0.0105 \\ 
 & -0.0201$\pm$0.0027$\pm$0.0002 & -0.0088$\pm$0.0007$\pm$0.0001 & -0.0081$\pm$0.0276$\pm$0.0001 & -0.0153$\pm$0.0054$\pm$0.0002 & -0.0012$\pm$0.0057$\pm$0.0000 \\ \hline
Polydeuces & 0.6016$\pm$0.0090$\pm$0.0786 & 0.6016$\pm$0.0121$\pm$0.0786 & 0.6375$\pm$0.0212$\pm$0.0833 & 0.5867$\pm$0.0087$\pm$0.0767 & 0.6201$\pm$0.0090$\pm$0.0811 \\ 
 & -0.0074$\pm$0.0016$\pm$0.0010 & -0.0075$\pm$0.0006$\pm$0.0010 & -0.0071$\pm$0.0031$\pm$0.0009 & -0.0099$\pm$0.0017$\pm$0.0013 & -0.0310$\pm$0.0022$\pm$0.0041 \\ \hline
Rhea & 0.7104$\pm$0.0056$\pm$0.0006 & 0.6928$\pm$0.0066$\pm$0.0005 & 0.7894$\pm$0.0046$\pm$0.0006 & 0.7106$\pm$0.0042$\pm$0.0006 & 0.6286$\pm$0.0031$\pm$0.0005 \\ 
 & -0.0112$\pm$0.0011$\pm$0.0000 & -0.0111$\pm$0.0011$\pm$0.0000 & -0.0094$\pm$0.0008$\pm$0.0000 & -0.0121$\pm$0.0011$\pm$0.0000 & -0.0067$\pm$0.0004$\pm$0.0000 \\ \hline
%Iapetus & 0.1793$\pm$0.0203$\pm$0.0007 & 0.1438$\pm$0.0183$\pm$0.0005 & 0.0976$\pm$0.0018$\pm$0.0004 & 0.0000$\pm$0.0000$\pm$0.0000 & 0.4564$\pm$0.0035$\pm$0.0017 \\ 
% &  0.0029$\pm$0.0033$\pm$0.0000 &  0.0051$\pm$0.0034$\pm$0.0000 &  0.0043$\pm$0.0002$\pm$0.0000 &  0.0000$\pm$0.0000$\pm$0.0000 & -0.0020$\pm$0.0013$\pm$0.0000 \\ \hline
\end{tabular}}}
\end{table}

\begin{table}
{\caption{Brightness Coefficients for Saturn's Moons based on data between 20$^\circ$ and 40$^\circ$ phase assuming a Minnaert surface scattering law with $k=0.50$. The first error bars are statistical, second are systematic errors due to uncertainties in the object's average size.}}
\label{brighttab50}
\centerline{\resizebox{6.5in}{!}{\begin{tabular}{lccccc}\hline
$B$ at $\alpha=30^\circ$ & All Data & Sub-Saturn Quadrant & Leading Quadrant & Anti-Saturn Quadrant & Trailing Quadrant \\
$C_B=dB/d\alpha$ (per degree) & & & & & \\ \hline
Pan & 0.6181$\pm$0.0066$\pm$0.0135 & 0.6099$\pm$0.0346$\pm$0.0134 & 0.5976$\pm$0.0068$\pm$0.0131 & 0.5935$\pm$0.0172$\pm$0.0130 & 0.6444$\pm$0.0099$\pm$0.0141 \\ 
 & -0.0093$\pm$0.0012$\pm$0.0002 & -0.0174$\pm$0.0061$\pm$0.0004 & -0.0079$\pm$0.0013$\pm$0.0002 & -0.0072$\pm$0.0030$\pm$0.0002 & -0.0067$\pm$0.0018$\pm$0.0001 \\ \hline
Atlas & 0.5903$\pm$0.0095$\pm$0.0079 & 0.5682$\pm$0.0141$\pm$0.0076 & 0.6134$\pm$0.0124$\pm$0.0082 & 0.5808$\pm$0.0207$\pm$0.0078 & 0.5595$\pm$0.0201$\pm$0.0075 \\ 
 & -0.0075$\pm$0.0016$\pm$0.0001 & -0.0202$\pm$0.0024$\pm$0.0003 & -0.0066$\pm$0.0019$\pm$0.0001 & -0.0080$\pm$0.0039$\pm$0.0001 &  0.0006$\pm$0.0034$\pm$0.0000 \\ \hline
Prometheus & 0.7362$\pm$0.0130$\pm$0.0120 & 0.7796$\pm$0.0255$\pm$0.0128 & 0.7167$\pm$0.0207$\pm$0.0117 & 0.7280$\pm$0.0242$\pm$0.0119 & 0.7486$\pm$0.0375$\pm$0.0122 \\ 
 & -0.0138$\pm$0.0025$\pm$0.0002 & -0.0192$\pm$0.0066$\pm$0.0003 & -0.0096$\pm$0.0035$\pm$0.0002 & -0.0160$\pm$0.0048$\pm$0.0003 & -0.0225$\pm$0.0081$\pm$0.0004 \\ \hline
Pandora & 0.7299$\pm$0.0057$\pm$0.0055 & 0.7540$\pm$0.0330$\pm$0.0057 & 0.7349$\pm$0.0072$\pm$0.0055 & 0.7105$\pm$0.0217$\pm$0.0053 & 0.7329$\pm$0.0074$\pm$0.0055 \\ 
 & -0.0082$\pm$0.0010$\pm$0.0001 & -0.0118$\pm$0.0056$\pm$0.0001 & -0.0082$\pm$0.0013$\pm$0.0001 & -0.0051$\pm$0.0047$\pm$0.0000 & -0.0086$\pm$0.0013$\pm$0.0001 \\ \hline
Janus & 0.5143$\pm$0.0086$\pm$0.0026 & 0.5589$\pm$0.0146$\pm$0.0029 & 0.4478$\pm$0.0055$\pm$0.0023 & 0.4791$\pm$0.0110$\pm$0.0025 & 0.5672$\pm$0.0060$\pm$0.0029 \\ 
 & -0.0082$\pm$0.0015$\pm$0.0000 & -0.0091$\pm$0.0026$\pm$0.0000 & -0.0074$\pm$0.0009$\pm$0.0000 & -0.0061$\pm$0.0018$\pm$0.0000 & -0.0112$\pm$0.0014$\pm$0.0001 \\ \hline
Epimetheus & 0.4494$\pm$0.0061$\pm$0.0025 & 0.4579$\pm$0.0149$\pm$0.0026 & 0.4336$\pm$0.0040$\pm$0.0024 & 0.4272$\pm$0.0075$\pm$0.0024 & 0.5027$\pm$0.0108$\pm$0.0028 \\ 
 & -0.0056$\pm$0.0011$\pm$0.0000 & -0.0047$\pm$0.0025$\pm$0.0000 & -0.0050$\pm$0.0007$\pm$0.0000 & -0.0088$\pm$0.0014$\pm$0.0000 & -0.0082$\pm$0.0020$\pm$0.0000 \\ \hline
Aegaeon & 0.1712$\pm$0.0048$\pm$0.0311 & 0.1578$\pm$0.0077$\pm$0.0287 & 0.1440$\pm$0.0044$\pm$0.0262 & 0.1245$\pm$0.0952$\pm$0.0226 & 0.2322$\pm$0.0132$\pm$0.0422 \\ 
 & -0.0007$\pm$0.0008$\pm$0.0001 & -0.0015$\pm$0.0008$\pm$0.0003 & -0.0040$\pm$0.0007$\pm$0.0007 & -0.0119$\pm$0.0189$\pm$0.0022 & -0.0053$\pm$0.0020$\pm$0.0010 \\ \hline
Mimas & 0.7362$\pm$0.0028$\pm$0.0015 & 0.7391$\pm$0.0036$\pm$0.0015 & 0.7147$\pm$0.0015$\pm$0.0014 & 0.7438$\pm$0.0079$\pm$0.0015 & 0.7640$\pm$0.0041$\pm$0.0015 \\ 
 & -0.0077$\pm$0.0006$\pm$0.0000 & -0.0099$\pm$0.0008$\pm$0.0000 & -0.0085$\pm$0.0004$\pm$0.0000 & -0.0088$\pm$0.0013$\pm$0.0000 & -0.0080$\pm$0.0008$\pm$0.0000 \\ \hline
Methone & 0.5396$\pm$0.0043$\pm$0.0112 & 0.5621$\pm$0.0102$\pm$0.0116 & 0.5395$\pm$0.0056$\pm$0.0112 & 0.5228$\pm$0.0073$\pm$0.0108 & 0.5282$\pm$0.0086$\pm$0.0109 \\ 
 & -0.0085$\pm$0.0008$\pm$0.0002 & -0.0054$\pm$0.0004$\pm$0.0001 & -0.0072$\pm$0.0011$\pm$0.0001 & -0.0109$\pm$0.0013$\pm$0.0002 & -0.0080$\pm$0.0015$\pm$0.0002 \\ \hline
Pallene & 0.5070$\pm$0.0063$\pm$0.0159 & 0.5436$\pm$0.0134$\pm$0.0171 & 0.4538$\pm$0.0049$\pm$0.0142 & 0.5336$\pm$0.0087$\pm$0.0168 & 0.5098$\pm$0.0057$\pm$0.0160 \\ 
 & -0.0086$\pm$0.0011$\pm$0.0003 & -0.0055$\pm$0.0005$\pm$0.0002 & -0.0080$\pm$0.0010$\pm$0.0003 & -0.0107$\pm$0.0016$\pm$0.0003 & -0.0066$\pm$0.0009$\pm$0.0002 \\ \hline
Enceladus & 1.1264$\pm$0.0029$\pm$0.0009 & 1.1200$\pm$0.0036$\pm$0.0009 & 1.0949$\pm$0.0044$\pm$0.0009 & 1.1310$\pm$0.0048$\pm$0.0009 & 1.1433$\pm$0.0036$\pm$0.0009 \\ 
 & -0.0120$\pm$0.0005$\pm$0.0000 & -0.0118$\pm$0.0007$\pm$0.0000 & -0.0107$\pm$0.0010$\pm$0.0000 & -0.0139$\pm$0.0008$\pm$0.0000 & -0.0115$\pm$0.0006$\pm$0.0000 \\ \hline
Tethys & 0.9152$\pm$0.0045$\pm$0.0010 & 0.9205$\pm$0.0048$\pm$0.0010 & 0.9330$\pm$0.0034$\pm$0.0011 & 0.9474$\pm$0.0109$\pm$0.0011 & 0.8694$\pm$0.0063$\pm$0.0010 \\ 
 & -0.0103$\pm$0.0008$\pm$0.0000 & -0.0112$\pm$0.0008$\pm$0.0000 & -0.0097$\pm$0.0006$\pm$0.0000 & -0.0093$\pm$0.0023$\pm$0.0000 & -0.0089$\pm$0.0010$\pm$0.0000 \\ \hline
Telesto & 0.8680$\pm$0.0107$\pm$0.0212 & 0.9083$\pm$0.0154$\pm$0.0222 & 0.8019$\pm$0.0064$\pm$0.0196 & 0.9258$\pm$0.0201$\pm$0.0226 & 0.8096$\pm$0.0149$\pm$0.0197 \\ 
 & -0.0144$\pm$0.0020$\pm$0.0004 & -0.0086$\pm$0.0006$\pm$0.0002 & -0.0106$\pm$0.0010$\pm$0.0003 & -0.0193$\pm$0.0047$\pm$0.0005 & -0.0103$\pm$0.0029$\pm$0.0003 \\ \hline
Calypso & 1.0692$\pm$0.0227$\pm$0.0450 & 1.0550$\pm$0.0296$\pm$0.0444 & 1.0738$\pm$0.0437$\pm$0.0452 & 1.0868$\pm$0.0374$\pm$0.0458 & 1.1901$\pm$0.0861$\pm$0.0501 \\ 
 & -0.0142$\pm$0.0041$\pm$0.0006 & -0.0128$\pm$0.0014$\pm$0.0005 & -0.0150$\pm$0.0083$\pm$0.0006 & -0.0196$\pm$0.0095$\pm$0.0008 & -0.0032$\pm$0.0162$\pm$0.0001 \\ \hline
Dione & 0.6632$\pm$0.0097$\pm$0.0005 & 0.6517$\pm$0.0103$\pm$0.0005 & 0.7462$\pm$0.0092$\pm$0.0005 & 0.6868$\pm$0.0149$\pm$0.0005 & 0.5435$\pm$0.0139$\pm$0.0004 \\ 
 & -0.0094$\pm$0.0017$\pm$0.0000 & -0.0046$\pm$0.0019$\pm$0.0000 & -0.0072$\pm$0.0014$\pm$0.0000 & -0.0069$\pm$0.0032$\pm$0.0000 & -0.0055$\pm$0.0020$\pm$0.0000 \\ \hline
Helene & 0.9935$\pm$0.0151$\pm$0.0110 & 1.0543$\pm$0.0174$\pm$0.0116 & 0.7958$\pm$0.2351$\pm$0.0088 & 1.0264$\pm$0.0275$\pm$0.0113 & 0.9315$\pm$0.0215$\pm$0.0103 \\ 
 & -0.0204$\pm$0.0028$\pm$0.0002 & -0.0094$\pm$0.0007$\pm$0.0001 & -0.0090$\pm$0.0320$\pm$0.0001 & -0.0139$\pm$0.0052$\pm$0.0002 & -0.0029$\pm$0.0048$\pm$0.0000 \\ \hline
Polydeuces & 0.5879$\pm$0.0090$\pm$0.0769 & 0.5925$\pm$0.0123$\pm$0.0774 & 0.6206$\pm$0.0215$\pm$0.0811 & 0.5705$\pm$0.0084$\pm$0.0746 & 0.6065$\pm$0.0060$\pm$0.0793 \\ 
 & -0.0079$\pm$0.0016$\pm$0.0010 & -0.0079$\pm$0.0006$\pm$0.0010 & -0.0073$\pm$0.0032$\pm$0.0010 & -0.0106$\pm$0.0016$\pm$0.0014 & -0.0322$\pm$0.0014$\pm$0.0042 \\ \hline
Rhea & 0.6984$\pm$0.0055$\pm$0.0005 & 0.6811$\pm$0.0065$\pm$0.0005 & 0.7760$\pm$0.0044$\pm$0.0006 & 0.6985$\pm$0.0041$\pm$0.0005 & 0.6179$\pm$0.0031$\pm$0.0005 \\ 
 & -0.0116$\pm$0.0011$\pm$0.0000 & -0.0115$\pm$0.0011$\pm$0.0000 & -0.0100$\pm$0.0008$\pm$0.0000 & -0.0126$\pm$0.0011$\pm$0.0000 & -0.0071$\pm$0.0004$\pm$0.0000 \\ \hline
%Iapetus & 0.1758$\pm$0.0199$\pm$0.0007 & 0.1412$\pm$0.0179$\pm$0.0005 & 0.0955$\pm$0.0018$\pm$0.0004 & 0.0000$\pm$0.0000$\pm$0.0000 & 0.4482$\pm$0.0034$\pm$0.0017 \\ 
% &  0.0027$\pm$0.0032$\pm$0.0000 &  0.0048$\pm$0.0033$\pm$0.0000 &  0.0041$\pm$0.0002$\pm$0.0000 &  0.0000$\pm$0.0000$\pm$0.0000 & -0.0025$\pm$0.0013$\pm$0.0000 \\ \hline
\end{tabular}}}
\end{table}

\clearpage

\pagebreak

\section{Brightness Trends among Saturn's small moons}
\label{trends}

The GEPM appears to provide reasonably consistent surface brightness estimates for moons with a range of sizes and shapes, so we will now compare these brightness estimates for the different moons and identify trends that might clarify the processes responsible for determining the moons' visible surface brightness. We begin by describing the specific brightness values we will use for this project in Section~\ref{estimates}. Section~\ref{ering}  then shows how these brightness parameters for the mid-sized moons correlate with the local flux of E-ring particles, which is consistent with prior work. Next, Section~\ref{radiation} examines why some of the small moons are darker than expected given their locations within the E ring, and shows that high-energy proton radiation probably also influences the moons' surface brightness. Finally, Section~\ref{trojans} discusses  the small moons that are brighter than one might expect given their locations.

\subsection{Comparable brightness estimates for Saturn's moons}
\label{estimates}

The calculations described in Section~\ref{model} demonstrate that our model is a sensible way to quantify the surface brightness of elongated moons. However, the absolute values of $B$ derived by this method are not directly comparable to traditional parameters like geometric or {Bond} albedos previously reported in the literature. In principle, the derived values of $B$ as functions of phase angle could be extrapolated to zero, and then we could translate that value of $B$ to a geometric albedo for an equivalent spherical object. However, in practice performing such an extrapolation would be problematic because the small moons were rarely observed by Cassini at phase angles  below 10$^\circ$, so we have no information about the magnitude or shape of the opposition surges for these bodies. Also, since these objects are not spherical, the amount of light scattered will depend on the orientation of the body, which complicates the classical definition of albedo. On the other hand, it is relatively straightforward to estimate $B$ for Saturn's other moons, and so we have chosen that approach for this analysis.

In order to compare the reflectances of these objects, we only consider the data obtained between phase angles of 20$^\circ$ and 40$^\circ$, and fit the brightness coefficients in this phase range to a linear function of  phase angle $\alpha$:
\begin{equation} B(\alpha) =  B_0 +  (\alpha-30^\circ) C_B  \end{equation} 
Note that $B_0$ is the brightness coefficient at 30$^\circ$ phase, which is in the middle of the range of observed phase angles and so does not depend on any questionable extrapolations.  For a spherical object of radius $r$ observed at $0^\circ$ and $30^\circ$ phase, $a_{\rm pred}=2.12 r^2$ and $1.86 r^2$ respectively for a Lambertian scattering law {(for a Minnaert scattering law with $k=0.5$, these numbers become $a_{\rm pred}=2.12 r^2$ and $1.91 r^2$, respectively)}. Hence the average reflectance the object would have at 30$^\circ$ if it were a sphere is $<I/F>_{\rm sphere}=0.59 B_0$ for a Lambertian scattering law {(or  $<I/F>_{\rm sphere}=0.61 B_0$ for a Minnaert $k=0.5$ scattering law)}. The corresponding geometric albedo of  the object assuming that the linear phase function could be extrapolated to an {appropriately small} phase angle $\alpha_0$ would be\footnote{{Note that at small phase angles the brightness differences for Minnaert and Lambertian scattering laws are negligible since the incidence and emission angles are nearly identical.}}
\begin{equation}
\mathcal{A}_{\rm sphere}=0.67[B_0 +  (\alpha_0-30^\circ) C_B]
\end{equation}
For the mid-sized moons Mimas, Enceladus, Tethys, Dione and Rhea, this expression yields $\mathcal{A}_{\rm sphere}=0.60, 0.92, 0.75, 057$ and 0.63, respectively for $\alpha_0=10^\circ$, which are 0.97, 0.92, 0.90, 0.87 and 1.00 times the central values reported by \citet{Thomas18}, so at this phase angle the calculations are reasonably consistent. However, if we extrapolate to zero degrees phase we obtain  $\mathcal{A}_{\rm sphere}= 0.65, 0.99, 0.82, 0.63$ and 0.55, which are  0.68, 0.72, 0.67, 0.63 and 0.74 times the geometric albedos measured by \citet{Verbiscer07}. This discrepancy most likely reflects the fact that this formula does not account for the opposition surges associated with these moons. 

\begin{figure*}
\centerline{\resizebox{5.5in}{!}{\includegraphics{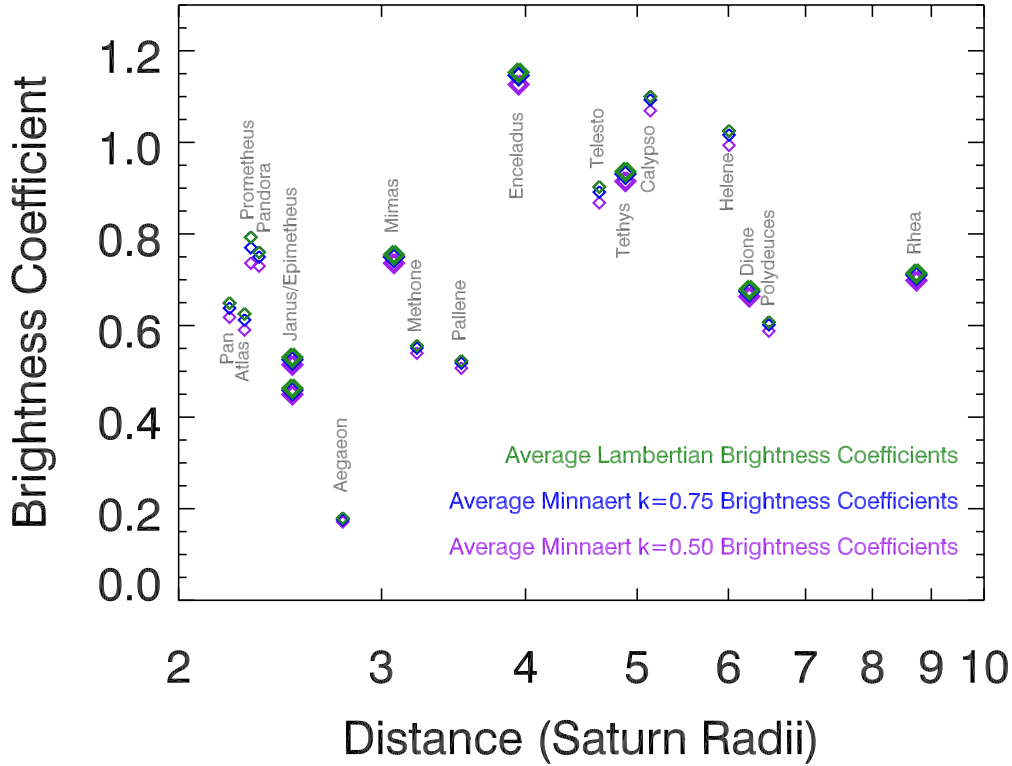}}}
\caption{{Comparisons of the average brightness coefficients of the various moons derived assuming different Minnaert scattering laws (see also Tables~\ref{brighttab}-~\ref{brighttab50}). While models assuming lower values of $k$ do yield systematically lower brightness coefficients, the differences between the different models are fairly subtle and do not alter the trends among the different moons.}}
\label{brightcomp}
\end{figure*}

\begin{figure*}
\centerline{\resizebox{5.5in}{!}{\includegraphics{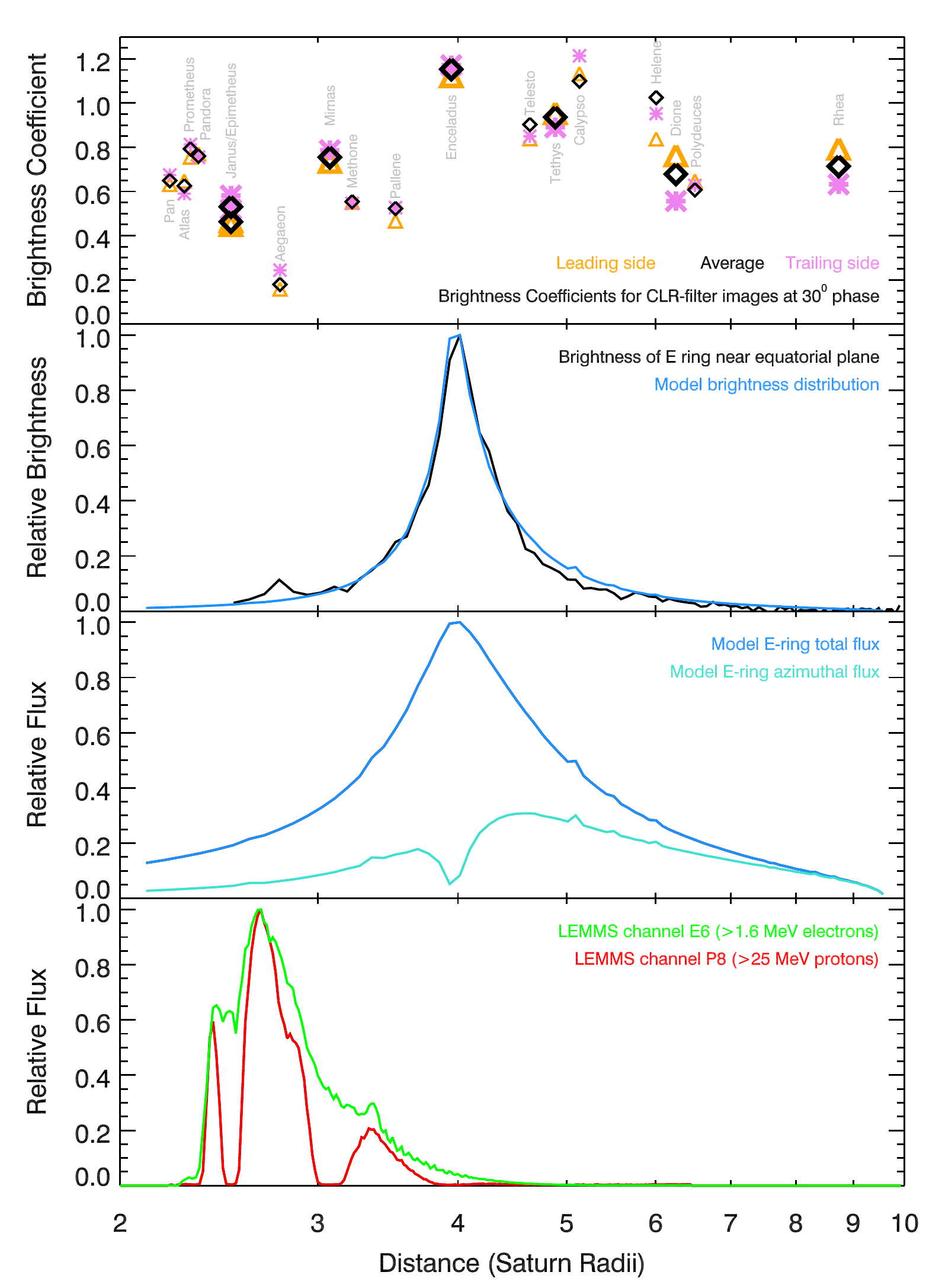}}}
\caption{Correlations between the moon's brightness and their environment. The top panel shows the brightness coefficients of the moons computed assuming a Lambertian scattering law as a function of distance from the planet's center. Note that the locations of the trojan moons Telesto, Calypso, Polydeuces and Helene are offset slightly for the sake of clarity. The next two panels show the E-ring's relative brightness density and the relative flux of E-ring particles into the moons (i.e. the parameters $F$ and $|\delta F_\lambda|$ see text). The bottom panel shows the flux of high-energy protons and electrons derived by the MIMI LEMMS instrument from one early pass through this entire region.}
\label{eringcomp}
\end{figure*}

For the sake of simplicity we will just use the parameters $B_0$  in our analyses below. {Tables~\ref{brighttab}-~\ref{brighttab50} tabulate $B_0$ and $C_B$  for all of Saturn's moons examined here assuming either a Lambertian scattering law or a Minnaert scattering law with $k=0.75$ or $k=0.50$}. These tables also includes statistical errors based on the scatter of the points around the mean trend, and a systematic error which is based on the uncertainty in the mean size of the object. Aditional systematic uncertainties associated with the size of the pixel response function are of order 5\% and are not included here because these systematic uncertainties are common to all the brightness estimates and so do not affect the moons' relative brightness.  
{Figure~\ref{brightcomp} compares the values of $B_0$ derived assuming the different scattering laws for each of the moons. This plot shows that models assuming lower values of the Minnaert $k$ parameter yield systematically lower estimates of $B_0$, but that these differences are rather subtle and have little effect on the relative brightnesses of the various moons. Hence, for the sake of simplicity we will only consider the estimates of $B_0$ computed assuming a Lambertian scattering law from here on.} Figure~\ref{eringcomp} plots these central brightness values as a function of distance from the center of Saturn, along with estimates of the relative E-ring particle density and fluxes derived from images, as well as the relative energetic proton and electron fluxes (see below).

\subsection{Correlations between the moons' brightnesses and the E ring particle flux}
\label{ering}

Figure~\ref{eringcomp} shows that the brightness coefficients for the largest moons (Janus, Epimetheus, Mimas, Tethys, Dione and Rhea) follow the same basic trends as previously reported for these moon's geometric albedos \citep{Verbiscer07, Verbiscer18}. The brightness of these larger moons falls off with distance from Enceladus' orbit, which strongly suggests that the E-ring plays an important role in determining these moons' surface brightness. Furthermore, the leading-trailing asymmetries in these moons' surface brightness are also consistent with how E-ring particles are expected to strike the moons \citep{Buratti98, Schenk11}. Most of the visible E ring particles have orbital semi-major axes close to Enceladus and a range of eccentricities \citep{Horanyi92, HB94}. The E-ring particles outside Enceladus' orbit are therefore closer to the apocenter of their eccentric orbits and so are moving slower around the planet than the moons are. Hence Tethys, Dione and Rhea mostly overtake the E-ring particles and the corresponding flux onto these moons is larger on their leading sides. On the other hand, the E-ring particles inside Enceladus' orbit are closer to their orbital pericenters, and so tend to move faster around the planet than the moons. Hence the E-ring particles tend to preferentially strike the trailing sides of Janus, Epimetheus and Mimas. For all of these moons, the side that is preferentially struck by E-ring particles is brighter, as one would expect if the E-ring flux was responsible for brightening the moons.

The connection between the E-ring and the moons' brightness can be made more quantitative by estimating the flux of E-ring particles onto each of the moons. The E ring consists of particles with a wide range of orbital elements, so detailed numerical simulations will likely be needed to accurately compute the fluxes onto each moon. Such simulations are beyond the scope of this report, and so we will here simply approximate the particle flux based on simplified analytical models of the E ring motivated by the available remote-sensing and in-situ observations.

Prior analyses of E-ring images obtained by the Cassini spacecraft provided maps of the local brightness density within this ring as functions of radius and height above Saturn's equatorial plane. These brightness densities are  proportional to the local number density of particles times the size-dependent scattering efficiency for those particles, so these maps provide relatively direct estimates of the particle density in the vicinity of the moons. For this particular study, we will focus on the E-ring density profile shown in Figure~\ref{eringcomp}, which  is derived from the E130MAP observation made on day 137 of 2006, and is described in detail in \citet{Hedman12}. This observation included a set of wide-angle camera images that provided an extensive and high signal-to-noise edge-on map of the E ring at phase angles around 130$^\circ$ (W1526532467, W1526536067, W1526539667, W1526543267, W1526546867, W1526550467 and W1526554067). These images were assembled into a single mosaic of the edge-on ring, and then onion-peeled  to transform the observed integrated brightness map into a map of the local ring brightness density in a vertical cut through the ring \citep{Hedman12}. The profile of the brightness density near Saturn's equatorial plane was extracted from this map as the average brightness in regions between 1000 and 2000 km from Saturn's equator plane after removing a background level based on the average brightness 20,000-30,000 km away from that plane. Note the region used here deliberately excluded the region within 1000 km of the equator plane to minimize contamination from the G ring interior to 175,000 km. The vertical scale height of the E ring is sufficiently large that including data in the 1000-2000 km range should provide a reasonable estimate of the density in the plane containing all these moons. 

Consistent with previous studies, the E-ring brightness density is strongly peaked around Enceladus' orbit. However, this profile also differs from the profiles reported elsewhere in the literature. For example,  this profile is much more strongly peaked than that the profile used by \citet{Verbiscer07}. This is because the \citet{Verbiscer07} profile was of the E-ring's vertically integrated brightness, rather than the brightness density close to the equatorial plane. The latter quantity is more sharply peaked because the E-ring's vertical thickness increases with distance from Enceladus' orbit, and because the ring is warped so that its peak brightness shifts away from Saturn's equator plane far from Enceladus' orbit \citep{Kempf08, Hedman12, Ye16}. Since all the moons orbit close to the planet's equator plane, the profile shown in Figure~\ref{eringcomp} is a better representation of the relative dust density surrounding the moons.  On the other hand, the profile shown in Figure~\ref{eringcomp} is also a bit broader than in-situ measurements would predict. These data have generally been fit to models where the particle density falls off with distance from Enceladus' orbit like power laws with very large indices \citep{Kempf08, Ye16}.  Assuming that the brightness density is proportional to the particle number density, these models tend to underpredict the signals seen exterior to 5 $R_S$ and interior to 3 $R_S$. These discrepancies arise in part because the in-situ instruments are only sensitive to grains with radii larger than 0.9 microns, while the images are sensitive to somewhat smaller particles that probably have a broader spatial distribution \citep{Horanyi08, Hedman12}. 
%For reference, the in-situ instruments find peak particle densities (for particles more than 1 micron in radius) between 0.02 and 0.2 particles per cubic meter \citep{Kempf08, Ye16}, and size distributions with differential slopes around -4 \citep{Ye14}.  
The profile shown in Figure~\ref{eringcomp} therefore should provide a reasonable estimate of the relative densities of the particles larger than 0.5 $\mu$m across within this ring.

However, for the purposes of understanding how the E-ring affects the moons' surface properties, we are primarily interested in the E-ring particle {flux} into the moons, which depends not only on the particle density, but also on the particles' velocity distribution. Since a detailed investigation of the E-ring's orbital element distribution is beyond the scope of this paper,  we will here consider a simple analytical model of the E-ring particles' orbital properties that can provide useful rough estimates of the particle flux into the various moons. 

Prior analyses of E-ring images indicate that most of the visible particles in this ring have semi-major axes close to Enceladus' orbit, and that the large size of this ring is because the particles have a wide range of eccentricities and inclinations. Furthermore, the images indicate that mean eccentricities, inclinations and semi-major axes were strongly correlated with each other \citep{Hedman12}. We therefore posit that the E-ring density distribution seen in Figure~\ref{eringcomp} primarily reflects a distribution of eccentricities $\mathcal{F}(e)$. We also allow the semi-major axes of the particles to vary, but for the sake of simplicity we assume that the semi-major axes of these particles $a$ is a strict function of $e$: $a=[240,0000+(e/0.1)\times5000]$ km, which is consistent with the imaging data \citep{Hedman12}. Since we are only concerned with particles near Saturn's equatorial plane, we will not consider the inclination distributions here. 

For a given eccentricity distribution $\mathcal{F}(e)$, the particle density as a function of radius $d(r)$ should be given by the following integral:
\begin{equation}
d(r)=\int_{e_{min}}^1 \frac{2}{|v_r| T} \mathcal{F}(e) de,
\end{equation}
where $v_r$ is the radial velocity of the particles and $T$ is their orbital period, and the factor of two arises because particles with a finite eccentricity passes through each radius twice. The integral's lower limit is $e_{min}=|1-r/a|$ because only particles with eccentricities greater than this can reach the observed $r$.
Similarly, the average flux of particles onto a body on a circular orbit at a given radius is given by the integral.
\begin{equation}
{F}(r)=\int_{e_{min}}^1 \frac{1}{2|v_r| T} \mathcal{F}(e)\sqrt{v_r^2+\delta v_\lambda^2} de,
\end{equation}
where $v_r$ is the particle's radial velocity as defined above, and $\delta v_\lambda$ is the particle's azimuthal velocity measured relative to the local circular orbital velocity. Note that a factor of $1/4$ arises from averaging over the moon's surface. Since the radial motions of particles on any given orbit are symmetric inwards and outwards, in this model there are no differences in the fluxes on the sub-Saturn and anti-Saturn sides of the moon. However, because the particles at a given radius can have different average azimuthal velocities, there can be asymmetries in the fluxes on the moon's leading and trailing sides. These asymmetries can be parametrized by the following integral:
\begin{equation}
\delta F_\lambda(r)=\int_{e_{min}}^1 \frac{1}{2|v_r| T} \mathcal{F}(e)\delta v_\lambda de,
\end{equation}
which is {\bf half} the difference between the fluxes on the leading and trailing points. 

In order to evaluate these integrals, recall the standard expressions for  $v_r$  and $\delta v_\lambda$ in terms of orbital elements  \citep{MD99}:
\begin{equation}
v_r=\frac{na}{\sqrt{1-e^2}}e\sin f
\end{equation}
\begin{equation}
\delta v_\lambda=\frac{na}{\sqrt{1-e^2}}(1+e\cos f)-na\sqrt{a/r}
\end{equation}
where $n=2\pi/T$ is the mean motion of the particle and $f$ is the true anomaly of the particle's orbit. At a given radius $r$ we will be observing particles at different mean anomalies given by the standard expression:
\begin{equation}
r=\frac{a(1-e^2)}{1+e\cos f}
\end{equation}
so we can re-express $|v_r|$ and $\delta v_\lambda$ in terms of $r$
\begin{equation}
|v_r|=\frac{na^2}{r}\sqrt{e^2-(r/a-1)^2},
\end{equation}
\begin{equation}
\delta v_\lambda=\frac{na^2}{r}\left[\sqrt{1-e^2}-\sqrt{r/a}\right].
\label{vlambda}
\end{equation}
Inserting these expressions into the above integrals, we find the density, flux and flux asymmetries can all be expressed as the following integrals over the eccentricity distribution (remember that $a$ is implicitly a function of $e$):
\begin{equation}
d(r)=\int_{e_{min}}^1 \frac{r\mathcal{F}(e)}{\pi a^2\sqrt{e^2-(r/a-1)^2}} de,
\end{equation}
\begin{equation}
F(r)=\int_{e_{min}}^1 \frac{n\mathcal{F}(e)}{4\pi}\sqrt{1+\frac{\left(\sqrt{1-e^2}-\sqrt{r/a}\right)^2}{e^2-(r/a-1)^2}} de,
\end{equation}
\begin{equation}
F_\lambda(r)=\int_{e_{min}}^1 \frac{n\mathcal{F}(e)(\sqrt{1-e^2}-\sqrt{r/a})}{4\pi\sqrt{e^2-(r/a-1)^2}} de.
\end{equation}

It turns out that the  observed density distribution  shown in Figure~\ref{eringcomp} can be  matched reasonably well by assuming the eccentricity is a Lorentzian times a regularization term to avoid singularities at $e=0$:
\begin{equation}
\mathcal{F}(e)=\frac{\mathcal{F}_0}{1+e^2/e_0^2}[1-\exp(-e/e_c)]
\end{equation}
with constants $\mathcal{F}_0$, $e_0$ and $e_c$. The specific model density distribution shown in Figure~\ref{eringcomp} has $e_0=0.17$ and $e_c=0.01$. Note that $e_c$ basically only affects the density levels close to Enceladus' orbit, while $e_0$ determines how quickly the signal falls off away from the E-ring core.

Using this same ansatz for $\mathcal{F}(e)$, we obtain estimates for how both the flux and flux asymmetry vary with radius, which are shown in the  third panel of Figure~\ref{eringcomp}. First note that the flux is much less sharply peaked than the density. This occurs because the particles are moving faster relative to the local circular velocity at larger distances from Enceladus' orbit thanks to their higher eccentricities. Also note that azimuthal component of the flux is larger exterior to Enceladus' orbit than it is interior to that moon. This asymmetry arises because exterior to Enceladus the particles are all moving slower than the local circular velocity, while interior to Enceladus, the particles are all moving faster than the local circular speed, but the azimuthal components of their velocity can still fall below the local circular velocity depending on their true anomaly. 

Finally, we  can translate these trends in the relative flux into rough estimates of the absolute flux of E-ring particles using recent in-situ measurements of the particle number density in the E-ring core. In principle, the brightness density seen in images can be translated into estimates of the particle number density, but in practice the relevant conversion factor depends on the size-dependent light scattering efficiency of the particles. By contrast, the peak E-ring particle density has been directly measured to be between 0.02 and 0.2 /m$^3$ with various in-situ measurements \citep{Kempf08, Ye16}.  Now, as mentioned above, these densities are for the particles larger than  the detection thresholds of these instruments, which is slightly larger than the sizes of the particles seen in remote-sensing images. Hence the number density of visible particles could be somewhat larger than these measurements. Nevertheless, these are still the best estimates of the absolute particle density near the core of the E ring, and they should still provide a useful order-of-magnitude estimate of the visible particles. For the sake of concreteness, we will here assume a peak number density of 0.03 particles/m$^3$, which is the value measured by \citet{Ye16} for particles larger than 1 $\mu$m in radius. 

\begin{figure}
\resizebox{6in}{!}{\includegraphics{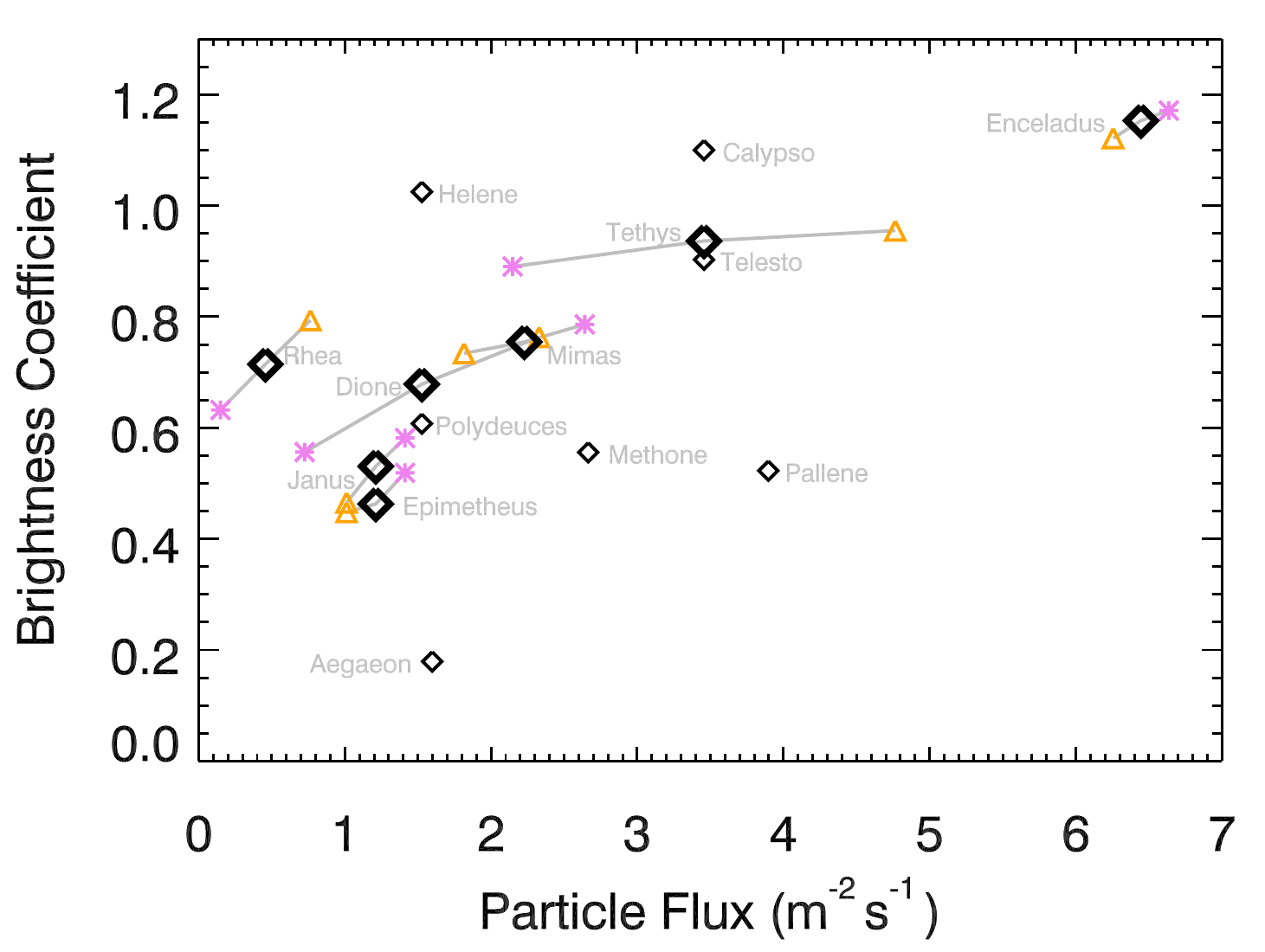}}
\caption{Moon brightness coefficients versus computed E-ring particle flux, assuming peak number density of 0.03/m$^3$, $e_0=0.17$ and $e_{min}=0.01$. The diamonds are the average values for each moon, while the  orange triangles and magenta stars indicate the values for the leading and trailing sides of the mid-sized moons.}
\label{eringflux}
\end{figure}

Figure~\ref{eringflux} and Table~\ref{fluxtab} give the brightness coefficients and estimated E-ring particle fluxes for all the moons known to be embedded in the E ring. For the mid-sized moons and the co-orbitals, there is a very good correlation between the moon's brightness coefficients and the estimated fluxes. This correlation not only applies to the average values for each moon, but also to their leading and trailing sides, where the estimated flux values are taken to be $F\pm\delta F_\lambda/\sqrt{2}$ (the factor of $\sqrt{2}$ accounts for the fact that the observations span a full quadrant of each body). The one mid-sized moon that falls noticeably off the main trend is Rhea. {This could be related to other unusual aspects of Rhea's surface. Rhea's visible spectrum is redder than any of the other mid-sized moons \citep{Thomas18}, and its near-infrared spectrum exhibits stronger ice bands than Dione \citep{Filacchione12, Scipioni14}.  Furthermore, radio-wave data indicate that Rhea has a higher albedo at centimeter wavelengths than Dione \citep{Black07, LeGall19}. Various explanations have been put forward to these anomalies, including differences in the moons' regolith structure \citep{Ostro10}, and variations in the energetic particle and/or ring particle flux \citep{Scipioni14, Thomas18}. The latter option  is probably the more likely explanation for Rhea's excess brightness at visible wavelengths, since the moon is in a region where both the predicted fluxes of  E-ring grains and energetic particles are expected to be very low (see also below). Indeed, our simple E-ring model is probably not completely accurate near Rhea's orbit. Detailed numerical simulations of the E-ring particles reveal that there could be a population of sub-micron particles orbiting in the outskirts of the E ring close to Rhea's orbit \citep{Horanyi08}. This population could potentially increase the particle flux into Rhea, moving that data point closer to the trend defined by Mimas and Dione, but it is not clear whether it could also explain that moon's spectral properties.} In any case, overall these data confirm that the E-ring particle flux is an important factor for determining the surface brightness of the mid-sized and co-orbital satellites.

Turning to the smaller moons, however, the situation is more complicated. Two of the co-orbital moons --Telesto and Polydeuces-- fall close to the same trend as their larger companions. However, the other two co-orbitals --Calypso and Helene-- fall noticeably above this trend, Meanwhile, Aegaeon, Methone and Pallene all fall well below the trend for the mid-sized moons. This implies that something besides the E-ring flux is affecting the brightnesses of these small moons.

\begin{table}
\caption{Particle and radiation fluxes}
\label{fluxtab}
\centerline{\begin{tabular}{| l | c c c | c c c |  r |}\hline
Moon & \multicolumn{3}{c|}{$B_0^a$} & \multicolumn{3}{c|}{E-ring flux$^b$} &  \multicolumn{1}{c|}{Radiation flux$^c$} \\
 & \multicolumn{3}{c|}{ }  &\multicolumn{3}{c|}{(part. m$^{-2}$sec$^{-1})$} & \multicolumn{1}{c|}{(protons cm$^{-2}$sec$^{-1})$}  \\
 & Ave. & Lead & Trail & Ave. & Lead & Trail &  \multicolumn{1}{c|}{ in range 25-59 MeV} \\ \hline
%Pan             &  0.69 & 0.68 & 0.72 & 0.92 & 0.77 & 1.06 & ---  & & --- \\     
%Atlas            & 0.67 & 0.69 & 0.63 & 0.97 & 0.81 & 1.13 &  --- & & --- \\ 
%Prometheus & 0.84 & 0.80 & 0.86 & 1.00 & 0.84 & 1.15 & ---- & & --- \\ 
%Pandora       & 0.81 & 0.82 & 0.80 &  1.04 & 0.87 & 1.21 & ---- & & --- \\ 
Janus           & 0.53 & 0.47 & 0.58  &  1.21 & 1.01 & 1.41 &     5.4 \\
Epimetheus  & 0.46 & 0.45 & 0.52 &  1.21 & 1.01 & 1.41 &    7.4 \\
Aegaeon      & 0.18 & 0.16 & 0.24 &  1.60 & 1.32 & 1.88 & 920.4 \\
Mimas          & 0.75 & 0.73 & 0.79 &  2.23 & 1.81 & 2.64 &    5.6 \\
Methone       & 0.56 & 0.55 & 0.55 &  2.66 & 2.15 & 3.17 &   140.2 \\
Pallene         & 0.53 &  0.47 & 0.53 & 3.89 & 3.19 & 4.60 &   162.0 \\
Enceladus    & 1.15 & 1.12 & 1.17 &  6.44 & 6.25 & 6.63 &       4.4 \\
Tethys           & 0.94 & 0.96 & 0.89 &  3.46 & 4.77 & 2.15 &     3.2 \\
Telesto          & 0.90 & 0.84 & 0.84 &  3.46 & 4.77 & 2.15 &    3.2 \\
Calypso        &  1.10 & 1.13 & 1.21 &  3.46 &4.77 & 2.15 &      3.2 \\
Dione            & 0.68 & 0.76 & 0.56 & 1.53 & 2.33 & 0.72 &       3.1 \\
Helene          & 1.02 & 0.84 & 0.95 & 1.53 & 2.33 & 0.72 &      3.1 \\
Polydeuces   & 0.61 & 0.65 & 0.63 & 1.53 & 2.33 & 0.72 &      3.1 \\
Rhea$^d$      & 0.71 & 0.79 & 0.63 & 0.46 & 0.76 & 0.15 &    3.1 \\
\hline
\end{tabular}}
$^a$ Brightness coefficient at 30$^\circ$ computed assuming surface follows Lambertian scattering law, see Table~\ref{brighttab}.

$^b$ E-ring flux computed assuming $e_0=0.17$, $e_c=0.01$ and a peak number density of 0.03 particles per cubic meter. Leading and trailing side values given by $F\pm \delta F_\lambda/\sqrt{2}$

$^c$ Radiation fluxes computed assuming $\pi$ steradian viewing angle.

$^d$ Radiation fluxes at Rhea assumed to be the same as those for Dione
\end{table}

%for i=0,18 do print, moons(i), brights(i,0,0), eringn(i)*1e12, (brights(i,4,0)-brights(i,2,0))/brights(i,0,0), eringhf(i)/eringf(i), radf(i), radfn(i)/eringn(i)*1e8

\subsection{Darkening Aegaeon, Methone and Pallene with radiation}
\label{radiation}

\begin{figure*}
\resizebox{6.5in}{!}{\includegraphics{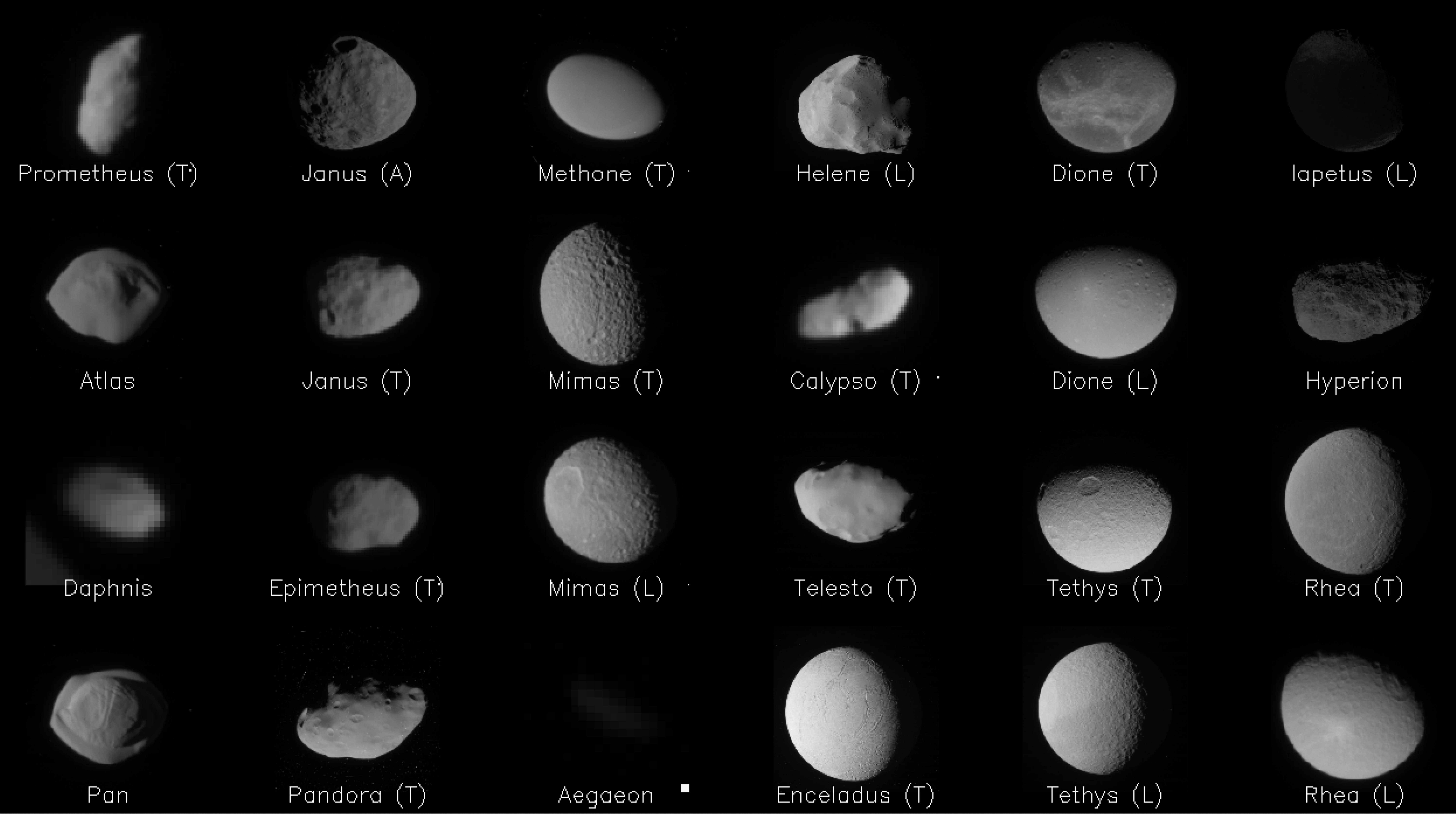}}
\caption{Resolved images of Saturn's moons at phase angles between 50$^\circ$ and 60$^\circ$, shown with a common stretch with a gamma of 0.75. Letters following the names indicate the hemisphere observed (T=trailing, L=leading, A=anti-Saturn). This gallery clearly demonstrates that Aegaeon is anomalously dark and has a comparable surface brightness to Iapetus' dark side. {Image files used for this mosaic are: N1643265020 = Aegaeon,  N1870695654 = Atlas, 
N1506184171 = Calypso (T), N1656997840 = Daphnis, N1597675353 = Dione (T), N1606072353 = Dione (L), 
N1824725635 = Enceladus (T), N1521539514 = Epimetheus (T), N1687121104 = Helene (L), N1669612923 = Hyperion, N1510339442 = Iapetus (L), N1521539514 = Janus  (T),  N1627323065 = Janus (A), N1716192290 = Methone (T), N1484509816 = Mimas (L), N1855894795 = Mimas (T), N1867602424 = Pan, N1504612286 = Pandora (T), N1853392689 = Prometheus (T), N1505772509 = Rhea (L), N1591815592 = Rhea (T), N1514163567 = Telesto (T), N1855868106 = Tethys (L), N1563723519 = Tethys (T).}
}
\label{ims}
\end{figure*}

Aegaeon, Methone and Pallene are all considerably darker than one would have predicted given their locations and the observed trend between Janus, Epimetheus, Mimas and Enceladus. Aegaeon in particular is exceptionally dark, with a surface brightness coefficient  around 0.2.  This conclusion is consistent with the few resolved images of these bodies, which show that only the dark side of Iapetus has a comparably low surface brightness as Aegaeon (see Figure~\ref{ims}). In principle, the darkness of these moons could be due to a number of different factors, including their small size and their associations with other dusty rings, but the most likely explanation involves their locations within Saturn's radiation belts.

Aegaeon, Methone and Pallene are much smaller than the mid-sized moons, and so  their small sizes could somehow be responsible for their low surface brightness. Indeed, based on preliminary investigations of Aegaeon's low surface brightness, \citet{Hedman10} suggested that a recent collision catastrophically disrupted the body, stripping off its bright outer layers to produce the debris that is now the G ring, and leaving behind a dark remnant. Deeper examination, however, does not support this model. For example, if a collision was sufficient to expose the dark core of a kilometer-scale body, then small impacts should expose dark materials on the other moons, but no such dark patches have been seen in any of the high-resolution images obtained by Cassini. More fundamentally, attributing Aegaeon's extreme darkness to some discrete event in the past is difficult because the E-ring should be able to brighten this moon relatively quickly. Assuming typical particle sizes of around 1 $\mu$m, the above estimates of the E-ring flux implies that Aegaeon would acquire a layer of E-ring particles 10 $\mu$m thick in only 100,000 years. This calculation neglects mixing within Aegaeon's regolith and any material ejected from Aegaeon, but even this basic calculation shows that E-ring particles would cause Aegaeon to brighten on very short timescales, which strongly suggests that whatever process is making Aegaeon dark is actively ongoing. 

Another possibility is that Aegaeon, Methone and Pallene are dark because they are embedded not only in the E ring, but also in narrower rings like the G ring.  However, since these rings likely consist of debris knocked off the moons' surfaces, this does not necessarily explain why this material would be so dark. Another problem is that Prometheus and Pandora are rather bright, even though these moons are close to the F ring and therefore likely coated in F-ring material (see below). Furthermore, since the rings/arcs associated with  Methone and Pallene are over an order of magnitude fainter than the G ring \citep{Hedman18}, it seems unlikely that these tenuous rings could significantly affect the surface properties of those two moons. Thus the G ring and other dusty rings do not appear to provide a natural explanation for the darkness of these small moons. 

A more plausible explanation for the low surface brightnesses of Aegaeon, Methone and Pallene is that these moons orbit Saturn within regions that have unique radiation environments. The bottom panel of Figure~\ref{eringcomp} shows profiles of energetic proton and electron fluxes from two of the highest energy channels of Cassini's Low-Energy Magnetospheric Measurement System (LEMMS) instrument \citep{Krimigis04}. Note that while the electron flux has a broad peak centered around the orbit of Aegaeon,  the high-energy proton flux is confined to three belts between the main rings and the orbits of Janus, Mimas and Enceladus \citep{Roussos08}. This disjoint distribution of high-energy protons arises because energetic protons circle the planet under the influence of the corotation electric field, but their longitudinal motion is enhanced by gradient-curvature drifts, which act in the same direction as corotation.  These particles therefore orbit the planet with a period of one to a few hours.  Combined with the alignment of the spin and dipole axes, this drift allows these protons to re-encounter the inner moons frequently, causing a permanent flux decrease (macrosignature) along their orbits for sufficiently energetic protons.  It is important to realize that these macrosignatures are not just depressions in the energetic proton density, but are also regions where the {\em flux} of protons is greatly decreased.  Proton macrosignatures along the moon orbits are visible in the data at energies of $>$ 300 keV or so, but are only expected to exist at much higher energies for electrons. Since Janus and Mimas are exposed to comparable amounts of energetic electrons as Aegaeon, Methone and Pallene, the electron flux is not a natural explanation for the smaller moons' low surface brightness. However,  Aegaeon, Methone and Pallene are exposed to much higher fluxes of high-energy protons than any of Saturn's other moons, so this is a plausible potential explanation for the darkness of these three moons.

Table~\ref{fluxtab} provides more quantitative estimates of the proton fluxes into the different moons. Here we use the protons observed with the P8 channel on the LEMMS instrument as a proxy for the total flux of high-energy protons \citep{Krimigis04}. In the radiation belts, this channel's sensitivity is mainly to protons of 25.2 to 59 MeV, and the numbers reported in Table~\ref{fluxtab} are for an acceptance solid angle of 1 steradian to facilitate comparisons with the E-ring particle flux. Since Mimas, Janus and Epimetheus occupy narrow gaps in the radiation belts, the fluxes provided in this table are averaged over the moons' true anomalies. Furthermore, for Janus and Epimetheus we also average over the semi-major axis variations associated with these moons' co-orbital interactions. Also note that while estimates of the Galactic Cosmic Ray background have been removed from these data, the fluxes for the moons beyond Enceladus are probably upper limits. Nevertheless, these numbers clearly show that Methone, Pallene and Aegaeon experience exceptionally high proton fluxes.
 
 \begin{figure}
 \resizebox{6in}{!}{\includegraphics{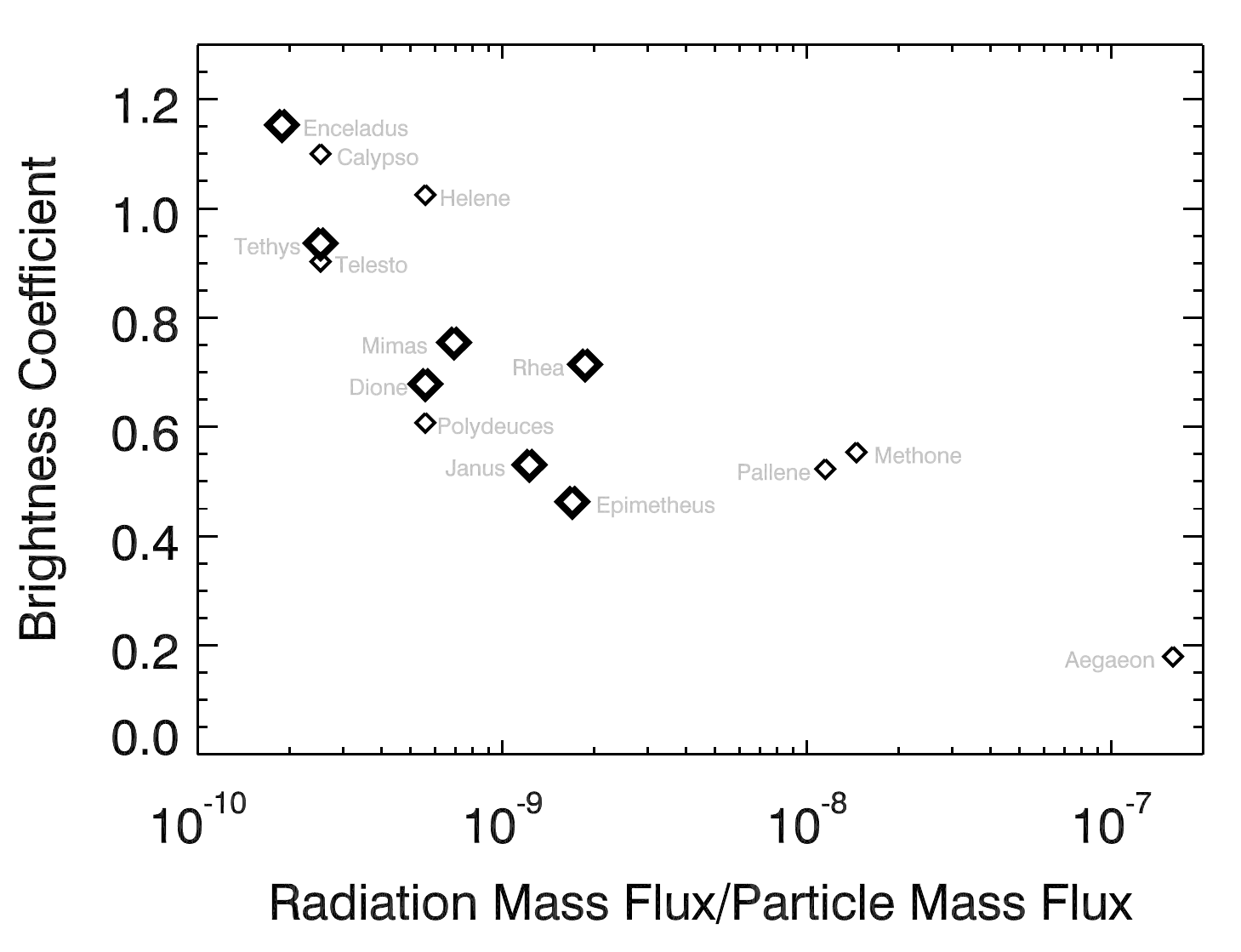}}
 \caption{Moon  brightness versus the ratio of the radiation mass flux to the E-ring particle flux. The radiation mass flux here is flux of protons with energies between 25 and 50 MeV, while the E-ring mass flux is the mass of particles larger than 1 micron. Note that the radiation fluxes for Methone, Pallene and Aegaeon are measured, while for the other moons they may represent upper limits. Still, the overall trend of decreasing brightness with increasing flux ratio is clear.}
 \label{radflux}
 \end{figure}
 
If radiation damage from high-energy protons is the dominant darkening process and the E-ring particles are the dominant brightening process, then we would expect the moons' equilibrium surface brightness to depend the ratio of the radiation flux to the E-ring particle flux. To make this ratio properly unitless, we consider the  mass flux ratio for high-energy protons and E-ring particles. This quantity is computed by taking the number flux of protons given in Table~\ref{fluxtab} multiplied by the proton mass and dividing that number by the average flux of E-ring particles given in Table~\ref{fluxtab} multiplied by  $m_{\rm eff}$,  the effective average mass of particles with radii larger than 1 $\mu$m. Specifically, we assume  $m_{\rm eff}=4\pi \ln(100)\times 10^{-15}$~kg, which is consistent with  a power-law size distribution with a differential index of -4 \citep{Ye14} extending between 1 and 100 microns, and a particle mass density of 1 g/cm$^3$.  Figure~\ref{radflux} shows that the moons' brightnesses do indeed systematically decrease as this ratio increases. Furthermore, Methone, Pallene and Aegaeon fall along the same basic trend in this plot as the mid-sized and co-orbital moons. This match provides strong evidence that radiation damage is the dominant process responsible for making Methone, Pallene and Aegaeon dark.

At the moment, we do not know precisely which components of the radiation flux are responsible for the darkening. The macrosignatures are most prominent in the highest-energy proton channels, which suggests energetic protons are the dominant agent, but other agents (e.g., electrons with even higher energies) may also contribute at some level. If the darkening agent is protons with energies greater than 1 MeV, there are several ways that they could be altering the surface. \citet{Howett11} suggested that energetic electrons ``sinter" or fuse grains in the uppermost layer, which can affect the reflectance properties at a range of wavelengths \citep{Schenk11, Howett18}.  Electrons are mainly slowed down by interactions with electrons in materials.  Protons are initially slowed down mainly by interactions with electrons in materials, although as they slow down, they interact more with nuclei. So it is possible very energetic protons (in the first fraction of their mean range) are causing the same kinds of changes to the ice as energetic electrons. Furthermore, they have large gyroradii and so they can often affect the whole body, i.e. they don't weather the surface differentially in many cases.  Radiation damage can also change the chemical properties of surface materials, {such as generating complex organics \citep{Hapke86, Poston18} or producing color centers in salts  \citep{Hand15, Hibbitts19, Trumbo19}, which could darken the surface and produce distinctive spectral signatures. }

\begin{figure}
\resizebox{6.5in}{!}{\includegraphics{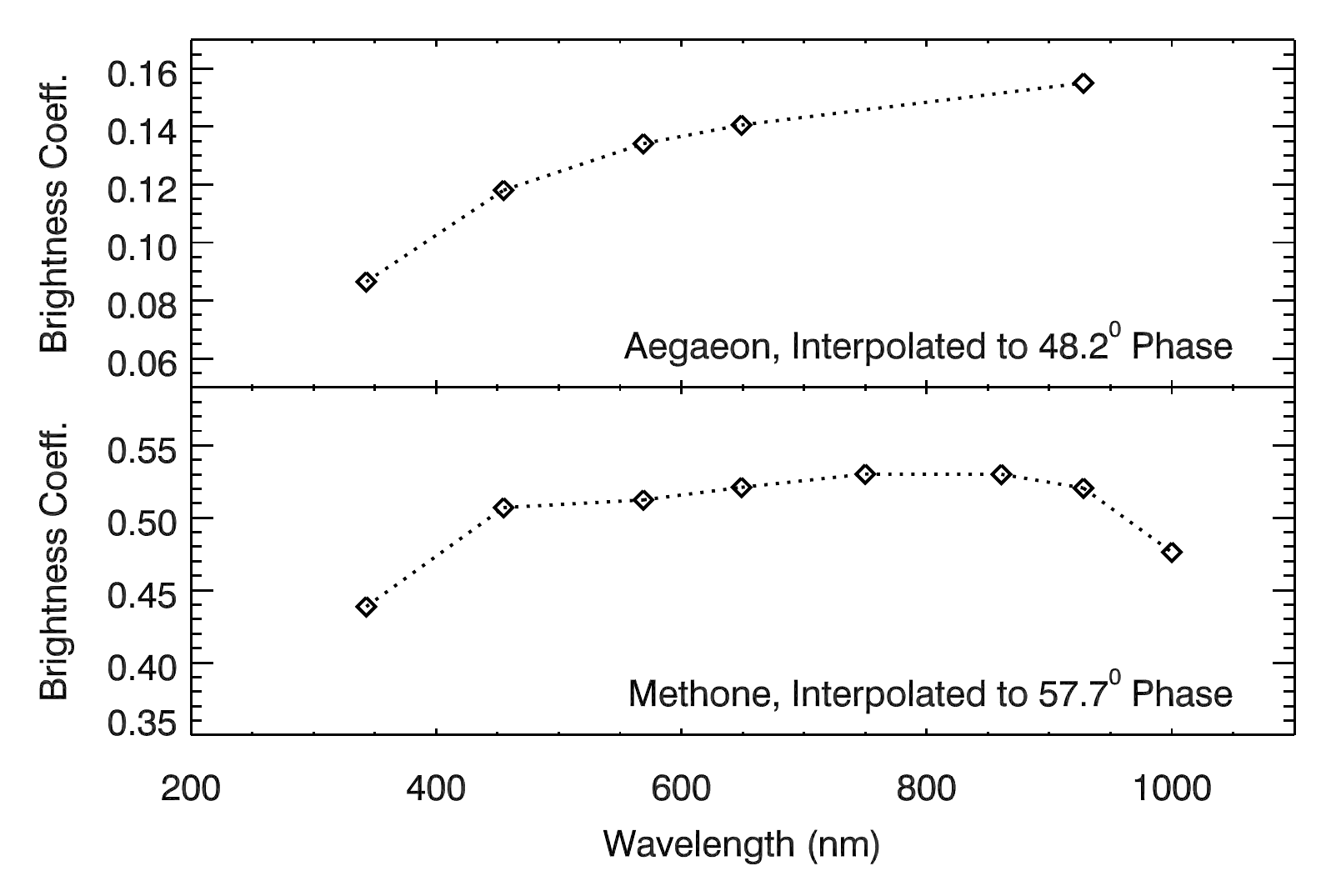}}
\caption{Visible spectra of Aegaeon and Methone obtained from resolved images during Cassini's closest encounters with these moons. The only clear spectral feature is the break in slope around 500 nm, which is commonly found in spectra of the icy moons.}
\label{colors}
\end{figure}

\begin{figure}
\resizebox{6.5in}{!}{\includegraphics{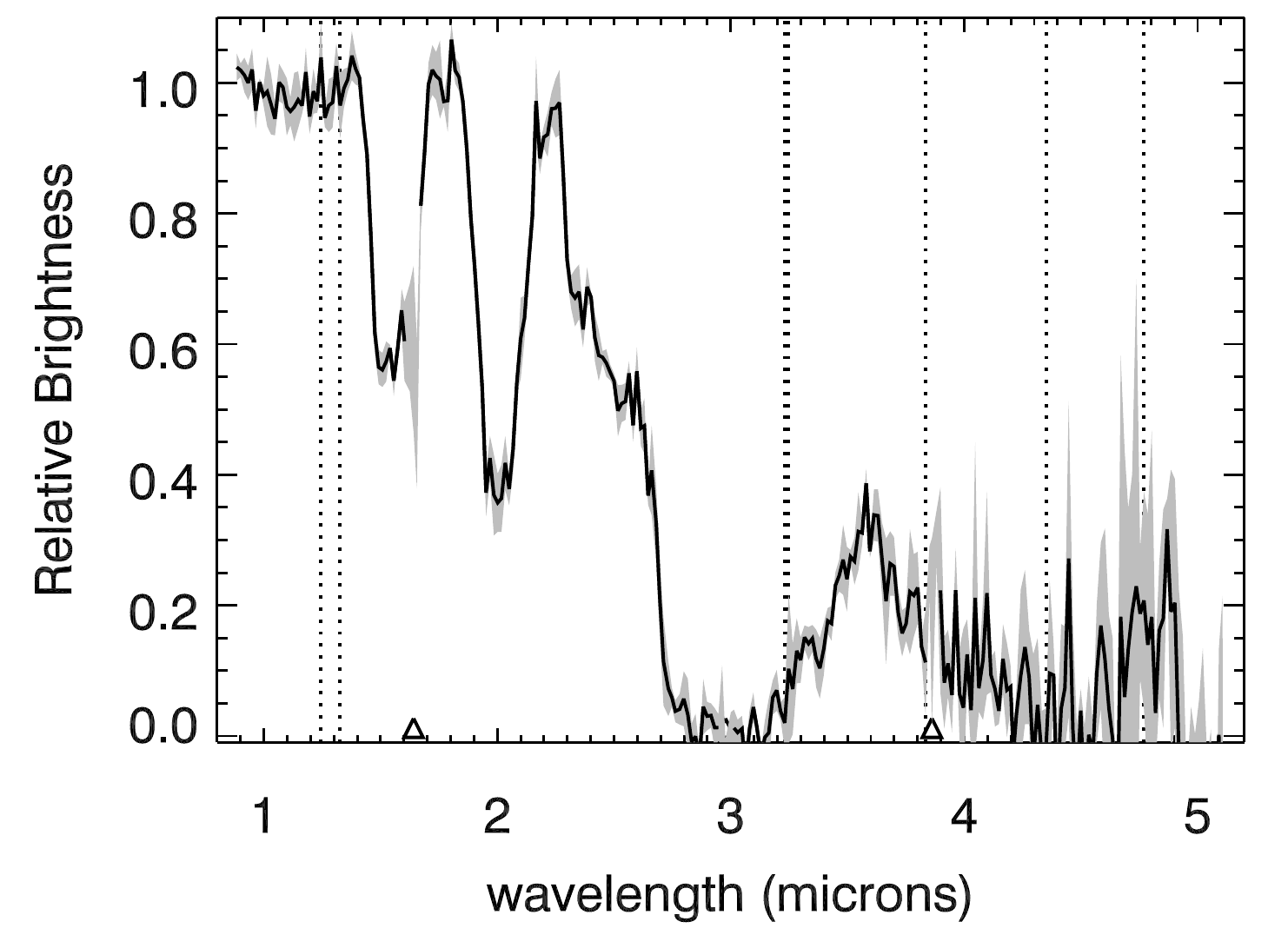}}
\caption{Near infrared spectrum of Methone derived from observations from the VIMS instrument. The brightness is shown at an arbitrarily normalized scale. Note the bands at 1.5, 2 and 3.1 microns are typical of water ice, and the overall shape of the spectrum is comparable to that of other small moons.}
\label{metspec}
\end{figure}

During its closest flybys of Aegaeon and Methone, Cassini's NAC obtained images of both moons through multiple filters. We processed these  high-resolution color images with the same methods as described above for the mid-sized moons, yielding the brightness coefficients given in {Tables~\ref{aegaeoncoltab} and~\ref{methonecoltab}}. We then interpolated these measurements to a single phase angle to obtain the visible spectra shown in Figure~\ref{colors}. Furthermore, during the close encounter with Methone, the Visual and Infrared Mapping Spectrometer (VIMS) \citep{Brown04} was able to obtain near-infrared spectra of the moon.  While VIMS was unable to spatially resolve Methone, VIMS acquired five ``cubes'' with good signal-to-noise (V1716191964, V1716192051, V1716192374, V1716192461 and V1716192872) that could yield decent disk-integrated spectra. We calibrated these cubes with the standard pipelines \citep[][flux calibration RC19]{Brown04,Clark12}, co-added the signals in the pixels containing the moon signal, removed backgrounds based on adjacent pixels, normalized the resulting spectra and averaged them together to produce a single mean spectrum shown in Figure~\ref{metspec} (Note that this spectrum is arbitrarily normalized). The near-infrared spectrum of Methone is dominated by water-ice, while the visible spectra of the two moons are generally flat with a spectral break around 0.5 microns.  These moon's spectral properties are therefore very similar to those of the rest of Saturn's moons \citep{Filacchione12, Buratti10, Buratti19}. {Furthermore, the depths of the water-ice bands at 1.5 and 2.0 microns (computed following the procedures laid out in \citet{Filacchione12}) are 0.45$\pm$0.04 and 0.60$\pm$0.06, respectively. These values are very similar to those previously found for Janus, Epimetheus, Telesto, Dione and Rhea \citep{Filacchione12}.}  Hence there is no evidence that these bodies have surfaces with distinctive spectrally active components, like irradiated salts \citep{Hibbitts19}. Instead, the radiation appears to change the surface brightness relatively uniformly at all wavelengths. 

An interesting point of comparison for these spectra is Jupiter's moon Callisto. Callisto and Aegaeon are probably exposed to similar fluxes of high-energy protons  \citep[roughly 10/(cm$^2$ s sr keV)  around energies of 1 MeV, ][]{Paranicas18}, and both objects are comparably dark (Callisto has a geometric albedo around 0.2 between 0.5 and 2.5 microns \citep{Calvin95}, and extrapolating the available data suggests that Aegaeon's equivalent spherical albedo is around 0.15). Like Aegaeon, Callisto has a visible spectrum that is flat longwards of 0.6 microns and has a downturn at shorter wavelengths \citep{Calvin95, Moore07}. The structure in Callisto's spectrum in the near-infrared primarily consists of comparatively weak water-ice features, along with a weak carbon-dioxide band at 4.2 $\mu$m \citep{Moore07}, which is consistent with the compositional features seen in Methone and Saturn's other moons \citep{Clark08, Hendrix18, Buratti19}. It may therefore be that both Callisto and Aegaeon represent end-members of what ice-rich surfaces look like when exposed to large doses of high-energy radiation. If correct, it could mean that even if we had near-IR spectra of Aegaeon, they would not contain  diagnostic spectral features of the chemicals responsible for making its surface dark. Whether the observed amount of darkening is consistent with structural changes in the regolith \citep{Howett11, Schenk11} or chemical changes in a more spectrally neutral component \citep[such as the non sulfur-bearing ices in][]{Poston18}, will require detailed spectral models that are beyond the scope of this work.

\subsection{The excess brightness of  Prometheus, Pandora, Calypso and Helene}
\label{trojans}

After Aegaeon, Methone and Pallene, the next most obvious anomalies in Figure~\ref{eringflux} are Helene and Calypso. Calypso is brighter than both Tethys and Telesto, while Helene is brighter than both Dione and Polydeuces. These findings are actually consistent with prior estimates of these moons' geometric albedos summarized in \citet{Verbiscer07, Verbiscer18}. At the same time, it is worth noting that despite being in a region where the E-ring flux should be extremely low, Pan, Atlas, Prometheus and Pandora are all reasonably bright, and in particular Prometheus and Pandora are  brighter than any of the other moons in their vicinity. {The excess brightness of these moons is probably related to distinctive aspects of their near-infrared spectra. For the mid-sized moons, the depths of the moon's water-ice bands are reasonably well correlated with their visible brightness \citep{Filacchione12}. However, Calypso, Prometheus and Pandora all have deeper water-ice bands than Enceladus, while Atlas, Telesto and Helene have water-ice bands intermediate in strength between Tethys and Dione \citep{Buratti10, Filacchione12, Filacchione13}. Note that even though Helene does not appear to have as exceptionally deep water-ice bands as Calypso, both Calypso and Helene have deeper bands than their respective large companions Tethys and Dione.   The anomalous spectral and photometric properties of Prometheus, Pandora, Calypso and Helene} could be attributed to a number of different factors, including the moons' size and differences in the local radiation environment. In practice, localized enhancements in the particle flux appears to be the most likely explanation, but the excess particle flux responsible for the high observed brightnesses of Helene and Calypso is far from clear.

In principle, the surface brightness of small moons could be different from the brightness of large moons in the same environment because their reduced surface gravity affects the porosity and structure of their regolith. In fact, one aspect of these moons' surface properties probably can be attributed to their small size: their general lack of a leading-trailing brightness asymmetries. In particular, all four of the trojan moons appear to lack the strong leading-trailing brightness asymmetries seen on Tethys and Dione.\footnote{{\citet{Hirata14} also found that there was not a strong leading-trailing asymmetry in the morphology of deposits on Telesto and Calypso. However, they also found that Helene appears to have a higher crater density on its trailing side, perhaps implying a thicker mantling deposit on its leading side.}} For Tethys and Dione, this asymmetry is thought to arise because of asymmetries in the E-ring flux, so it is perhaps  surprising that similar asymmetries are not seen in the smaller moons. However, the smaller size and much lower surface gravity of these moons means that secondary ejecta from the E-ring impacts can be more easily globally dispersed over their surfaces, providing a natural explanation for why these moons have more uniform surface properties.

While detailed modeling of impact debris transport around these small moons is beyond the scope of this paper, we can provide rough order-of-magnitude calculations that are sufficient to demonstrate the feasibility of this idea. Assuming that most of the E-ring particles striking these moons are near the apocenters of their eccentric orbits, we can assume $r \simeq (1+e) a$ in Equation~\ref{vlambda} and estimate the E-ring particle's impact velocities at the orbits of Tethys and Dione to be 1.4 km/s and 3.6 km/s, respectively. Experimental studies of impacts into ice-rich targets by 20-100 $\mu$m glass beads found impact yields at these velocities of order 100 \citep{Koschny01}, and assuming roughly comparable values for micron-sized E-ring grains would imply an effective average ejection velocity of at most  14 m/s and 36 m/s, respectively. These speeds are much less than the escape speeds from Tethys and Dione (which are 390 and 510 m/s respectively), but are comparable to the escape speeds of Telesto, Calypso, and Helene (which are 7, 19 and 9 m/s, respectively). Hence it is reasonable to think that the debris from E-ring particle impacts is more uniformly distributed on the smaller trojans than they are on Tethys and Dione, making their surfaces more uniform in brightness.

However, while their size is likely responsible for their lack of leading-trailing brightness asymmetries,  it is unlikely that size-related phenomena  can explain the overall high brightness of Prometheus, Pandora, Calypso and Helene. The problem with such ideas is that there are no clear trends between size and brightness excess. Telesto is intermediate in size between Helene and Calypso, but is roughly the same brightness as Tethys. Also, Prometheus and Pandora are themselves intermediate in size between Epimetheus and Atlas. This lack of obvious trends with mass or size suggests that the brightness excess of these moons most likely reflects something different about their environments. 

Given the previous analyses of the trends among the other moons,  the easiest way to increase the brightness of  these moons would be either by decreasing the radiation flux or increasing the particle flux.  Decreasing the radiation flux is an unlikely explanation because the high-energy proton flux onto many of the moons is already very low (see Table~\ref{fluxtab}). Furthermore, for the radiation flux to be lower at Calypso than at Telesto, or lower at Helene than at Polydeuces, there would need to a strong asymmetry in the radiation flux on either side of Tethys and Dione. While both Tethys and Dione have been observed to produce localized reductions in plasma known as microsignatures, these do not extend all the way to the trojan moons. Also, given that the plasma is bound to the planet's magnetic field, which rotates faster than the moons orbit around the planet, it is hard to imagine any interaction with charged particles that would similarly affect Calypso (which trails Tethys) and Helene (which leads Dione). 
 
The above considerations leave variations in the particle flux as the best remaining option. For Prometheus and Pandora, this is a perfectly reasonable explanation, since these moons fall within the outskirts of the F ring, which is a natural source of dust flux into these moons. If an excess flux of F-ring particles is indeed responsible for the high brightness of Prometheus and Pandora, this is interesting because it means that the dust impacting the moons does not have to be the relatively fresh ice found in the E ring to produce increases in brightness. In this case, the lower-opacity dusty rings surrounding Pan and Atlas could be keeping those moons somewhat brighter than they would otherwise be given their locations well interior to the E ring.  

Excess dust fluxes are also an attractive explanation for the excess brightness of Calypso and Helene because these moons are nearly as bright as Enceladus, an object whose brightness is certainly due to a high particle flux. However, there is no obvious source for the particles that would strike Calypso and Helene more than their co-orbital companions. The brightness differences cannot be easily attributed to differences in the E-ring flux, since Telesto, Calypso and Tethys are at basically the same orbital distance from Saturn, as are Helene, Polydeuces and Dione.  Hence the E-ring flux into Telesto, Tethys and Calypso (or Helene, Dione and Polydeuces) should be nearly the same. 

In principle, asymmetries in the particle flux could arise from either subtle interactions between the larger moons and the E-ring grains, or more local particle populations sourced from the larger moons. For example, particles on nearly circular orbits launched from Tethys that are carried outward by plasma drag would preferentially fall behind Tethys, and thus potentially be more likely to strike Calypso than Telesto.  The problem with such explanations is that for Calypso and Helene to fall along the same trend as the other moons in Figure~\ref{eringflux}, the excess particle flux into these moons needs to be a substantial fraction of the E ring flux, which would imply a large excess density of particles around these moons compared to Tethys, Dione, Telesto or Polydeuces. These particles should produce localized brightness enhancements and/or longitudinal brightness asymmetries in the regions around Tethys' and Dione's orbits, and no such features have yet been seen in the Cassini images of the E ring. In principle, there could be some subtle interactions between E ring particles and the relevant moons that would cause the requisite variations in the particle flux without producing detectable asymmetries in the density of the rings. Fully exploring such possibilities would likely require numerical simulations that are beyond the scope of this work. 

Given the above challenges with identifying a suitable dust population around these moons during the Cassini mission, it is worth considering whether Calypso's and Helene's current brightness are not their  steady-state values, but are instead transient phenomena caused by some event in the relatively recent past. Specifically, we can consider the possibility that a recent impact into Calypso and/or Helene released enough material to affect the global surface properties of both these moons. A thorough quantitative evaluation of such scenarios is beyond the scope of this paper. Instead we will just provide some order-of-magnitude calculations which demonstrate that this is a viable possibility.

The total amount of material needed to affect the overall surface brightness of Calypso and Helene is simply the amount of material needed to coat these moons with a layer thick enough to determine its optical properties at optical wavelengths. Since light can only penetrate a few wavelengths through the surface regolith, we may conservatively estimate that a 10-$\mu$m thick layer of material will be sufficient for these purposes. Given that Calypso and Helene  have surface areas of around 400 km$^2$ and 1400 km$^2$ respectively, the total volume of material needed to coat these moons is 4000 m$^3$ and 14,000 m$^3$. Conservatively assuming this is ice-rich material of negligible porosity, these volumes correspond to total debris masses of 4$\times10^{6}$ and 14$\times 10^{6}$ kg, respectively. 

The flux of objects large enough to produce these sorts of debris clouds is fairly well constrained thanks to observations of comparably-massive impact-generated debris clouds above Saturn's rings described by \citet{Tiscareno13}.  The masses of those clouds are uncertain and depend on the assumed size distribution of the debris, but the feature designated Bx probably had a mass between $2\times10^{5}$ kg and $2\times10^{7}$ kg, which is comparable to that required to brighten Calypso or Helene. The estimated flux of impactors able of producing this amount of debris is $3\times10^{-20}$/m$^2$/s \citep{Tiscareno13}. Assuming cross-sections consistent with the observed sizes of Calypso and Helene, this means the  mean time between impacts should be between 1000 and 10,000 years.

The other relevant timescale in this problem is how long it would take E-ring material to coat these moons and erase any transient signals from an impact. Using the fluxes provided in Table~\ref{fluxtab}, and assuming a typical E-ring particle radius of around 1 $\mu$m, we estimate that E-ring material would accumulate on the surfaces at rates of between 0.2 nm/year for Helene and 0.5 nm/year for Calypso. It would therefore take 20,000 years for Calypso and  50,000 years for Helene to accumulate a 10-micron-thick layer of E-ring particles. These numbers could potentially be reduced by a factor of $\sim$100 if we account for the yield of secondary debris from each E-ring particle impact (see above). The resurfacing time due to E-ring particles is therefore likely comparable to the mean time between impacts capable of covering these entire moons in fresh impact debris. It is therefore not unreasonable that Calypso and Helene both experienced a recent impact that brightened their surfaces, while Telesto and Polydeuces managed to avoid such a recent collision. In principle, some of the material from a collision into Helene or Calypso could have been transported to the other moon, allowing a single impact to brighten both moons, but to properly explore the relative likelihood of such an event will require numerical simulations of both dynamics of the impact debris similar to those done by \citet{Dobrovolskis09}, perhaps including non-gravitational forces.

Note that this explanation for Calypso's and Helene's brightness means that impacts tend to increase the moons' brightness, rather than darken them. This might at first be counterintuitive because cometary debris in the outer solar system is generally assumed to be much darker than Saturn's moons. However, it is important to realize that most of the debris produced by the impacts would come from Helene and Calypso, not the impactor. Indeed, the expected impact yields for such bodies is of order 10,000 \citep{CE98, Tiscareno13}, and so the impactor required to produce $4-14 \times 10^{6}$ kg of debris would be only $4-14 \times 10^{2}$ kg, which corresponds to an ice-rich object with a radius between 0.5 and 0.7 meters. Thus most of the debris falling on the moons would be pure ice, and should be able to brighten moon surfaces like the E and F rings are apparently able to do. This bright deposit would also be much thinner than the deposits previously identified in high-resolution images, which affect both the morphology and the color of the surface \citep{Thomas13, Hirata14}. While such a thin deposit  would probably not produce obvious morphological structures (even the original crater would be near the resolution limit of the best images), it is not clear whether a fresh global covering of fine debris is consistent with the color variations currently observed on these moons \citep{Thomas13}. Also, it is not obvious whether recent impacts are consistent with near-infrared spectra of these moons, which show that Prometheus, Pandora and Calypso have deeper ice bands than Telesto and Helene \citep{Filacchione12}. Even so, the above considerations indicate that a recent impact is a possible explanation for Calypso's and Helene's high surface brightness and so merits further investigation.

\section{Summary}
\label{summary}

In conclusion, we have used a new photometric model for non-spherical objects to obtain surface brightness estimates that are comparable to those for the mid-sized satellites. Applying this model to  Saturn's moons has revealed a number of interesting features and trends, which are summarized below in order of the moon's increasing distance from Saturn.
\begin{itemize}
\item Prometheus and Pandora are brighter than moons orbiting exterior and interior to them, suggesting that  dust from the F ring is brightening these moons.
\item Janus and Epimetheus have brighter trailing hemispheres than leading hemispheres, which is consistent with the expected pattern for impacting E-ring particles.
\item  Aegaeon, Methone and Pallene are all darker than expected given their location within the E ring. This is most likely due to the high flux of high-energy radiation into these moons.
\item The spectral data for Aegaeon and Methone indicate that whatever material is responsible for making these moons dark reduces their brightness over a broad range of wavelengths and does not have obvious spectral signatures.
\item The photometric data indicate that Anthe probably has a shape similar to Methone.
\item Pallene's leading side is slightly darker than its trailing side. 
\item Telesto and Polydeuces have surface brightnesses similar to their larger orbital companions (Tethys and Dione, respectively).
\item Calypso is substantially brighter than Tethys and Telesto, while Helene is substantially brighter than Dione and Polydeuces. These phenomena could either be due to an asymmetric flux of E-ring particles, or recent collisions with larger impactors.

\end{itemize}

\section{Acknowledgements}
Authors MMH and PH offer thanks to NASA's Cassini Mission Project during various stages of this study. MMH acknowledges support from the Cassini Data Analysis Program Grant NNX15AQ67G. PH gratefully acknowledges that photometric modeling enhancements, as well as refinements and testing of computer software that are used and first-reported in this paper were developed under the auspices of NASA Planetary Geology and Geophysics Program grant NNX14AN04G.  MH also wishes to acknowledge that the initial analyses of the small moons' brightnesses were done by M. Rehnberg as part of the REU program, and the calculations regarding the deposition rates on the various small moons were inspired by conservations with S. Morrison, S.G. Zaidi and H. Sharma. We thank the anonymous reviewer for their helpful comments on an earlier draft of this manuscript.
\pagebreak

%\bibliographystyle{icarus}
%\bibliography{refs.bib} 

\begin{thebibliography}{}

\bibitem[{Acton}(1996){Acton}]{Acton96}
{Acton}, C.~H., 1996.
\newblock {Ancillary data services of NASA's Navigation and Ancillary
  Information Facility}.
\newblock \planss~44, 65--70.

\bibitem[{Black} et~al.(2007){Black}, {Campbell}, and {Carter}]{Black07}
{Black}, G.~J., {Campbell}, D.~B., {Carter}, L.~M., 2007.
\newblock {Arecibo radar observations of Rhea, Dione, Tethys, and Enceladus}.
\newblock Icarus~191 (2), 702--711.

\bibitem[{Brown} et~al.(2004){Brown}, {Baines}, {Bellucci}, {Bibring},
  {Buratti}, {Capaccioni}, {Cerroni}, {Clark}, {Coradini}, {Cruikshank},
  {Drossart}, {Formisano}, {Jaumann}, {Langevin}, {Matson}, {McCord},
  {Mennella}, {Miller}, {Nelson}, {Nicholson}, {Sicardy}, and {Sotin}]{Brown04}
{Brown}, R.~H., {Baines}, K.~H., {Bellucci}, G., {Bibring}, J.-P., {Buratti},
  B.~J., {Capaccioni}, F., {Cerroni}, P., {Clark}, R.~N., {Coradini}, A.,
  {Cruikshank}, D.~P., {Drossart}, P., {Formisano}, V., {Jaumann}, R.,
  {Langevin}, Y., {Matson}, D.~L., {McCord}, T.~B., {Mennella}, V., {Miller},
  E., {Nelson}, R.~M., {Nicholson}, P.~D., {Sicardy}, B., {Sotin}, C., 2004.
\newblock {The Cassini Visual And Infrared Mapping Spectrometer (Vims)
  Investigation}.
\newblock SSR~115, 111--168.

\bibitem[{Buratti} and {Veverka}(1984){Buratti} and {Veverka}]{BV84}
{Buratti}, B., {Veverka}, J., 1984.
\newblock {Voyager photometry of Rhea, Dione, Tethys, Enceladus and Mimas}.
\newblock \icarus~58 (2), 254--264.

\bibitem[{Buratti}(1984){Buratti}]{Buratti84}
{Buratti}, B.~J., 1984.
\newblock {Voyager disk resolved photometry of the Saturnian satellites}.
\newblock Icarus~59 (3), 392--405.

\bibitem[{Buratti} et~al.(2010){Buratti}, {Bauer}, {Hicks}, {Mosher},
  {Filacchione}, {Momary}, {Baines}, {Brown}, {Clark}, and
  {Nicholson}]{Buratti10}
{Buratti}, B.~J., {Bauer}, J.~M., {Hicks}, M.~D., {Mosher}, J.~A.,
  {Filacchione}, G., {Momary}, T., {Baines}, K.~H., {Brown}, R.~H., {Clark},
  R.~N., {Nicholson}, P.~D., 2010.
\newblock {Cassini spectra and photometry 0.25-5.1 {\ensuremath{\mu}}m of the
  small inner satellites of Saturn}.
\newblock \icarus~206 (2), 524--536.

\bibitem[{Buratti} et~al.(1990){Buratti}, {Mosher}, and {Johnson}]{Buratti90}
{Buratti}, B.~J., {Mosher}, J.~A., {Johnson}, T.~V., 1990.
\newblock {Albedo and color maps of the Saturnian satellites}.
\newblock \icarus~87 (2), 339--357.

\bibitem[{Buratti} et~al.(1998){Buratti}, {Mosher}, {Nicholson}, {McGhee}, and
  {French}]{Buratti98}
{Buratti}, B.~J., {Mosher}, J.~A., {Nicholson}, P.~D., {McGhee}, C.~A.,
  {French}, R.~G., 1998.
\newblock {Near-Infrared Photometry of the Saturnian Satellites during Ring
  Plane Crossing}.
\newblock \icarus~136 (2), 223--231.

\bibitem[Buratti et~al.(2019)Buratti, Thomas, Roussos, Howett, Sei{\ss},
  Hendrix, Helfenstein, Brown, Clark, Denk, Filacchione, Hoffmann, Jones,
  Khawaja, Kollmann, Krupp, Lunine, Momary, Paranicas, Postberg, Sachse, Spahn,
  Spencer, Srama, Albin, Baines, Ciarniello, Economou, Hsu, Kempf, Krimigis,
  Mitchell, Moragas-Klostermeyer, Nicholson, Porco, Rosenberg, Simolka, and
  Soderblom]{Buratti19}
Buratti, B.~J., Thomas, P.~C., Roussos, E., Howett, C., Sei{\ss}, M., Hendrix,
  A.~R., Helfenstein, P., Brown, R.~H., Clark, R.~N., Denk, T., Filacchione,
  G., Hoffmann, H., Jones, G.~H., Khawaja, N., Kollmann, P., Krupp, N., Lunine,
  J., Momary, T.~W., Paranicas, C., Postberg, F., Sachse, M., Spahn, F.,
  Spencer, J., Srama, R., Albin, T., Baines, K.~H., Ciarniello, M., Economou,
  T., Hsu, H.-W., Kempf, S., Krimigis, S.~M., Mitchell, D.,
  Moragas-Klostermeyer, G., Nicholson, P.~D., Porco, C.~C., Rosenberg, H.,
  Simolka, J., Soderblom, L.~A., 2019.
\newblock Close cassini flybys of saturn{\textquoteright}s ring moons pan,
  daphnis, atlas, pandora, and epimetheus.
\newblock Science~364 (6445).

\bibitem[{Calvin} et~al.(1995){Calvin}, {Clark}, {Brown}, and
  {Spencer}]{Calvin95}
{Calvin}, W.~M., {Clark}, R.~N., {Brown}, R.~H., {Spencer}, J.~R., 1995.
\newblock {Spectra of the icy Galilean satellites from 0.2 to 5
  {\ensuremath{\mu}}m: A compilation, new observations, and a recent summary}.
\newblock \jgr~100 (E9), 19041--19048.

\bibitem[{Clark} et~al.(2012){Clark}, {Cruikshank}, {Jaumann}, {Brown},
  {Stephan}, {Dalle Ore}, {Eric Livo}, {Pearson}, {Curchin}, {Hoefen},
  {Buratti}, {Filacchione}, {Baines}, and {Nicholson}]{Clark12}
{Clark}, R.~N., {Cruikshank}, D.~P., {Jaumann}, R., {Brown}, R.~H., {Stephan},
  K., {Dalle Ore}, C.~M., {Eric Livo}, K., {Pearson}, N., {Curchin}, J.~M.,
  {Hoefen}, T.~M., {Buratti}, B.~J., {Filacchione}, G., {Baines}, K.~H.,
  {Nicholson}, P.~D., 2012.
\newblock {The surface composition of Iapetus: Mapping results from Cassini
  VIMS}.
\newblock \icarus~218 (2), 831--860.

\bibitem[{Clark} et~al.(2008){Clark}, {Curchin}, {Jaumann}, {Cruikshank},
  {Brown}, {Hoefen}, {Stephan}, {Moore}, {Buratti}, {Baines}, {Nicholson}, and
  {Nelson}]{Clark08}
{Clark}, R.~N., {Curchin}, J.~M., {Jaumann}, R., {Cruikshank}, D.~P., {Brown},
  R.~H., {Hoefen}, T.~M., {Stephan}, K., {Moore}, J.~M., {Buratti}, B.~J.,
  {Baines}, K.~H., {Nicholson}, P.~D., {Nelson}, R.~M., 2008.
\newblock {Compositional mapping of Saturn's satellite Dione with Cassini VIMS
  and implications of dark material in the Saturn system}.
\newblock \icarus~193 (2), 372--386.

\bibitem[{Cuzzi} and {Estrada}(1998){Cuzzi} and {Estrada}]{CE98}
{Cuzzi}, J.~N., {Estrada}, P.~R., 1998.
\newblock {Compositional Evolution of Saturn's Rings Due to Meteoroid
  Bombardment}.
\newblock \icarus~132 (1), 1--35.

\bibitem[{Dobrovolskis} et~al.(2010){Dobrovolskis}, {Alvarellos}, {Zahnle}, and
  {Lissauer}]{Dobrovolskis09}
{Dobrovolskis}, A.~R., {Alvarellos}, J.~L., {Zahnle}, K.~J., {Lissauer}, J.~J.,
  2010.
\newblock {Exchange of ejecta between Telesto and Calypso: Tadpoles,
  horseshoes, and passing orbits}.
\newblock \icarus~210 (1), 436--445.

\bibitem[{Filacchione} et~al.(2012){Filacchione}, {Capaccioni}, {Ciarniello},
  {Clark}, {Cuzzi}, {Nicholson}, {Cruikshank}, {Hedman}, {Buratti}, and
  {Lunine}]{Filacchione12}
{Filacchione}, G., {Capaccioni}, F., {Ciarniello}, M., {Clark}, R.~N., {Cuzzi},
  J.~N., {Nicholson}, P.~D., {Cruikshank}, D.~P., {Hedman}, M.~M., {Buratti},
  B.~J., {Lunine}, J.~I., 2012.
\newblock {Saturn's icy satellites and rings investigated by Cassini-VIMS: III
  - Radial compositional variability}.
\newblock Icarus~220 (2), 1064--1096.

\bibitem[{Filacchione} et~al.(2013){Filacchione}, {Capaccioni}, {Clark},
  {Nicholson}, {Cruikshank}, {Cuzzi}, {Lunine}, {Brown}, {Cerroni}, {Tosi},
  {Ciarniello}, {Buratti}, {Hedman}, and {Flamini}]{Filacchione13}
{Filacchione}, G., {Capaccioni}, F., {Clark}, R.~N., {Nicholson}, P.~D.,
  {Cruikshank}, D.~P., {Cuzzi}, J.~N., {Lunine}, J.~I., {Brown}, R.~H.,
  {Cerroni}, P., {Tosi}, F., {Ciarniello}, M., {Buratti}, B.~J., {Hedman},
  M.~M., {Flamini}, E., 2013.
\newblock {The Radial Distribution of Water Ice and Chromophores across
  Saturn's System}.
\newblock \apj~766 (2), 76.

\bibitem[{Franz} and {Millis}(1975){Franz} and {Millis}]{Franz75}
{Franz}, O.~G., {Millis}, R.~L., 1975.
\newblock {Photometry of Dione, Tethys, and Enceladus on the UBV System}.
\newblock \icarus~24 (4), 433--442.

\bibitem[{Hamilton} and {Burns}(1994){Hamilton} and {Burns}]{HB94}
{Hamilton}, D.~P., {Burns}, J.~A., 1994.
\newblock {Origin of Saturn's E Ring: Self-Sustained, Naturally}.
\newblock Science~264 (5158), 550--553.

\bibitem[{Hand} and {Carlson}(2015){Hand} and {Carlson}]{Hand15}
{Hand}, K.~P., {Carlson}, R.~W., 2015.
\newblock {Europa's surface color suggests an ocean rich with sodium chloride}.
\newblock \grl~42 (9), 3174--3178.

\bibitem[{Hapke}(1986){Hapke}]{Hapke86}
{Hapke}, B., 1986.
\newblock {On the sputter alteration of regoliths of outer solar system
  bodies}.
\newblock \icarus~66 (2), 270--279.

\bibitem[{Hedman} et~al.(2012){Hedman}, {Burns}, {Hamilton}, and
  {Showalter}]{Hedman12}
{Hedman}, M.~M., {Burns}, J.~A., {Hamilton}, D.~P., {Showalter}, M.~R., 2012.
\newblock {The three-dimensional structure of Saturn's E ring}.
\newblock \icarus~217 (1), 322--338.

\bibitem[{Hedman} et~al.(2010){Hedman}, {Cooper}, {Murray}, {Beurle}, {Evans},
  {Tiscareno}, and {Burns}]{Hedman10}
{Hedman}, M.~M., {Cooper}, N.~J., {Murray}, C.~D., {Beurle}, K., {Evans},
  M.~W., {Tiscareno}, M.~S., {Burns}, J.~A., 2010.
\newblock {Aegaeon (Saturn LIII), a G-ring object}.
\newblock \icarus~207, 433--447.

\bibitem[{Hedman} et~al.(2018){Hedman}, {Postberg}, {Hamilton}, {Renner}, and
  {Hsu}]{Hedman18}
{Hedman}, M.~M., {Postberg}, F., {Hamilton}, D.~P., {Renner}, S., {Hsu}, H.~W.,
  2018.
\newblock {\em {Dusty Rings}}, pp.\  308--337.

\bibitem[{Helfenstein} and {Veverka}(1989){Helfenstein} and {Veverka}]{HV89}
{Helfenstein}, P., {Veverka}, J., 1989.
\newblock {Physical characterization of asteroid surfaces from photometric
  analysis}.
\newblock In: {Binzel}, R.~P., {Gehrels}, T., {Matthews}, M.~S. (Eds.),
  Asteroids II, pp.\  557--593.

\bibitem[{Hendrix} et~al.(2018){Hendrix}, {Buratti}, {Cruikshank}, {Clark},
  {Scipioni}, and {Howett}]{Hendrix18}
{Hendrix}, A.~R., {Buratti}, B.~J., {Cruikshank}, D.~P., {Clark}, R.~N.,
  {Scipioni}, F., {Howett}, C.~J.~A., 2018.
\newblock {\em {Surface Composition of Saturn's Icy Moons}}, pp.\  307.

\bibitem[{Hibbitts} et~al.(2019){Hibbitts}, {Stockstill-Cahill}, {Wing}, and
  {Paranicas}]{Hibbitts19}
{Hibbitts}, C.~A., {Stockstill-Cahill}, K., {Wing}, B., {Paranicas}, C., 2019.
\newblock {Color centers in salts - Evidence for the presence of sulfates on
  Europa}.
\newblock \icarus~326, 37--47.

\bibitem[{Hirata} et~al.(2014){Hirata}, {Miyamoto}, and {Showman}]{Hirata14}
{Hirata}, N., {Miyamoto}, H., {Showman}, A.~P., 2014.
\newblock {Particle deposition on the saturnian satellites from ephemeral
  cryovolcanism on Enceladus}.
\newblock \grl~41 (12), 4135--4141.

\bibitem[{Horanyi} et~al.(1992){Horanyi}, {Burns}, and {Hamilton}]{Horanyi92}
{Horanyi}, M., {Burns}, J.~A., {Hamilton}, D.~P., 1992.
\newblock {The dynamics of Saturn's E ring particles}.
\newblock \icarus~97 (2), 248--259.

\bibitem[{Hor{\'a}nyi} et~al.(2008){Hor{\'a}nyi}, {Juh{\'a}sz}, and
  {Morfill}]{Horanyi08}
{Hor{\'a}nyi}, M., {Juh{\'a}sz}, A., {Morfill}, G.~E., 2008.
\newblock {Large-scale structure of Saturn's E-ring}.
\newblock \grl~35 (4), L04203.

\bibitem[{Howett} et~al.(2018){Howett}, {Hendrix}, {Nordheim}, {Paranicas},
  {Spencer}, and {Verbiscer}]{Howett18}
{Howett}, C.~J.~A., {Hendrix}, A.~R., {Nordheim}, T.~A., {Paranicas}, C.,
  {Spencer}, J.~R., {Verbiscer}, A.~J., 2018.
\newblock {\em {Ring and Magnetosphere Interactions with Satellite Surfaces}},
  pp.\  343.

\bibitem[{Howett} et~al.(2011){Howett}, {Spencer}, {Schenk}, {Johnson},
  {Paranicas}, {Hurford}, {Verbiscer}, and {Segura}]{Howett11}
{Howett}, C.~J.~A., {Spencer}, J.~R., {Schenk}, P., {Johnson}, R.~E.,
  {Paranicas}, C., {Hurford}, T.~A., {Verbiscer}, A., {Segura}, M., 2011.
\newblock {A high-amplitude thermal inertia anomaly of probable magnetospheric
  origin on Saturn's moon Mimas}.
\newblock \icarus~216 (1), 221--226.

\bibitem[{Kempf} et~al.(2008){Kempf}, {Beckmann}, {Moragas-Klostermeyer},
  {Postberg}, {Srama}, {Economou}, {Schmidt}, {Spahn}, and {Gr{\"u}n}]{Kempf08}
{Kempf}, S., {Beckmann}, U., {Moragas-Klostermeyer}, G., {Postberg}, F.,
  {Srama}, R., {Economou}, T., {Schmidt}, J., {Spahn}, F., {Gr{\"u}n}, E.,
  2008.
\newblock {The E ring in the vicinity of Enceladus. I. Spatial distribution and
  properties of the ring particles}.
\newblock Icarus~193 (2), 420--437.

\bibitem[{Koschny} and {Gr{\"u}n}(2001){Koschny} and {Gr{\"u}n}]{Koschny01}
{Koschny}, D., {Gr{\"u}n}, E., 2001.
\newblock {Impacts into Ice-Silicate Mixtures: Crater Morphologies, Volumes,
  Depth-to-Diameter Ratios, and Yield}.
\newblock \icarus~154 (2), 391--401.

\bibitem[{Krimigis} et~al.(2004){Krimigis}, {Mitchell}, {Hamilton}, {Livi},
  {Dandouras}, {Jaskulek}, {Armstrong}, {Boldt}, {Cheng}, {Gloeckler}, {Hayes},
  {Hsieh}, {Ip}, {Keath}, {Kirsch}, {Krupp}, {Lanzerotti}, {Lundgren}, {Mauk},
  {McEntire}, {Roelof}, {Schlemm}, {Tossman}, {Wilken}, and
  {Williams}]{Krimigis04}
{Krimigis}, S.~M., {Mitchell}, D.~G., {Hamilton}, D.~C., {Livi}, S.,
  {Dandouras}, J., {Jaskulek}, S., {Armstrong}, T.~P., {Boldt}, J.~D., {Cheng},
  A.~F., {Gloeckler}, G., {Hayes}, J.~R., {Hsieh}, K.~C., {Ip}, W.~H., {Keath},
  E.~P., {Kirsch}, E., {Krupp}, N., {Lanzerotti}, L.~J., {Lundgren}, R.,
  {Mauk}, B.~H., {McEntire}, R.~W., {Roelof}, E.~C., {Schlemm}, C.~E.,
  {Tossman}, B.~E., {Wilken}, B., {Williams}, D.~J., 2004.
\newblock {Magnetosphere Imaging Instrument (MIMI) on the Cassini Mission to
  Saturn/Titan}.
\newblock \ssr~114 (1-4), 233--329.

\bibitem[Le~Gall et~al.(2019)Le~Gall, West, and Bonnefoy]{LeGall19}
Le~Gall, A., West, R.~D., Bonnefoy, L.~E., 2019.
\newblock Dust and snow cover on saturn's icy moons.
\newblock Geophysical Research Letters~n/a (n/a).

\bibitem[{McEwen}(1991){McEwen}]{McEwen91}
{McEwen}, A.~S., 1991.
\newblock {Photometric functions for photoclinometry and other applications}.
\newblock \icarus~92 (2), 298--311.

\bibitem[{Minnaert}(1941){Minnaert}]{Minnaert41}
{Minnaert}, M., 1941.
\newblock {The reciprocity principle in lunar photometry}.
\newblock \apj~93, 403--410.

\bibitem[{Moore} et~al.(2007){Moore}, {Chapman}, {Bierhaus}, {Greeley},
  {Chuang}, {Klemaszewski}, {Clark}, {Dalton}, {Hibbitts}, and
  {Schenk}]{Moore07}
{Moore}, J.~M., {Chapman}, C.~R., {Bierhaus}, E.~B., {Greeley}, R., {Chuang},
  F.~C., {Klemaszewski}, J., {Clark}, R.~N., {Dalton}, J.~B., {Hibbitts},
  C.~A., {Schenk}, P.~M., 2007.
\newblock {\em {Callisto}}, pp.\  397.

\bibitem[{Muinonen} and {Lumme}(2015){Muinonen} and {Lumme}]{ML15}
{Muinonen}, K., {Lumme}, K., 2015.
\newblock {Disk-integrated brightness of a Lommel-Seeliger scattering
  ellipsoidal asteroid}.
\newblock \aap~584, A23.

\bibitem[{Murray} and {Dermott}(1999){Murray} and {Dermott}]{MD99}
{Murray}, C.~D., {Dermott}, S.~F., 1999.
\newblock {\em {Solar system dynamics}}.

\bibitem[{Noland} et~al.(1974){Noland}, {Veverka}, {Morrison}, {Cruikshank},
  {Lazarewicz}, {Morrison}, {Elliot}, {Goguen}, and {Burns}]{Noland74}
{Noland}, M., {Veverka}, J., {Morrison}, D., {Cruikshank}, D.~P., {Lazarewicz},
  A.~R., {Morrison}, N.~D., {Elliot}, J.~L., {Goguen}, J., {Burns}, J.~A.,
  1974.
\newblock {Six-Color Photometry of Iapetus, Titan, Rhea, Dione, and Tethys}.
\newblock \icarus~23 (3), 334--354.

\bibitem[{Ostro} et~al.(2010){Ostro}, {West}, {Wye}, {Zebker}, {Janssen},
  {Stiles}, {Kelleher}, {Anderson}, {Boehmer}, {Callahan}, {Gim}, {Hamilton},
  {Johnson}, {Veeramachaneni}, {Lorenz}, and {The Cassini Radar Team}]{Ostro10}
{Ostro}, S.~J., {West}, R.~D., {Wye}, L.~C., {Zebker}, H.~A., {Janssen}, M.~A.,
  {Stiles}, B., {Kelleher}, K., {Anderson}, Y.~Z., {Boehmer}, R.~A.,
  {Callahan}, P., {Gim}, Y., {Hamilton}, G.~A., {Johnson}, W.~T.~K.,
  {Veeramachaneni}, C., {Lorenz}, R.~D., {The Cassini Radar Team}, 2010.
\newblock {New Cassini RADAR results for Saturn{\textquoteright}s icy
  satellites}.
\newblock \icarus~206 (2), 498--506.

\bibitem[{Paranicas} et~al.(2018){Paranicas}, {Hibbitts}, {Kollmann}, {Ligier},
  {Hendrix}, {Nordheim}, {Roussos}, {Krupp}, {Blaney}, and
  {Cassidy}]{Paranicas18}
{Paranicas}, C., {Hibbitts}, C.~A., {Kollmann}, P., {Ligier}, N., {Hendrix},
  A.~R., {Nordheim}, T.~A., {Roussos}, E., {Krupp}, N., {Blaney}, D.,
  {Cassidy}, T.~A., 2018.
\newblock {Magnetospheric considerations for solar system ice state}.
\newblock \icarus~302, 560--564.

\bibitem[{Pitman} et~al.(2010){Pitman}, {Buratti}, and {Mosher}]{Pitman10}
{Pitman}, K.~M., {Buratti}, B.~J., {Mosher}, J.~A., 2010.
\newblock {Disk-integrated bolometric Bond albedos and rotational light curves
  of saturnian satellites from Cassini Visual and Infrared Mapping
  Spectrometer}.
\newblock \icarus~206 (2), 537--560.

\bibitem[{Porco} et~al.(2004){Porco}, {West}, {Squyres}, {McEwen}, {Thomas},
  {Murray}, {Del Genio}, {Ingersoll}, {Johnson}, {Neukum}, {Veverka}, {Dones},
  {Brahic}, {Burns}, {Haemmerle}, {Knowles}, {Dawson}, {Roatsch}, {Beurle}, and
  {Owen}]{Porco04}
{Porco}, C.~C., {West}, R.~A., {Squyres}, S., {McEwen}, A., {Thomas}, P.,
  {Murray}, C.~D., {Del Genio}, A., {Ingersoll}, A.~P., {Johnson}, T.~V.,
  {Neukum}, G., {Veverka}, J., {Dones}, L., {Brahic}, A., {Burns}, J.~A.,
  {Haemmerle}, V., {Knowles}, B., {Dawson}, D., {Roatsch}, T., {Beurle}, K.,
  {Owen}, W., 2004.
\newblock {Cassini Imaging Science: Instrument Characteristics And Anticipated
  Scientific Investigations At Saturn}.
\newblock \ssr~115, 363--497.

\bibitem[{Poston} et~al.(2018){Poston}, {Mahjoub}, {Ehlmann}, {Blacksberg},
  {Brown}, {Carlson}, {Eiler}, {Hand}, {Hodyss}, and {Wong}]{Poston18}
{Poston}, M.~J., {Mahjoub}, A., {Ehlmann}, B.~L., {Blacksberg}, J., {Brown},
  M.~E., {Carlson}, R.~W., {Eiler}, J.~M., {Hand}, K.~P., {Hodyss}, R., {Wong},
  I., 2018.
\newblock {Visible Near-infrared Spectral Evolution of Irradiated Mixed Ices
  and Application to Kuiper Belt Objects and Jupiter Trojans}.
\newblock \apj~856 (2), 124.

\bibitem[{Roussos} et~al.(2008){Roussos}, {Krupp}, {Armstrong}, {Paranicas},
  {Mitchell}, {Krimigis}, {Jones}, {Dialynas}, {Sergis}, and
  {Hamilton}]{Roussos08}
{Roussos}, E., {Krupp}, N., {Armstrong}, T.~P., {Paranicas}, C., {Mitchell},
  D.~G., {Krimigis}, S.~M., {Jones}, G.~H., {Dialynas}, K., {Sergis}, N.,
  {Hamilton}, D.~C., 2008.
\newblock {Discovery of a transient radiation belt at Saturn}.
\newblock \grl~35 (22), L22106.

\bibitem[{Schenk} et~al.(2011){Schenk}, {Hamilton}, {Johnson}, {McKinnon},
  {Paranicas}, {Schmidt}, and {Showalter}]{Schenk11}
{Schenk}, P., {Hamilton}, D.~P., {Johnson}, R.~E., {McKinnon}, W.~B.,
  {Paranicas}, C., {Schmidt}, J., {Showalter}, M.~R., 2011.
\newblock {Plasma, plumes and rings: Saturn system dynamics as recorded in
  global color patterns on its midsize icy satellites}.
\newblock \icarus~211 (1), 740--757.

\bibitem[{Schr{\"o}der} et~al.(2014){Schr{\"o}der}, {Mottola}, {Keller},
  {Raymond}, and {Russell}]{Schroeder14}
{Schr{\"o}der}, S.~E., {Mottola}, S., {Keller}, H.~U., {Raymond}, C.~A.,
  {Russell}, C.~T., 2014.
\newblock {Reprint of: Resolved photometry of Vesta reveals physical properties
  of crater regolith}.
\newblock \planss~103, 66--81.

\bibitem[{Scipioni} et~al.(2013){Scipioni}, {Tosi}, {Stephan}, {Filacchione},
  {Ciarniello}, {Capaccioni}, and {Cerroni}]{Scipioni13}
{Scipioni}, F., {Tosi}, F., {Stephan}, K., {Filacchione}, G., {Ciarniello}, M.,
  {Capaccioni}, F., {Cerroni}, P., 2013.
\newblock {Spectroscopic classification of icy satellites of Saturn I:
  Identification of terrain units on Dione}.
\newblock \icarus~226 (2), 1331--1349.

\bibitem[{Scipioni} et~al.(2014){Scipioni}, {Tosi}, {Stephan}, {Filacchione},
  {Ciarniello}, {Capaccioni}, and {Cerroni}]{Scipioni14}
{Scipioni}, F., {Tosi}, F., {Stephan}, K., {Filacchione}, G., {Ciarniello}, M.,
  {Capaccioni}, F., {Cerroni}, P., 2014.
\newblock {Spectroscopic classification of icy satellites of Saturn II:
  Identification of terrain units on Rhea}.
\newblock Icarus~234, 1--16.

\bibitem[{Shkuratov} et~al.(2011){Shkuratov}, {Kaydash}, {Korokhin},
  {Velikodsky}, {Opanasenko}, and {Videen}]{Shkuratov11}
{Shkuratov}, Y., {Kaydash}, V., {Korokhin}, V., {Velikodsky}, Y., {Opanasenko},
  N., {Videen}, G., 2011.
\newblock {Optical measurements of the Moon as a tool to study its surface}.
\newblock \planss~59 (13), 1326--1371.

\bibitem[Thomas and Helfenstein(2019)Thomas and Helfenstein]{Thomas19}
Thomas, P., Helfenstein, P., 2019.
\newblock The small inner satellites of saturn: Shapes, structures and some
  implications.
\newblock Icarus.

\bibitem[{Thomas} et~al.(2018){Thomas}, {Tiscareno}, and
  {Helfenstein}]{Thomas18}
{Thomas}, P., {Tiscareno}, M.~S., {Helfenstein}, P., 2018.
\newblock The inner small satellites of saturn, and hyperion.
\newblock In: {Schenk}, P., {Clark}, R., {Howett}, C. J.~A., {Verbiscer}, A.,
  {Waite}, J.~H., {Dotson}, R. (Eds.), Enceladus and the Icy Moons of Saturn,
  pp.\  387--408. University of Arizona Press.

\bibitem[{Thomas}(2010){Thomas}]{Thomas10}
{Thomas}, P.~C., 2010.
\newblock {Sizes, shapes, and derived properties of the saturnian satellites
  after the Cassini nominal mission}.
\newblock \icarus~208, 395--401.

\bibitem[{Thomas} et~al.(2013){Thomas}, {Burns}, {Hedman}, {Helfenstein},
  {Morrison}, {Tiscareno}, and {Veverka}]{Thomas13}
{Thomas}, P.~C., {Burns}, J.~A., {Hedman}, M., {Helfenstein}, P., {Morrison},
  S., {Tiscareno}, M.~S., {Veverka}, J., 2013.
\newblock {The inner small satellites of Saturn: A variety of worlds}.
\newblock \icarus~226, 999--1019.

\bibitem[{Tiscareno} et~al.(2013){Tiscareno}, {Mitchell}, {Murray}, {Di Nino},
  {Hedman}, {Schmidt}, {Burns}, {Cuzzi}, {Porco}, {Beurle}, and
  {Evans}]{Tiscareno13}
{Tiscareno}, M.~S., {Mitchell}, C.~J., {Murray}, C.~D., {Di Nino}, D.,
  {Hedman}, M.~M., {Schmidt}, J., {Burns}, J.~A., {Cuzzi}, J.~N., {Porco},
  C.~C., {Beurle}, K., {Evans}, M.~W., 2013.
\newblock {Observations of Ejecta Clouds Produced by Impacts onto
  Saturn{\textquoteright}s Rings}.
\newblock Science~340 (6131), 460--464.

\bibitem[{Trumbo} et~al.(2019){Trumbo}, {Brown}, and {Hand}]{Trumbo19}
{Trumbo}, S.~K., {Brown}, M.~E., {Hand}, K.~P., 2019.
\newblock {Sodium chloride on the surface of Europa}.
\newblock Science~5 (6), aaw7123.

\bibitem[{Verbiscer} et~al.(2007){Verbiscer}, {French}, {Showalter}, and
  {Helfenstein}]{Verbiscer07}
{Verbiscer}, A., {French}, R., {Showalter}, M., {Helfenstein}, P., 2007.
\newblock {Enceladus: Cosmic Graffiti Artist Caught in the Act}.
\newblock Science~315, 815.

\bibitem[{Verbiscer} et~al.(2018){Verbiscer}, P.{Helfenstein}, {Buratti}, and
  {Royer}]{Verbiscer18}
{Verbiscer}, A., P.{Helfenstein}, {Buratti}, B., {Royer}, E., 2018.
\newblock Surface properties of saturn's moons from optical remonte sensing.
\newblock In: {Schenk}, P., {Clark}, R., {Howett}, C. J.~A., {Verbiscer}, A.,
  {Waite}, J.~H., {Dotson}, R. (Eds.), Enceladus and the Icy Moons of Saturn,
  pp.\  323--341. University of Arizona Press.

\bibitem[{Verbiscer} and {Veverka}(1989){Verbiscer} and {Veverka}]{Verbiscer89}
{Verbiscer}, A.~J., {Veverka}, J., 1989.
\newblock {Albedo dichotomy of Rhea: Hapke analysis of Voyager photometry}.
\newblock \icarus~82 (2), 336--353.

\bibitem[{Verbiscer} and {Veverka}(1992){Verbiscer} and {Veverka}]{Verbiscer92}
{Verbiscer}, A.~J., {Veverka}, J., 1992.
\newblock {Mimas: Photometric roughness and albedo map}.
\newblock \icarus~99 (1), 63--69.

\bibitem[{Verbiscer} and {Veverka}(1994){Verbiscer} and {Veverka}]{Verbiscer94}
{Verbiscer}, A.~J., {Veverka}, J., 1994.
\newblock {A Photometric Study of Enceladus}.
\newblock \icarus~110 (1), 155--164.

\bibitem[{Vickers}(1996){Vickers}]{Vickers96}
{Vickers}, G., 1996.
\newblock {The projected areas of ellipsoids and cylinders}.
\newblock Powder Technology~86, 195--200.

\bibitem[{West} et~al.(2010){West}, {Knowles}, {Birath}, {Charnoz}, {Di Nino},
  {Hedman}, {Helfenstein}, {McEwen}, {Perry}, {Porco}, {Salmon}, {Throop}, and
  {Wilson}]{West10}
{West}, R., {Knowles}, B., {Birath}, E., {Charnoz}, S., {Di Nino}, D.,
  {Hedman}, M., {Helfenstein}, P., {McEwen}, A., {Perry}, J., {Porco}, C.,
  {Salmon}, J., {Throop}, H., {Wilson}, D., 2010.
\newblock {In-flight calibration of the Cassini imaging science sub-system
  cameras}.
\newblock \planss~58, 1475--1488.

\bibitem[{Ye} et~al.(2016){Ye}, {Gurnett}, and {Kurth}]{Ye16}
{Ye}, S.~Y., {Gurnett}, D.~A., {Kurth}, W.~S., 2016.
\newblock {In-situ measurements of Saturn's dusty rings based on dust impact
  signals detected by Cassini RPWS}.
\newblock Icarus~279, 51--61.

\bibitem[{Ye} et~al.(2014){Ye}, {Gurnett}, {Kurth}, {Averkamp}, {Kempf}, {Hsu},
  {Srama}, and {Gr{\"u}n}]{Ye14}
{Ye}, S.~Y., {Gurnett}, D.~A., {Kurth}, W.~S., {Averkamp}, T.~F., {Kempf}, S.,
  {Hsu}, H.~W., {Srama}, R., {Gr{\"u}n}, E., 2014.
\newblock {Properties of dust particles near Saturn inferred from voltage
  pulses induced by dust impacts on Cassini spacecraft}.
\newblock Journal of Geophysical Research (Space Physics)~119 (8), 6294--6312.

\end{thebibliography}

\section*{Appendix A: Python code to evaluate predicted areas}

Note, the inputs to this program are the sub-solar longitude {\tt slon}, the sub-solar latitude {\tt slat} , the sub-observer longitude {\tt clon}, the sub-observer latitude {\tt clat}, the object's shape parameters {\tt a,b,c}  and the Minnaert parameter {\tt k}

\bigskip

{\scriptsize \tt \noindent def  objmod (slon,slat,clon,clat,a,b,c,k):

 \indent        import numpy as np
 
 \indent       theta=np.pi/180*np.array(range(181))
 
 \indent     phi=np.pi/180*np.array(range(361))
 
\indent        mat=np.array(np.ones((361,181),float))\

\indent        thx=mat*1.0

\indent         phix=mat*1.0

\indent         for i in range(361):

\indent \indent                for j in range(181):

\indent \indent                       thx[i,j]=np.pi*j/180

\indent   \indent                     phix[i,j]=np.pi*i/180

\indent        areaf=np.sqrt(a**2*b**2*np.cos(thx)**2+c**2*(b**2*np.cos(phix)**2 +a**2*np.sin(phix)**2)*np.sin(thx)**2)

\indent        rf=np.sqrt(areaf)

\indent        xf=np.sin(thx)*np.cos(phix)/a

\indent        yf=np.sin(thx)*np.sin(phix)/b

\indent        zf=np.cos(thx)/c

\indent        nx=xf/np.sqrt(xf**2+yf**2+zf**2)

\indent        ny=yf/np.sqrt(xf**2+yf**2+zf**2)

\indent        nz=zf/np.sqrt(xf**2+yf**2+zf**2)

\indent        sln=slon*np.pi/180

\indent        slt=(90-slat)*np.pi/180

\indent        cln=clon*np.pi/180

\indent        clt=(90-clat)*np.pi/180

\indent        cosi=nx*np.sin(slt)*np.cos(sln)+ny*np.sin(slt)*np.sin(sln)+nz*np.cos(slt)

\indent        cose=nx*np.sin(clt)*np.cos(cln)+ny*np.sin(clt)*np.sin(cln)+nz*np.cos(clt)

\indent        cosi=.5*(cosi+abs(cosi))

\indent        cose=.5*(cose+abs(cose))

\indent        bright=cosi**k*cose**(1-k)

\indent        brfact=bright*np.sin(thx)*areaf*cose*np.pi/180*np.pi/180

\indent        pred=np.sum(brfact)

\indent        return pred}
 
 \normalsize
 
 \section*{Appendix B: A Revised Shape for Polydeuces}
 
\begin{figure}
\centerline{\resizebox{4.5in}{!}{\includegraphics{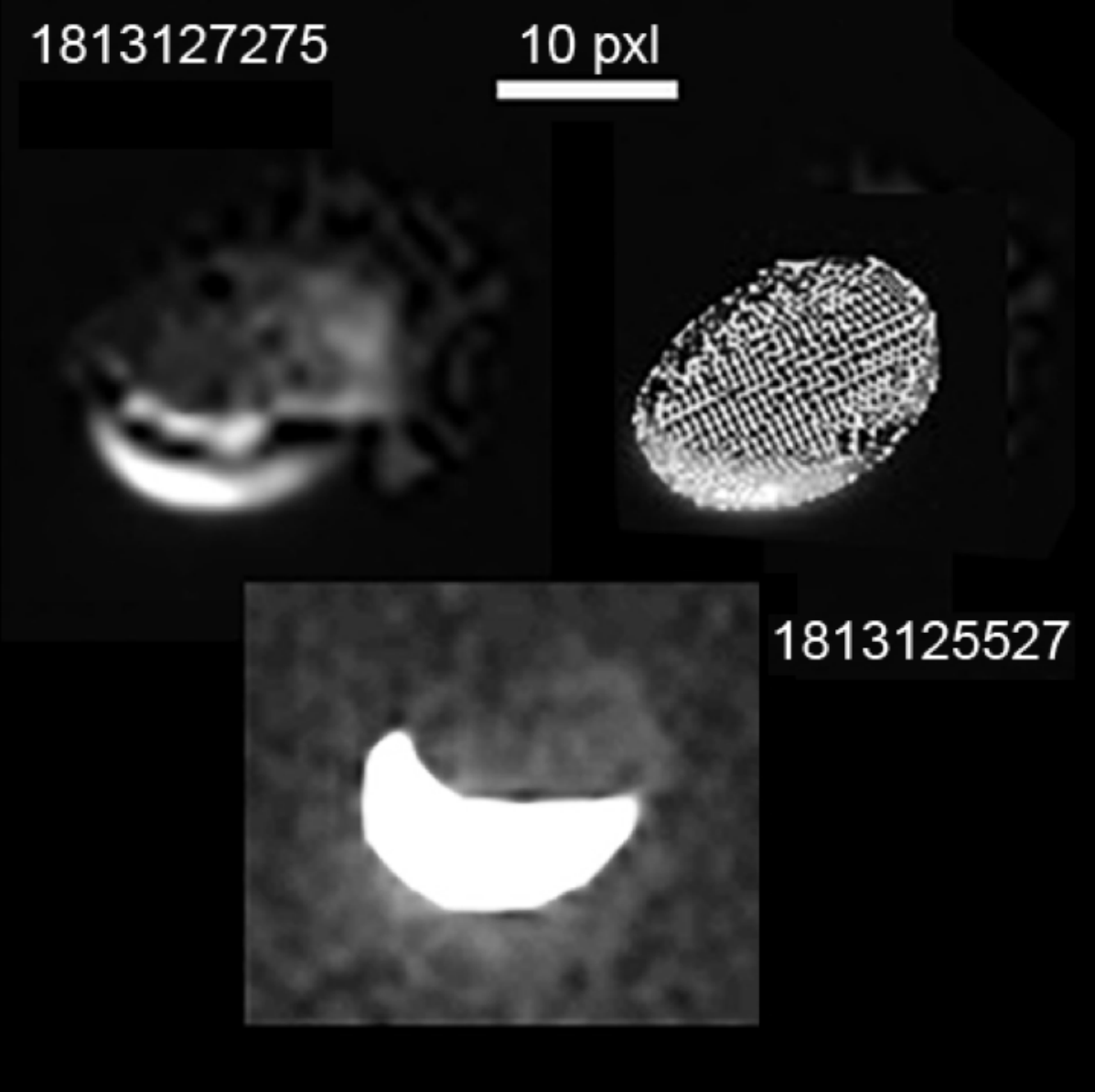}}}
\caption{Shape analysis of Polydeuces from image N1813125527.  Top left: Image with regions under direct illumination and Saturn-shine illumination stretched separately.  Top right: Final shape model overlaid on image. Bottom: Image with a single stretch.}
\label{polyim}
\end{figure}

\begin{figure}
\centerline{\resizebox{4.5in}{!}{\includegraphics{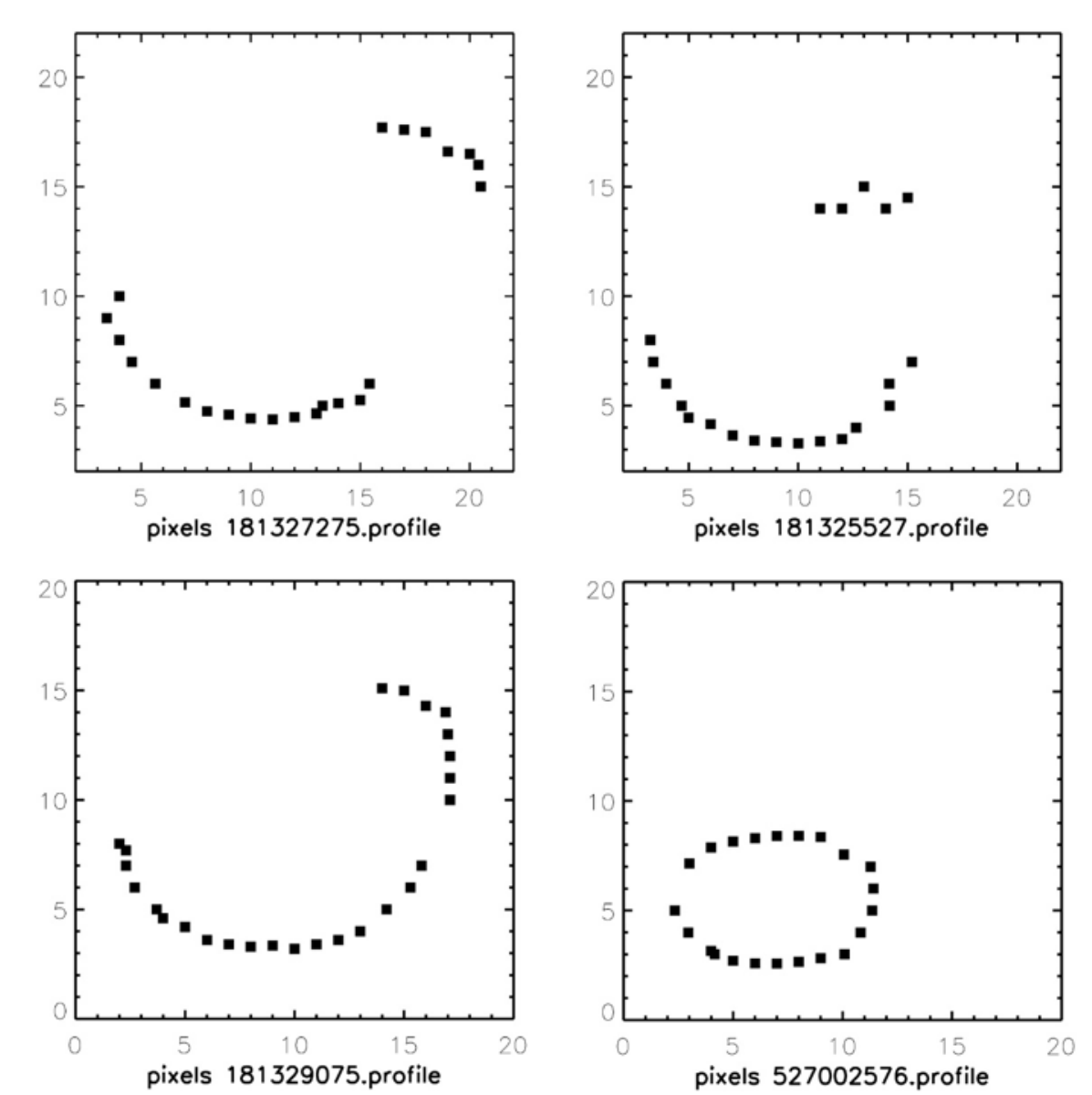}}}
\caption{Estimated limb locations from representative images of Polydeuces.}
\label{polylimb}
\end{figure}

The shape model of Polydeuces reported by \citet{Thomas13} has relatively large uncertainties because it was based on a very limited number of resolved images. Fortunately, late in the Cassini mission there were two additional flybies of Polydeuces where the moon was over 10 pixels wide in the Narrow Angle Camera. Furthermore, in several of these images a large fraction of the limb is visible thanks to the combination of direct illumination and Saturn shine (see Figures~\ref{polyim}-~\ref{polylimb}). The full set of resolved images listed in Table~\ref{polyshape} were then manually fit to a ( 5 $^\circ\times 5^\circ$ )  shape model.  Saturn shine was crucial in getting centers. Shape control is good for the $a$ and $c$ axes; $a$ is the best constrained because of the Saturn shine images taken from near the intermediate axis showing the full projection of $a$.  The $b$ axis is poorly constrained, with the terminator in some images being the primary source of information about this dimension.  Note the limb positions shown in Figure~\ref{polylimb} are relatively smooth, but they do not approximate simple ellipsoids.
 
 \begin{table}
 \caption{Resolved images of Polydeuces used to derive its shape}
 \label{polyshape}
 \centerline{\resizebox{6.5in}{!}{\begin{tabular}{l c c c c c c c c }\hline
  Image  &   Sub-spacecraft &   Sub-spacecraft  &   Sub-solar &   Sub-solar  &   Range &  Pole$^a$  & $x_c$ &     $y_c$  \\
  & Lat. (deg) & Long. (deg) & Lat. (deg) & Long. (deg) & (km) & (deg) & (pixels) & (pixels) \\
  \hline
 N1526998156 &  -0.69 & 300.81 & -17.38 & 265.61 & 70745.54 & 359.92   & 504.9 & 512.4 \\ 
 N1527002576  &-0.51 & 282.09 & -17.38 & 272.34  &64087.42 & 359.92   & 524.1 & 511.2  \\
 N1527002708  &-0.50 &  281.47 & -17.38 &  272.54 & 64101.53 & 359.92   & 527.4 & 511.9  \\
 N1527006179  & -0.28 & 265.98 & -17.38 & 277.82  & 68549.73 & 359.92   & 552.1 & 511.8  \\
 N1809910321  & -0.02 & 30.47  & 24.54 & 300.35  & 47250.34  & 99.94    & 511.1 & 575.9  \\
 N1809910848  & -0.05 & 34.59  & 24.54 & 301.16  & 44829.21  & 99.92    & 520.7 & 580.0  \\
 N1813120961 &  0.76 & 221.86 & 24.76 & 146.90  & 54004.30 & 179.92   & 495.6 & 519.7  \\
 N1813121872  & 0.87 & 227.85 & 24.76 & 148.29 & 49544.58 & 179.92   & 511.8 & 418.4  \\
 N1813121938  & 0.88 & 228.31 & 24.76 & 148.39 & 49237.54 & 179.92  &  511.3 & 407.1  \\
 N1813123674  & 1.13&  242.24 & 24.76 &151.03  & 41948.07 & 179.92  & 539.3 & 571.7  \\
 N1813125527  & 1.41& 261.21 & 24.76 & 153.85  & 36595.63 & 151.89  & 533.3 & 414.4  \\
 N1813125560  & 1.42 & 261.59 & 24.76 & 153.90 & 36528.47 & 151.89  & 533.8 & 413.8  \\
 N1813127275  & 1.60 & 282.26 & 24.76 & 156.51  & 34808.05 & 153.29  & 546.6 & 574.0  \\
 N1813129075 &  1.61 &303.92 & 24.76 & 159.25 & 36813.49 & 157.30  & 543.7&  424.7  \\
 N1813129108  & 1.61 & 304.29 & 24.76 & 159.30 & 36884.91 & 157.39  & 541.3 & 410.2 \\
 \hline
 \end{tabular}}}

 $^a$ Predicted positive pole azimuth in images, clockwise from up. 
\end{table}

\clearpage 

\pagebreak

\section*{ Appendix C: Data Tables for individual observations}

{In these tables, ``Range" indicates the distance between the spacecraft and the body center. ``Phase" is the solar phase angle at body center, ``Sub S/C Latitude/Longitude" are the sub-spacecraft planetocentric latitude and (west) longitude, while ``Sub-Solar Latitude/Longitude" are the sub-solar planetocentric latitude and (west) longitude. The value of $A_{\rm eff}$ is computed as described in the text, and the statistical error estimate is based on scatter of brightness values in regions near the planet (note this uncertainty is very small and so is not included for the larger moons. $A_{\rm phys}$ is the geometrical cross section of the moon from the perspective of the spacecraft, and the three values of $a_{\rm pred}$ are computed assuming Minnaert scattering laws with $k=1.0, 0.75$ and 0.50 assuming the shape parameters provided in Table~\ref{shapes}. Note that since the shape of Anthe is not well determined and so these parameters could not be computed for that moon. All images were obtained using the Cassini ISS NAC clear filters expect for those given in the last two tables. All these data tables are provided in machine-readable format in supplementary information associated with this article.}

\startlongtable
\begin{deluxetable*}{ccccccccccccc}
\tablewidth{6.5in}
\tabletypesize{\tiny}
\tablecaption{Brightness estimates from clear-filter  images of Aegaeon.
\label{aegaeontab}}
\tablecolumns{13}
\tablehead{
\colhead{Filename} &  \colhead{Range} & \colhead{Phase} & \colhead{Sub-S/C} & \colhead{Sub-S/C} & \colhead{Sub-Solar} & \colhead{Sub-Solar} & \colhead{$A_{\rm eff}$} & \colhead{$A_{\rm eff}$} &\colhead{$A_{\rm phys}$} &  \colhead{$a_{\rm pred}$} & \colhead{$a_{\rm pred}$} & \colhead{$a_{\rm pred}$} \\
\colhead{} & \colhead{(km)} & \colhead{(deg)} & \colhead{Latitude} & \colhead{Longitude} & \colhead{Latitude} & \colhead{Longitude} & \colhead{Value} & \colhead{Error} & \colhead{} &
 \colhead{k=1} & \colhead{k=0.75} & \colhead{k=0.50} \\
 \colhead{} & \colhead{} & \colhead{} & \colhead{(deg)} & \colhead{(deg)} & \colhead{(deg)} & \colhead{(deg)} & \colhead{(km$^2$)}  & \colhead{(km$^2$)}   & \colhead{(km$^2$)}  & \colhead{(km$^2$)} 
 & \colhead{(km$^2$)}   & \colhead{(km$^2$)} 
\\}
\startdata
     N1560588018 &         1708521.8 &              42.9 &               0.5 &              94.9 &             -12.0 &             136.3 &           0.05365 &           0.00490 &            0.4384 &            0.2172 &            0.2323 &             0.2511\\
      N1597476237 &         1188834.9 &              28.2 &               5.0 &             264.3 &              -5.6 &             238.1 &           0.07027 &           0.00199 &            0.4390 &            0.2630 &            0.2720 &             0.2826\\
      N1597477967 &         1215347.6 &              28.2 &               5.2 &             273.2 &              -5.6 &             247.1 &           0.06765 &           0.00234 &            0.4404 &            0.2819 &            0.2871 &             0.2938\\
      N1598073885 &         1170967.9 &              12.9 &              -0.8 &              91.4 &              -5.5 &              79.3 &           0.08536 &           0.00193 &            0.4397 &            0.3073 &            0.3084 &             0.3098\\
      N1598104211 &         1179616.4 &              28.4 &               3.9 &             262.5 &              -5.5 &             235.7 &           0.05817 &           0.00382 &            0.4371 &            0.2573 &            0.2672 &             0.2787\\
      N1598106121 &         1209151.0 &              28.4 &               4.0 &             272.3 &              -5.5 &             245.5 &           0.06960 &           0.00162 &            0.4403 &            0.2803 &            0.2861 &             0.2933\\
      N1600657200 &         1205263.9 &              15.2 &               4.4 &              89.6 &              -5.0 &              77.6 &           0.07635 &           0.00125 &            0.4407 &            0.3020 &            0.3038 &             0.3063\\
\enddata
\end{deluxetable*}

\startlongtable
\begin{deluxetable*}{ccccccccc}
\tablewidth{6.5in}
\tabletypesize{\tiny}
\tablecolumns{9}
\tablecaption{Brightness estimates from clear-filter  images of Anthe.
\label{anthetab}}
\tablehead{
\colhead{Filename} &  \colhead{Range} & \colhead{Phase} & \colhead{Sub-S/C} & \colhead{Sub-S/C} & \colhead{Sub-Solar} & \colhead{Sub-Solar} & \colhead{$A_{\rm eff}$} & \colhead{$A_{\rm eff}$} \\
\colhead{} & \colhead{(km)} & \colhead{(deg)} & \colhead{Latitude} & \colhead{Longitude} & \colhead{Latitude} & \colhead{Longitude} & \colhead{Value} & \colhead{Error}  \\
 \colhead{} & \colhead{} & \colhead{} & \colhead{(deg)} & \colhead{(deg)} & \colhead{(deg)} & \colhead{(deg)} & \colhead{(km$^2$)}  & \colhead{(km$^2$)}  
\\}
\startdata
      N1575629162 &         1802888.5 &              14.3 &               4.6 &             314.9 &              -9.4 &             312.0 &           0.87980 &            0.00018\\
      N1579321873 &         1440656.1 &              11.3 &               2.1 &              36.8 &              -8.8 &              34.0 &           0.94105 &            0.00006\\
      N1579364158 &         1258508.1 &              17.9 &               6.6 &             213.0 &              -8.8 &             203.9 &           0.76642 &            0.00094\\
      N1580356175 &         1214750.8 &              20.0 &               3.0 &             247.6 &              -8.6 &             231.3 &           0.88177 &            0.00086\\
      N1581514393 &         1472790.1 &              21.2 &              12.5 &             210.3 &              -8.4 &             206.6 &           0.74765 &            0.00064\\
      N1591878477 &         1067058.5 &               9.7 &              -2.1 &             117.6 &              -6.6 &             109.0 &           1.11239 &            0.00026\\
      N1595481922 &          869685.1 &              30.3 &             -13.2 &             225.5 &              -5.9 &             195.6 &           0.56951 &            0.00006\\
      N1596338308 &         1242696.0 &              28.9 &              22.9 &              42.6 &              -5.8 &              38.7 &           0.57676 &            0.00029\\
\enddata
\end{deluxetable*}

\startlongtable
\begin{deluxetable*}{ccccccccccccc}
\tablewidth{6.5in}
\tabletypesize{\tiny}
\tablecolumns{13}
\tablecaption{Brightness estimates from clear-filter  images of Methone.
\label{methonetab}}
\tablehead{
\colhead{Filename} &  \colhead{Range} & \colhead{Phase} & \colhead{Sub-S/C} & \colhead{Sub-S/C} & \colhead{Sub-Solar} & \colhead{Sub-Solar} & \colhead{$A_{\rm eff}$} & \colhead{$A_{\rm eff}$} &\colhead{$A_{\rm phys}$} &  \colhead{$a_{\rm pred}$} & \colhead{$a_{\rm pred}$} & \colhead{$a_{\rm pred}$} \\
\colhead{} & \colhead{(km)} & \colhead{(deg)} & \colhead{Latitude} & \colhead{Longitude} & \colhead{Latitude} & \colhead{Longitude} & \colhead{Value} & \colhead{Error} & \colhead{} &
 \colhead{k=1} & \colhead{k=0.75} & \colhead{k=0.50} \\
 \colhead{} & \colhead{} & \colhead{} & \colhead{(deg)} & \colhead{(deg)} & \colhead{(deg)} & \colhead{(deg)} & \colhead{(km$^2$)}  & \colhead{(km$^2$)}   & \colhead{(km$^2$)}  & \colhead{(km$^2$)} 
 & \colhead{(km$^2$)}   & \colhead{(km$^2$)} 
\\}
\startdata
      N1502651083 &         2157110.5 &              61.1 &             -16.3 &             195.6 &             -20.7 &             131.0 &           0.85344 &           0.00034 &            5.3901 &            1.7843 &            1.7464 &             1.7810\\
      N1502796764 &         2182712.0 &              49.2 &             -16.1 &              63.9 &             -20.7 &              12.0 &           1.23989 &           0.00288 &            7.0567 &            2.7398 &            2.9195 &             3.1932\\
      N1505466661 &         2597383.5 &              75.2 &              -0.3 &             304.2 &             -20.4 &             229.8 &           1.08037 &           0.00377 &            6.7187 &            2.6551 &            2.7021 &             2.8410\\
      N1506716564 &         2357443.8 &              78.3 &              -0.3 &              65.0 &             -20.2 &             347.3 &           0.59662 &           0.00542 &            7.0143 &            1.2778 &            1.3753 &             1.5609\\
      N1513354056 &         2991500.0 &              75.9 &              -0.2 &              90.1 &             -19.3 &              14.9 &           0.92142 &           0.00364 &            7.3746 &            1.9798 &            2.1500 &             2.4335\\
      N1513572473 &         2639783.5 &              77.0 &              -0.3 &             272.5 &             -19.3 &             196.2 &           0.92816 &           0.00253 &            7.3726 &            1.9867 &            2.1553 &             2.4368\\
      N1513697664 &         2444116.8 &              64.6 &              -0.3 &              55.8 &             -19.3 &             352.7 &           0.46728 &           0.00244 &            6.7177 &            1.6638 &            1.7749 &             1.9790\\
\enddata
\end{deluxetable*}

\startlongtable
\begin{deluxetable*}{ccccccccccccc}
\tablewidth{6.5in}
\tabletypesize{\tiny}
\tablecolumns{13}
\tablecaption{Brightness estimates from clear-filter  images of Pallene.
\label{pallenetab}}
\tablehead{
\colhead{Filename} &  \colhead{Range} & \colhead{Phase} & \colhead{Sub-S/C} & \colhead{Sub-S/C} & \colhead{Sub-Solar} & \colhead{Sub-Solar} & \colhead{$A_{\rm eff}$} & \colhead{$A_{\rm eff}$} &\colhead{$A_{\rm phys}$} &  \colhead{$a_{\rm pred}$} & \colhead{$a_{\rm pred}$} & \colhead{$a_{\rm pred}$} \\
\colhead{} & \colhead{(km)} & \colhead{(deg)} & \colhead{Latitude} & \colhead{Longitude} & \colhead{Latitude} & \colhead{Longitude} & \colhead{Value} & \colhead{Error} & \colhead{} &
 \colhead{k=1} & \colhead{k=0.75} & \colhead{k=0.50} \\
 \colhead{} & \colhead{} & \colhead{} & \colhead{(deg)} & \colhead{(deg)} & \colhead{(deg)} & \colhead{(deg)} & \colhead{(km$^2$)}  & \colhead{(km$^2$)}   & \colhead{(km$^2$)}  & \colhead{(km$^2$)} 
 & \colhead{(km$^2$)}   & \colhead{(km$^2$)} 
\\}
\startdata
      N1502276381 &         2437835.2 &              78.9 &             -10.4 &             287.9 &             -20.8 &             205.9 &           2.01443 &           0.00006 &           16.3913 &            5.0531 &            5.3314 &             5.8471\\
      N1502795579 &         2300004.0 &              56.1 &             -15.3 &             339.6 &             -20.7 &             280.6 &           2.70475 &           0.00006 &           13.2926 &            6.5691 &            6.3338 &             6.2744\\
      N1505466121 &         2520183.0 &              76.3 &              -0.3 &             279.6 &             -20.4 &             204.1 &           1.68066 &           0.00263 &           16.5546 &            4.9522 &            5.2495 &             5.7894\\
      N1506716999 &         2173371.8 &              78.1 &              -0.3 &             118.7 &             -20.2 &              41.4 &           1.67143 &           0.00796 &           15.7004 &            5.3684 &            5.5332 &             5.9192\\
      N1506907435 &         2497063.0 &              71.6 &              -0.3 &              79.2 &             -20.2 &               8.8 &           1.63914 &           0.00849 &           16.5278 &            4.4971 &            4.8088 &             5.3639\\
      N1508412740 &         2300827.8 &              74.2 &              -0.4 &             117.9 &             -20.0 &              44.6 &           1.74305 &           0.00371 &           15.7529 &            5.8123 &            5.9700 &             6.3476\\
      N1511332959 &         2177036.2 &              70.4 &              -0.3 &             218.9 &             -19.6 &             149.7 &           1.25172 &           0.00081 &           14.0280 &            3.1824 &            3.2588 &             3.5060\\
\enddata
\end{deluxetable*}

\startlongtable
\begin{deluxetable*}{ccccccccccccc}
\tablewidth{6.5in}
\tabletypesize{\tiny}
\tablecolumns{13}
\tablecaption{Brightness estimates from clear-filter  images of Telesto.
\label{telestotab}}
\tablehead{
\colhead{Filename} &  \colhead{Range} & \colhead{Phase} & \colhead{\tiny Sub-S/C} & \colhead{\tiny Sub-S/C} & \colhead{\tiny Sub-Solar} & \colhead{\tiny Sub-Solar} & \colhead{$A_{\rm eff}$} & \colhead{$A_{\rm eff}$} &\colhead{$A_{\rm phys}$} &  \colhead{$a_{\rm pred}$} & \colhead{$a_{\rm pred}$} & \colhead{$a_{\rm pred}$} \\
\colhead{} & \colhead{(km)} & \colhead{(deg)} & \colhead{\tiny Latitude} & \colhead{\tiny Longitude} & \colhead{\tiny Latitude} & \colhead{\tiny Longitude} & \colhead{Value} & \colhead{Error} & \colhead{} &
 \colhead{k=1} & \colhead{k=0.75} & \colhead{k=0.50} \\
 \colhead{} & \colhead{} & \colhead{} & \colhead{(deg)} & \colhead{(deg)} & \colhead{(deg)} & \colhead{(deg)} & \colhead{(km$^2$)}  & \colhead{(km$^2$)}   & \colhead{(km$^2$)}  & \colhead{(km$^2$)} 
 & \colhead{(km$^2$)}   & \colhead{(km$^2$)} 
\\}
\startdata
     N1495206641 &         1158822.1 &              41.0 &             -20.6 &             283.1 &             -21.7 &             239.0 &         209.49303 &           0.00203 &          510.9615 &          291.6927 &          296.2795 &           304.6195\\
      N1495310692 &          533103.5 &              48.7 &             -16.1 &              57.3 &             -21.6 &             108.6 &         186.92639 &           0.00024 &          477.0852 &          267.6940 &          266.3135 &           269.2979\\
      N1513353874 &         3223840.5 &              76.1 &              -0.2 &              38.3 &             -19.3 &             323.0 &          55.22128 &           0.02512 &          418.3909 &           80.5283 &           81.8249 &            87.6863\\
      N1513525391 &         2910381.2 &              70.1 &              -0.2 &              50.4 &             -19.3 &             341.4 &          60.29562 &           0.02358 &          448.4575 &           98.7257 &          103.0927 &           113.0088\\
      N1515851740 &         1677181.9 &              64.8 &              -0.1 &             138.3 &             -19.0 &              75.1 &         125.14794 &           0.00486 &          424.5765 &          195.9870 &          191.5797 &           192.7610\\
      N1550920790 &         1675677.8 &              54.1 &              17.0 &              20.2 &             -13.6 &              65.3 &         118.35385 &           0.00402 &          402.1162 &          166.1985 &          162.1580 &           163.5099\\
      N1551094911 &         1848412.8 &              78.9 &              35.6 &              24.2 &             -13.6 &              89.6 &          67.92459 &           0.01501 &          474.1757 &          127.2060 &          127.8723 &           134.8444\\
\enddata
\end{deluxetable*}

\startlongtable
\begin{deluxetable*}{ccccccccccccc}
\tablewidth{6.5in}
\tabletypesize{\tiny}
\tablecolumns{13}
\tablecaption{Brightness estimates from clear-filter  images of Calypso.
\label{calypsotab}}
\tablehead{
\colhead{Filename} &  \colhead{Range} & \colhead{Phase} & \colhead{\tiny Sub-S/C} & \colhead{\tiny Sub-S/C} & \colhead{\tiny Sub-Solar} & \colhead{\tiny Sub-Solar} & \colhead{$A_{\rm eff}$} & \colhead{$A_{\rm eff}$} &\colhead{$A_{\rm phys}$} &  \colhead{$a_{\rm pred}$} & \colhead{$a_{\rm pred}$} & \colhead{$a_{\rm pred}$} \\
\colhead{} & \colhead{(km)} & \colhead{(deg)} & \colhead{\tiny Latitude} & \colhead{\tiny Longitude} & \colhead{\tiny Latitude} & \colhead{\tiny Longitude} & \colhead{Value} & \colhead{Error} & \colhead{} &
 \colhead{k=1} & \colhead{k=0.75} & \colhead{k=0.50} \\
 \colhead{} & \colhead{} & \colhead{} & \colhead{(deg)} & \colhead{(deg)} & \colhead{(deg)} & \colhead{(deg)} & \colhead{(km$^2$)}  & \colhead{(km$^2$)}   & \colhead{(km$^2$)}  & \colhead{(km$^2$)} 
 & \colhead{(km$^2$)}   & \colhead{(km$^2$)} 
\\}
\startdata
     N1495209871 &          890860.8 &              16.6 &             -26.6 &             140.6 &             -21.7 &             123.2 &         215.58528 &           0.00317 &          286.1855 &          174.0992 &          172.8976 &           172.3568\\
      N1495306082 &          744442.8 &               8.1 &             -13.8 &             337.8 &             -21.6 &             335.5 &         175.25667 &           0.00123 &          226.2691 &          127.2145 &          124.7317 &           122.5648\\
      N1511290086 &         2690833.8 &              65.2 &              -0.4 &              36.6 &             -19.6 &             332.9 &          17.58906 &           0.08132 &          232.8271 &           49.1007 &           49.7600 &            52.9677\\
      N1513526081 &         2894673.2 &              79.3 &              -0.2 &             306.3 &             -19.3 &             227.6 &          74.90391 &           0.00928 &          263.6748 &           94.7175 &           95.4810 &            99.6274\\
      N1513746037 &         2373977.8 &              60.5 &              -0.4 &              51.7 &             -19.3 &             353.0 &          46.70824 &           0.01034 &          260.2657 &           63.1165 &           66.3753 &            72.8773\\
      N1515907721 &         1464391.1 &              60.9 &              -0.2 &             142.6 &             -19.0 &              83.5 &          83.69676 &           0.00505 &          233.0732 &          113.9775 &          108.6222 &           106.2716\\
      N1552346494 &         1321233.1 &              61.8 &              27.5 &             159.6 &             -13.4 &             207.5 &          37.35155 &           0.00874 &          268.3173 &           44.9510 &           47.2656 &            52.3660\\
\enddata
\end{deluxetable*}

\startlongtable
\begin{deluxetable*}{ccccccccccccc}
\tablewidth{6.5in}
\tabletypesize{\tiny}
\tablecolumns{13}
\tablecaption{Brightness estimates from clear-filter  images of Helene.
\label{helenetab}}
\tablehead{
\colhead{Filename} &  \colhead{Range} & \colhead{Phase} & \colhead{\tiny Sub-S/C} & \colhead{\tiny Sub-S/C} & \colhead{\tiny Sub-Solar} & \colhead{\tiny Sub-Solar} & \colhead{$A_{\rm eff}$} & \colhead{$A_{\rm eff}$} &\colhead{$A_{\rm phys}$} &  \colhead{$a_{\rm pred}$} & \colhead{$a_{\rm pred}$} & \colhead{$a_{\rm pred}$} \\
\colhead{} & \colhead{(km)} & \colhead{(deg)} & \colhead{\tiny Latitude} & \colhead{\tiny Longitude} & \colhead{\tiny Latitude} & \colhead{\tiny Longitude} & \colhead{Value} & \colhead{Error} & \colhead{} &
 \colhead{k=1} & \colhead{k=0.75} & \colhead{k=0.50} \\
 \colhead{} & \colhead{} & \colhead{} & \colhead{(deg)} & \colhead{(deg)} & \colhead{(deg)} & \colhead{(deg)} & \colhead{(km$^2$)}  & \colhead{(km$^2$)}   & \colhead{(km$^2$)}  & \colhead{(km$^2$)} 
 & \colhead{(km$^2$)}   & \colhead{(km$^2$)} 
\\}
\startdata
      N1495206161 &         1294705.5 &              42.7 &             -18.2 &             304.9 &             -21.7 &             259.4 &         396.51160 &           0.00187 &          967.6930 &          517.9379 &          516.7247 &           523.9229\\
      N1495311312 &          759950.5 &              37.1 &             -11.6 &              23.3 &             -21.6 &              60.7 &         441.52603 &           0.00102 &          875.4782 &          473.1956 &          464.6283 &           463.5204\\
      N1518953101 &         2748696.2 &              79.6 &              -0.2 &             105.4 &             -18.5 &              26.3 &         112.52875 &           0.00832 &          936.0639 &          252.2294 &          259.4079 &           278.8613\\
      N1551142776 &         1958827.1 &              79.7 &              37.5 &             358.6 &             -13.6 &              64.0 &         131.82022 &           0.00863 &         1073.9272 &          202.1025 &          211.6382 &           233.8763\\
      N1552435415 &         1408418.5 &              71.1 &              39.7 &             182.1 &             -13.4 &             233.0 &         156.52940 &           0.00212 &         1089.9978 &          236.4063 &          248.8237 &           275.2325\\
      N1552559550 &         2128087.8 &              76.1 &              34.2 &             359.5 &             -13.4 &              62.1 &         153.74387 &           0.00710 &         1041.6052 &          216.2016 &          225.0033 &           246.5506\\
      N1553980614 &         2283324.2 &              73.3 &              30.3 &               5.3 &             -13.1 &              66.7 &         174.85187 &           0.01106 &         1003.6191 &          232.7841 &          239.9312 &           259.7333\\
\enddata
\end{deluxetable*}

\startlongtable
\begin{deluxetable*}{ccccccccccccc}
\tablewidth{6.5in}
\tabletypesize{\tiny}
\tablecolumns{13}
\tablecaption{Brightness estimates from clear-filter  images of Polydeuces.
\label{polydeucestab}}
\tablehead{
\colhead{Filename} &  \colhead{Range} & \colhead{Phase} & \colhead{Sub-S/C} & \colhead{Sub-S/C} & \colhead{Sub-Solar} & \colhead{Sub-Solar} & \colhead{$A_{\rm eff}$} & \colhead{$A_{\rm eff}$} &\colhead{$A_{\rm phys}$} &  \colhead{$a_{\rm pred}$} & \colhead{$a_{\rm pred}$} & \colhead{$a_{\rm pred}$} \\
\colhead{} & \colhead{(km)} & \colhead{(deg)} & \colhead{Latitude} & \colhead{Longitude} & \colhead{Latitude} & \colhead{Longitude} & \colhead{Value} & \colhead{Error} & \colhead{} &
 \colhead{k=1} & \colhead{k=0.75} & \colhead{k=0.50} \\
 \colhead{} & \colhead{} & \colhead{} & \colhead{(deg)} & \colhead{(deg)} & \colhead{(deg)} & \colhead{(deg)} & \colhead{(km$^2$)}  & \colhead{(km$^2$)}   & \colhead{(km$^2$)}  & \colhead{(km$^2$)} 
 & \colhead{(km$^2$)}   & \colhead{(km$^2$)} 
\\}
\startdata
      N1495210661 &          786145.6 &              20.5 &             -30.4 &             160.4 &             -21.7 &             139.7 &           2.90468 &           0.00001 &            7.0605 &            4.2419 &            4.2603 &             4.3066\\
      N1495210808 &          785039.1 &              20.5 &             -30.5 &             160.6 &             -21.7 &             139.9 &           2.91503 &           0.00050 &            7.0595 &            4.2381 &            4.2567 &             4.3035\\
      N1495306772 &          654423.7 &              23.8 &             -15.1 &             314.0 &             -21.6 &             289.8 &           3.43378 &           0.00023 &            6.9841 &            4.3321 &            4.3030 &             4.3052\\
      N1502650213 &         2702061.5 &              61.8 &             -13.0 &             352.9 &             -20.7 &             288.4 &           0.97387 &           0.00811 &            6.5699 &            2.6354 &            2.6208 &             2.6964\\
      N1502795114 &         1786225.5 &              57.4 &             -19.8 &             206.2 &             -20.7 &             144.6 &           0.98527 &           0.00242 &            6.8046 &            2.6539 &            2.6866 &             2.8104\\
      N1506004925 &         1776319.6 &              42.5 &              -0.3 &              24.1 &             -20.3 &             345.8 &           1.53273 &           0.00150 &            6.5743 &            3.0580 &            3.0755 &             3.1661\\
      N1506764534 &         2079087.6 &              74.1 &              -0.3 &             130.9 &             -20.2 &              57.7 &           1.00535 &           0.00187 &            6.8619 &            2.3180 &            2.3419 &             2.4618\\
\enddata
\end{deluxetable*}

\startlongtable
\begin{deluxetable*}{ccccccccccccc}
\tablewidth{6.5in}
\tabletypesize{\tiny}
\tablecolumns{13}
\tablecaption{Brightness estimates from clear-filter  images of Pan.
\label{pantab}}
\tablehead{
\colhead{Filename} &  \colhead{Range} & \colhead{Phase} & \colhead{\tiny Sub-S/C} & \colhead{\tiny Sub-S/C} & \colhead{\tiny Sub-Solar} & \colhead{\tiny Sub-Solar} & \colhead{$A_{\rm eff}$} & \colhead{$A_{\rm eff}$} &\colhead{$A_{\rm phys}$} &  \colhead{$a_{\rm pred}$} & \colhead{$a_{\rm pred}$} & \colhead{$a_{\rm pred}$} \\
\colhead{} & \colhead{(km)} & \colhead{(deg)} & \colhead{\tiny Latitude} & \colhead{\tiny Longitude} & \colhead{\tiny Latitude} & \colhead{\tiny Longitude} & \colhead{Value} & \colhead{Error} & \colhead{} &
 \colhead{k=1} & \colhead{k=0.75} & \colhead{k=0.50} \\
 \colhead{} & \colhead{} & \colhead{} & \colhead{(deg)} & \colhead{(deg)} & \colhead{(deg)} & \colhead{(deg)} & \colhead{(km$^2$)}  & \colhead{(km$^2$)}   & \colhead{(km$^2$)}  & \colhead{(km$^2$)} 
 & \colhead{(km$^2$)}   & \colhead{(km$^2$)} 
\\}
\startdata
     N1577141695 &         1928322.1 &              23.9 &              14.1 &             255.1 &              -9.2 &             260.7 &          222.3964 &            0.0696 &          578.4056 &          329.4306 &          331.6341 &           337.6532\\
      N1578630786 &         1758181.6 &              32.0 &              22.1 &             241.8 &              -8.9 &             249.6 &          184.8326 &            0.0461 &          584.9566 &          296.6020 &          301.0208 &           311.0923\\
      N1579656793 &         1773520.2 &              33.9 &              21.7 &             108.1 &              -8.7 &             123.4 &          178.0623 &            0.0538 &          593.9390 &          290.1136 &          297.4645 &           310.7608\\
      N1579750304 &         1820849.6 &              38.5 &              24.4 &              60.8 &              -8.7 &              80.9 &          166.6279 &            0.0497 &          592.5009 &          285.7020 &          290.0645 &           300.9773\\
      N1580566295 &         1680456.4 &              25.2 &              16.6 &             232.0 &              -8.6 &             233.0 &          195.7154 &            0.0450 &          555.9801 &          293.7928 &          297.1923 &           304.7319\\
      N1580653070 &         1733973.4 &              31.7 &              20.7 &             129.4 &              -8.6 &             141.7 &          163.4020 &            0.0510 &          565.7847 &          265.4792 &          272.7606 &           285.4930\\
      N1580766531 &         1705000.6 &              35.9 &              25.8 &             232.9 &              -8.5 &             243.8 &          166.9957 &            0.0418 &          585.8973 &          276.7147 &          282.0406 &           293.7582\\
\enddata
\end{deluxetable*}

\startlongtable
\begin{deluxetable*}{ccccccccccccc}
\tablewidth{6.5in}
\tabletypesize{\tiny}
\tablecolumns{13}
\tablecaption{Brightness estimates from clear-filter  images of Atlas.
\label{atlastab}}
\tablehead{
\colhead{Filename} &  \colhead{Range} & \colhead{Phase} & \colhead{\tiny Sub-S/C} & \colhead{\tiny Sub-S/C} & \colhead{\tiny Sub-Solar} & \colhead{\tiny Sub-Solar} & \colhead{$A_{\rm eff}$} & \colhead{$A_{\rm eff}$} &\colhead{$A_{\rm phys}$} &  \colhead{$a_{\rm pred}$} & \colhead{$a_{\rm pred}$} & \colhead{$a_{\rm pred}$} \\
\colhead{} & \colhead{(km)} & \colhead{(deg)} & \colhead{\tiny Latitude} & \colhead{\tiny Longitude} & \colhead{\tiny Latitude} & \colhead{\tiny Longitude} & \colhead{Value} & \colhead{Error} & \colhead{} &
 \colhead{k=1} & \colhead{k=0.75} & \colhead{k=0.50} \\
 \colhead{} & \colhead{} & \colhead{} & \colhead{(deg)} & \colhead{(deg)} & \colhead{(deg)} & \colhead{(deg)} & \colhead{(km$^2$)}  & \colhead{(km$^2$)}   & \colhead{(km$^2$)}  & \colhead{(km$^2$)} 
 & \colhead{(km$^2$)}   & \colhead{(km$^2$)} 
\\}
\startdata
      N1495208581 &         1200176.6 &              20.9 &             -19.6 &              50.0 &             -21.7 &              27.8 &          281.1014 &            0.0047 &          657.0602 &          362.1453 &          363.9504 &           367.7587\\
      N1503007485 &         1538741.6 &              34.7 &             -19.2 &              68.2 &             -20.7 &              31.3 &          223.5278 &            0.0023 &          668.9564 &          338.6541 &          343.4461 &           352.5163\\
      N1503007793 &         1533078.1 &              34.6 &             -19.2 &              70.2 &             -20.7 &              33.4 &          227.3121 &            0.0027 &          670.7019 &          341.2420 &          346.0067 &           355.0086\\
      N1503008105 &         1527283.9 &              34.5 &             -19.3 &              72.3 &             -20.7 &              35.5 &          228.6507 &            0.0051 &          672.3640 &          343.8194 &          348.5363 &           357.4478\\
      N1503008417 &         1521438.0 &              34.4 &             -19.4 &              74.3 &             -20.7 &              37.7 &          231.1872 &            0.0018 &          673.9126 &          346.3409 &          350.9915 &           359.7968\\
      N1503008885 &         1512585.0 &              34.2 &             -19.5 &              77.4 &             -20.7 &              40.9 &          232.9991 &            0.0036 &          676.0113 &          349.9915 &          354.5059 &           363.1126\\
      N1503009353 &         1503649.2 &              34.1 &             -19.6 &              80.6 &             -20.7 &              44.2 &          232.8272 &            0.0053 &          677.8268 &          353.4491 &          357.7854 &           366.1541\\
\enddata
\end{deluxetable*}

\startlongtable
\begin{deluxetable*}{cccccccccccc}
\tablewidth{6.5in}
\tabletypesize{\tiny}
\tablecolumns{12}
\tablecaption{Brightness estimates from clear-filter  images of Prometheus
\label{prometheustab}}
\tablehead{
\colhead{Filename} &  \colhead{Range} & \colhead{Phase} & \colhead{\tiny Sub-S/C} & \colhead{\tiny Sub-S/C} & \colhead{\tiny Sub-Solar} & \colhead{\tiny Sub-Solar} & \colhead{$A_{\rm eff}$} &\colhead{$A_{\rm phys}$} &  \colhead{$a_{\rm pred}$} & \colhead{$a_{\rm pred}$} & \colhead{$a_{\rm pred}$} \\
\colhead{} & \colhead{(km)} & \colhead{(deg)} & \colhead{\tiny Latitude} & \colhead{\tiny Longitude} & \colhead{\tiny Latitude} & \colhead{\tiny Longitude} & \colhead{Value} &  \colhead{} &
 \colhead{k=1} & \colhead{k=0.75} & \colhead{k=0.50} \\
 \colhead{} & \colhead{} & \colhead{} & \colhead{(deg)} & \colhead{(deg)} & \colhead{(deg)} & \colhead{(deg)} & \colhead{(km$^2$)}  & \colhead{(km$^2$)}    & \colhead{(km$^2$)} 
 & \colhead{(km$^2$)}   & \colhead{(km$^2$)} 
\\}
\startdata
     N1575914492 &         2251982.5 &              23.7 &              11.0 &             290.4 &              -9.4 &             302.5 &         3789.8745 &         5945.7690 &         3187.5754 &         3251.0505 &          3354.0364\\
      N1577051965 &         1729552.4 &              21.7 &              10.5 &             105.2 &              -9.2 &             114.1 &         2767.1677 &         6027.8188 &         3416.9006 &         3461.8240 &          3542.7866\\
      N1577098105 &         1950314.6 &              23.5 &              11.7 &              56.7 &              -9.2 &              67.5 &         3105.4451 &         5615.8433 &         3143.3091 &         3130.0500 &          3154.3108\\
      N1577215016 &         1997065.0 &              30.5 &              16.9 &             126.3 &              -9.2 &             142.4 &         1868.2275 &         5680.9424 &         2381.0442 &         2488.8511 &          2652.8489\\
      N1579655833 &         1695057.1 &              33.8 &              22.7 &             159.5 &              -8.7 &             172.1 &         1180.3311 &         4969.6963 &         1394.5358 &         1483.8430 &          1628.8689\\
      N1579750574 &         1792714.6 &              39.1 &              24.8 &              75.0 &              -8.7 &              95.8 &         2139.3582 &         6508.8945 &         3160.9849 &         3228.7004 &          3371.5254\\
      N1580566520 &         1695084.1 &              25.1 &              16.5 &             241.6 &              -8.6 &             242.1 &         2460.6575 &         5902.4258 &         3075.2437 &         3117.7126 &          3207.7449\\
      N1580613950 &         1675761.0 &              28.5 &              19.4 &             198.9 &              -8.6 &             204.3 &         1072.9626 &         4690.8086 &         1637.4780 &         1684.0397 &          1775.9514\\
\enddata
\end{deluxetable*}

\startlongtable
\begin{deluxetable*}{cccccccccccc}
\tablewidth{6.5in}
\tabletypesize{\tiny}
\tablecolumns{12}
\tablecaption{Brightness estimates from clear-filter  images of Pandora
\label{pandoratab}}
\tablehead{
\colhead{Filename} &  \colhead{Range} & \colhead{Phase} & \colhead{\tiny Sub-S/C} & \colhead{\tiny Sub-S/C} & \colhead{\tiny Sub-Solar} & \colhead{\tiny Sub-Solar} & \colhead{$A_{\rm eff}$} &\colhead{$A_{\rm phys}$} &  \colhead{$a_{\rm pred}$} & \colhead{$a_{\rm pred}$} & \colhead{$a_{\rm pred}$} \\
\colhead{} & \colhead{(km)} & \colhead{(deg)} & \colhead{\tiny Latitude} & \colhead{\tiny Longitude} & \colhead{\tiny Latitude} & \colhead{\tiny Longitude} & \colhead{Value} &  \colhead{} &
 \colhead{k=1} & \colhead{k=0.75} & \colhead{k=0.50} \\
 \colhead{} & \colhead{} & \colhead{} & \colhead{(deg)} & \colhead{(deg)} & \colhead{(deg)} & \colhead{(deg)} & \colhead{(km$^2$)}  & \colhead{(km$^2$)}    & \colhead{(km$^2$)} 
 & \colhead{(km$^2$)}   & \colhead{(km$^2$)} 
\\}
\startdata
      N1509277557 &          468022.3 &              22.9 &              -0.4 &             265.5 &             -19.9 &             253.3 &         2551.8445 &         5090.0981 &         3120.5850 &         3123.7810 &          3154.0962\\
      N1509278230 &          472646.5 &              22.6 &              -0.4 &             269.5 &             -19.9 &             257.8 &         2570.1541 &         5096.4727 &         3147.9783 &         3147.3135 &          3173.4172\\
      N1509279125 &          478756.8 &              22.3 &              -0.4 &             274.7 &             -19.9 &             263.7 &         2566.7014 &         5092.3550 &         3167.1782 &         3161.3831 &          3181.8542\\
      N1580567525 &         1640145.8 &              26.5 &              17.1 &             154.1 &              -8.6 &             160.8 &         1430.8479 &         4400.8467 &         2070.5557 &         2112.7646 &          2190.2957\\
      N1580765496 &         1904911.6 &              35.2 &              22.9 &              17.8 &              -8.5 &              34.0 &         1456.2356 &         4503.1094 &         1980.4275 &         2009.6714 &          2086.2158\\
      N1586165759 &         1350404.0 &              30.7 &              22.8 &             196.6 &              -7.6 &             192.5 &         1438.1946 &         4457.3887 &         1918.5052 &         1974.0736 &          2072.5266\\
      N1587847451 &         1348720.8 &              33.2 &              25.9 &             181.8 &              -7.3 &             180.1 &         1434.0127 &         4492.5977 &         1816.3823 &         1879.5070 &          1989.8180\\
      N1588712327 &         1356261.8 &              36.9 &              29.6 &             149.3 &              -7.1 &             152.9 &         1364.9861 &         4862.9883 &         1992.1320 &         2069.7925 &          2203.2756\\
\enddata
\end{deluxetable*}

\startlongtable
\begin{deluxetable*}{cccccccccccc}
\tablewidth{6.5in}
\tabletypesize{\tiny}
\tablecolumns{12}
\tablecaption{Brightness estimates from clear-filter  images of Janus.
\label{janustab}}
\tablehead{
\colhead{Filename} &  \colhead{Range} & \colhead{Phase} & \colhead{\tiny Sub-S/C} & \colhead{\tiny Sub-S/C} & \colhead{\tiny Sub-Solar} & \colhead{\tiny Sub-Solar} & \colhead{$A_{\rm eff}$} &\colhead{$A_{\rm phys}$} &  \colhead{$a_{\rm pred}$} & \colhead{$a_{\rm pred}$} & \colhead{$a_{\rm pred}$} \\
\colhead{} & \colhead{(km)} & \colhead{(deg)} & \colhead{\tiny Latitude} & \colhead{\tiny Longitude} & \colhead{\tiny Latitude} & \colhead{\tiny Longitude} & \colhead{Value} &  \colhead{} &
 \colhead{k=1} & \colhead{k=0.75} & \colhead{k=0.50} \\
 \colhead{} & \colhead{} & \colhead{} & \colhead{(deg)} & \colhead{(deg)} & \colhead{(deg)} & \colhead{(deg)} & \colhead{(km$^2$)}  & \colhead{(km$^2$)}    & \colhead{(km$^2$)} 
 & \colhead{(km$^2$)}   & \colhead{(km$^2$)} 
\\}
\startdata
    N1524964907 &          216295.5 &              32.9 &              -0.2 &             231.7 &             -17.7 &             260.0 &         7242.8188 &        23032.1738 &        13006.3779 &        12950.2812 &         13073.6650\\
      N1524965120 &          218050.4 &              33.1 &              -0.2 &             232.7 &             -17.7 &             261.3 &         7309.7822 &        23068.7754 &        13013.6660 &        12960.8555 &         13088.8672\\
      N1524965317 &          219704.9 &              33.2 &              -0.2 &             233.7 &             -17.7 &             262.4 &         7263.7207 &        23101.8184 &        13018.2275 &        12968.6035 &         13101.0693\\
      N1559172642 &         1524436.8 &              32.8 &               3.4 &             240.2 &             -12.3 &             269.3 &         7429.6016 &        23336.8145 &        13109.0371 &        13110.5742 &         13289.6104\\
      N1574754459 &         2196734.8 &              30.9 &               6.4 &             117.4 &              -9.6 &             143.9 &         5808.9624 &        23470.2305 &        12579.3428 &        12747.0762 &         13089.3350\\
      N1575011551 &         1781307.5 &              35.9 &               8.7 &             214.5 &              -9.5 &             245.6 &         6037.6162 &        22617.5801 &        12008.4717 &        11995.7139 &         12184.7939\\
      N1575756631 &         2123755.0 &              20.2 &               7.7 &              25.9 &              -9.4 &              36.7 &         7426.0449 &        22477.6230 &        13007.3262 &        13013.5850 &         13124.5449\\
      N1577098705 &         1741632.5 &              24.1 &              13.2 &             158.3 &              -9.2 &             167.3 &         5780.8052 &        22500.2676 &        12159.1025 &        12271.3994 &         12527.1123\\
\enddata
\end{deluxetable*}

\startlongtable
\begin{deluxetable*}{cccccccccccc}
\tablewidth{6.5in}
\tabletypesize{\tiny}
\tablecolumns{12}
\tablecaption{Brightness estimates from clear-filter  images of Epimetheus
\label{epimetheustab}}
\tablehead{
\colhead{Filename} &  \colhead{Range} & \colhead{Phase} & \colhead{\tiny Sub-S/C} & \colhead{\tiny Sub-S/C} & \colhead{\tiny Sub-Solar} & \colhead{\tiny Sub-Solar} & \colhead{$A_{\rm eff}$} &\colhead{$A_{\rm phys}$} &  \colhead{$a_{\rm pred}$} & \colhead{$a_{\rm pred}$} & \colhead{$a_{\rm pred}$} \\
\colhead{} & \colhead{(km)} & \colhead{(deg)} & \colhead{\tiny Latitude} & \colhead{\tiny Longitude} & \colhead{\tiny Latitude} & \colhead{\tiny Longitude} & \colhead{Value} &  \colhead{} &
 \colhead{k=1} & \colhead{k=0.75} & \colhead{k=0.50} \\
 \colhead{} & \colhead{} & \colhead{} & \colhead{(deg)} & \colhead{(deg)} & \colhead{(deg)} & \colhead{(deg)} & \colhead{(km$^2$)}  & \colhead{(km$^2$)}    & \colhead{(km$^2$)} 
 & \colhead{(km$^2$)}   & \colhead{(km$^2$)} 
\\}
\startdata
      N1564454340 &         3290726.2 &              35.7 &               0.2 &             102.2 &             -11.4 &             136.2 &         2377.4211 &        10843.6914 &         5896.0864 &         5991.3555 &          6178.0361\\
      N1575054751 &         1876191.9 &              36.5 &               8.1 &             305.2 &              -9.5 &             337.3 &         2441.1824 &        10606.0703 &         5452.1060 &         5553.2422 &          5752.8184\\
      N1575800491 &         1999183.2 &              23.9 &               9.4 &             116.1 &              -9.4 &             130.8 &         3085.0137 &        10715.6104 &         6366.2119 &         6423.5264 &          6536.9209\\
      N1577051695 &         1832468.5 &              21.1 &               9.9 &              62.7 &              -9.2 &              71.6 &         3373.8743 &        10739.0752 &         6656.8516 &         6666.2461 &          6722.8457\\
      N1577141905 &         1907909.6 &              24.0 &              14.3 &             248.9 &              -9.2 &             254.0 &         3560.7798 &        10821.0596 &         6570.8481 &         6597.0679 &          6682.7441\\
      N1578545720 &         1936707.5 &              26.9 &              16.3 &              20.4 &              -8.9 &              29.8 &         2513.5916 &        10150.9297 &         5696.2822 &         5726.6514 &          5825.5596\\
      N1579749784 &         1883149.0 &              37.0 &              23.5 &              30.9 &              -8.7 &              49.6 &         2117.9294 &        10441.0947 &         5433.1025 &         5473.6685 &          5617.6523\\
      N1580482099 &         1570688.6 &              21.3 &              12.1 &             115.5 &              -8.6 &             120.5 &         3110.6165 &        10744.9404 &         6555.0938 &         6591.6235 &          6677.4932\\
\enddata
\end{deluxetable*}

\startlongtable
\begin{deluxetable*}{cccccccccccc}
\tablewidth{6.5in}
\tabletypesize{\tiny}
\tablecolumns{12}
\tablecaption{Brightness estimates from clear-filter  images of Mimas.
\label{mimastab}}
\tablehead{
\colhead{Filename} &  \colhead{Range} & \colhead{Phase} & \colhead{\tiny Sub-S/C} & \colhead{\tiny Sub-S/C} & \colhead{\tiny Sub-Solar} & \colhead{\tiny Sub-Solar} & \colhead{$A_{\rm eff}$} &\colhead{$A_{\rm phys}$} &  \colhead{$a_{\rm pred}$} & \colhead{$a_{\rm pred}$} & \colhead{$a_{\rm pred}$} \\
\colhead{} & \colhead{(km)} & \colhead{(deg)} & \colhead{\tiny Latitude} & \colhead{\tiny Longitude} & \colhead{\tiny Latitude} & \colhead{\tiny Longitude} & \colhead{Value} &  \colhead{} &
 \colhead{k=1} & \colhead{k=0.75} & \colhead{k=0.50} \\
 \colhead{} & \colhead{} & \colhead{} & \colhead{(deg)} & \colhead{(deg)} & \colhead{(deg)} & \colhead{(deg)} & \colhead{(km$^2$)}  & \colhead{(km$^2$)}    & \colhead{(km$^2$)} 
 & \colhead{(km$^2$)}   & \colhead{(km$^2$)} 
\\}
\startdata
      N1481705814 &          901668.0 &              26.0 &               1.2 &             328.3 &             -23.1 &             318.9 &        58778.3984 &       120528.9844 &        72071.1719 &        72161.4062 &         72962.6797\\
      N1488814545 &         1607311.0 &              36.1 &              -0.0 &              92.4 &             -22.4 &              63.2 &        48915.5898 &       124417.2969 &        69472.8047 &        70052.2266 &         71725.5547\\
      N1488829675 &         1389412.9 &              38.3 &              -0.3 &             160.1 &             -22.4 &             127.9 &        43297.4219 &       118568.5938 &        64146.7969 &        64234.5859 &         65466.9531\\
      N1488881205 &         1509689.5 &              33.6 &               0.1 &              25.5 &             -22.4 &             359.8 &        47407.6680 &       119972.6641 &        66077.2109 &        66508.0078 &         67946.8750\\
      N1488897255 &         1261645.8 &              28.7 &              -0.1 &              87.2 &             -22.4 &              68.7 &        54479.7578 &       124463.9375 &        74076.8281 &        74477.5938 &         75695.7109\\
      N1488913065 &         1011333.3 &              30.2 &              -0.4 &             157.8 &             -22.4 &             136.5 &        51089.8398 &       118754.7812 &        68925.3750 &        68958.1172 &         69846.3750\\
      N1488923681 &          968930.9 &              34.8 &              -0.5 &             211.3 &             -22.4 &             183.4 &        47263.4258 &       119611.2031 &        64885.4414 &        65408.5273 &         66984.3828\\
      N1490685953 &         1217789.1 &              27.1 &              -0.1 &              73.3 &             -22.2 &              57.0 &        55973.9688 &       124187.1406 &        74193.6562 &        74603.1797 &         75775.0391\\
\enddata
\end{deluxetable*}

\startlongtable
\begin{deluxetable*}{cccccccccccc}
\tablewidth{6.5in}
\tabletypesize{\tiny}
\tablecolumns{12}
\tablecaption{Brightness estimates from clear-filter  images of Enceladus.
\label{enceladustab}}
\tablehead{
\colhead{Filename} &  \colhead{Range} & \colhead{Phase} & \colhead{\tiny Sub-S/C} & \colhead{\tiny Sub-S/C} & \colhead{\tiny Sub-Solar} & \colhead{\tiny Sub-Solar} & \colhead{$A_{\rm eff}$} &\colhead{$A_{\rm phys}$} &  \colhead{$a_{\rm pred}$} & \colhead{$a_{\rm pred}$} & \colhead{$a_{\rm pred}$} \\
\colhead{} & \colhead{(km)} & \colhead{(deg)} & \colhead{\tiny Latitude} & \colhead{\tiny Longitude} & \colhead{\tiny Latitude} & \colhead{\tiny Longitude} & \colhead{Value} &  \colhead{} &
 \colhead{k=1} & \colhead{k=0.75} & \colhead{k=0.50} \\
 \colhead{} & \colhead{} & \colhead{} & \colhead{(deg)} & \colhead{(deg)} & \colhead{(deg)} & \colhead{(deg)} & \colhead{(km$^2$)}  & \colhead{(km$^2$)}    & \colhead{(km$^2$)} 
 & \colhead{(km$^2$)}   & \colhead{(km$^2$)} 
\\}
\startdata
     N1477565247 &          765619.5 &              33.6 &               4.8 &             269.2 &             -23.6 &             250.8 &       129597.7109 &       200214.7812 &       114034.1094 &       114808.4453 &        117213.5234\\
      N1481509793 &         1980087.2 &              38.1 &              -3.6 &              37.2 &             -23.2 &               3.3 &       112414.9297 &       198977.5781 &       107244.1953 &       108142.8125 &        110943.3359\\
      N1481706474 &          671641.8 &              31.8 &               1.9 &             262.1 &             -23.2 &             242.0 &       134909.2031 &       200088.9375 &       115689.9297 &       116444.5000 &        118710.9219\\
      N1488878396 &         1267780.9 &              29.5 &              -0.2 &             107.2 &             -22.4 &              87.3 &       131631.8125 &       199811.5156 &       118290.2266 &       118853.9844 &        120768.7969\\
      N1490643112 &         1433963.0 &              30.3 &              -0.2 &              73.8 &             -22.2 &              52.4 &       131019.1172 &       200326.0312 &       117010.0703 &       117714.7188 &        119840.6875\\
      N1490687213 &          978323.2 &              38.4 &              -0.2 &             219.8 &             -22.2 &             187.5 &       113043.2344 &       197782.8125 &       105885.6484 &       106861.8672 &        109754.1641\\
      N1495039341 &         1616055.4 &              38.8 &             -20.1 &             140.7 &             -21.7 &              99.0 &       109264.4844 &       198342.5156 &       108068.6719 &       108744.9062 &        111325.3438\\
      N1499868413 &         1547410.6 &              29.0 &             -18.2 &              51.8 &             -21.1 &              21.1 &       133408.5625 &       200039.6406 &       117567.9141 &       118313.5156 &        120384.3125\\
\enddata
\end{deluxetable*}

\startlongtable
\begin{deluxetable*}{cccccccccccc}
\tablewidth{6.5in}
\tabletypesize{\tiny}
\tablecolumns{12}
\tablecaption{Brightness estimates from clear-filter  images of Tethys
\label{tethystab}}
\tablehead{
\colhead{Filename} &  \colhead{Range} & \colhead{Phase} & \colhead{\tiny Sub-S/C} & \colhead{\tiny Sub-S/C} & \colhead{\tiny Sub-Solar} & \colhead{\tiny Sub-Solar} & \colhead{$A_{\rm eff}$} &\colhead{$A_{\rm phys}$} &  \colhead{$a_{\rm pred}$} & \colhead{$a_{\rm pred}$} & \colhead{$a_{\rm pred}$} \\
\colhead{} & \colhead{(km)} & \colhead{(deg)} & \colhead{\tiny Latitude} & \colhead{\tiny Longitude} & \colhead{\tiny Latitude} & \colhead{\tiny Longitude} & \colhead{Value} &  \colhead{} &
 \colhead{k=1} & \colhead{k=0.75} & \colhead{k=0.50} \\
 \colhead{} & \colhead{} & \colhead{} & \colhead{(deg)} & \colhead{(deg)} & \colhead{(deg)} & \colhead{(deg)} & \colhead{(km$^2$)}  & \colhead{(km$^2$)}    & \colhead{(km$^2$)} 
 & \colhead{(km$^2$)}   & \colhead{(km$^2$)} 
\\}
\startdata
   N1477564227 &          668321.5 &              34.3 &               5.5 &             108.6 &             -23.6 &             127.2 &       467384.2500 &       888561.6875 &       501631.5938 &       505492.8750 &        516790.9688\\
      N1481443192 &         2092752.0 &              38.5 &              -5.2 &              82.3 &             -23.2 &              46.9 &       418260.5312 &       890977.5000 &       482936.9062 &       487384.6250 &        500435.1562\\
      N1488790300 &         1703386.8 &              35.0 &              -0.2 &              87.5 &             -22.4 &              59.7 &       453760.6875 &       890506.1250 &       500846.7188 &       504735.3750 &        516249.4062\\
      N1488792254 &         1675049.4 &              34.9 &              -0.2 &              91.7 &             -22.4 &              64.0 &       452431.1562 &       890194.3750 &       501912.9062 &       505747.7500 &        517133.7812\\
      N1488822445 &         1311927.4 &              38.4 &               0.0 &             162.6 &             -22.4 &             130.6 &       428045.9062 &       875014.5000 &       474645.6562 &       477596.4375 &        488958.7188\\
      N1488925841 &         1396850.2 &              28.3 &              -0.4 &              17.0 &             -22.4 &             358.8 &       497584.6562 &       882233.8125 &       521838.3750 &       524370.2500 &        532586.0625\\
      N1490557912 &         1967571.0 &              39.0 &              -0.3 &              33.5 &             -22.2 &               0.3 &       409800.2500 &       884963.8750 &       473073.0312 &       477229.9375 &        490086.0312\\
      N1493533631 &         1427707.1 &              28.7 &             -21.0 &             117.6 &             -21.9 &              86.8 &       514314.7812 &       887667.5625 &       530566.8750 &       532909.5000 &        540927.2500\\
\enddata
\end{deluxetable*}%

\end{document}